\documentclass[a4paper,accepted=2025-03-05]{quantumarticle}
\pdfoutput=1
\usepackage[utf8]{inputenc}
\usepackage[english]{babel}
\usepackage[T1]{fontenc}

\usepackage{amsmath}
\usepackage{amssymb}
\usepackage{graphicx}

\allowdisplaybreaks

\usepackage{hyperref}
\hypersetup{
 pdftitle={Alternatives of entanglement depth and metrological entanglement criteria},
 pdfauthor={Szilárd Szalay, Géza Tóth},
 pdfsubject={research article on multipartite entanglement and quantum metrology},
 pdfcreator={pdflatex}, pdfproducer={vim,overleaf},
 pdfkeywords={quantum entanglement} {multipartite} {mixed states} {entanglement depth} {entanglement depth of formation} {quantum metrology} {Fisher information},
 unicode=true, pdftoolbar=true, pdfmenubar=true, pdffitwindow=false, pdfstartview={FitH},
 pdfnewwindow=true, colorlinks=true, linktoc=page, linkcolor=blue, citecolor=blue, filecolor=blue, urlcolor=blue
}

\usepackage[numbers,sort&compress]{natbib}

\newcommand{\field}[1]{\mathbb{#1}}
\newcommand{\vs}[1]{\boldsymbol{#1}}
\newcommand{\vvs}[1]{\underline{\boldsymbol{#1}}}
\newcommand{\cket}[1]{\vert #1 \rangle}
\newcommand{\bra}[1]{\langle #1 \vert}

\newcommand{\proj}[1]{\cket{#1}\bra{#1}}

\newcommand{\ee}{\mathrm{e}}
\newcommand{\Id}{\mathrm{I}}
\newcommand{\cmpl}[1]{\overline{#1}}

\newcommand{\pconj}{\dagger}

\DeclareMathOperator{\Tr}{Tr}
\DeclareMathOperator{\Lin}{Lin}
\DeclareMathOperator{\Conv}{Conv}
\DeclareMathOperator{\Extr}{Extr}

\DeclareMathOperator{\Var}{Var}

\DeclareMathOperator{\var}{var}
\DeclareMathOperator{\avg}{avg}

\DeclareMathOperator{\Dim}{Dim}
\DeclareMathOperator{\DoF}{DoF}

\DeclareMathOperator{\downset}{\downarrow}
\DeclareMathOperator{\upset}{\uparrow}

\newcommand{\decomp}{\vdash}

\def\precdot{\mathrel{%
   \mathchoice{\PRECDOT}{\PRECDOT}{\scriptstyle\PRECDOT}{\scriptscriptstyle\PRECDOT}%
}}
\def\PRECDOT{{%
    \setbox0\hbox{$\prec$}%
    \rlap{\hbox to \wd0{\hss$\cdot$}}\box0
}}

\def\SUCCDOT{{%
    \setbox0\hbox{$\succ$}%
    \rlap{\hbox to \wd0{\hss$\cdot$}}\box0
}}

\newcommand{\finereq}{\preceq}
\newcommand{\nfinereq}{\npreceq}
\newcommand{\finer}{\prec}

\newcommand{\finerc}{\precdot}

\newcommand{\coarsereq}{\succeq}
\newcommand{\ncoarsereq}{\nsucceq}

\newcommand{\majordeq}{\leq}

\newcommand{\majordc}{\lessdot}

\newcommand{\majors}{>}

\newcommand{\majorsc}{\gtrdot}

\newcommand{\equals}{{\;\;=\;\;}}
\newcommand{\lequals}{{\;\;\leq\;\;}}
\newcommand{\gequals}{{\;\;\geq\;\;}}
\newcommand{\equalsref}[1]{\overset{\eqref{#1}}{\equals}}
\newcommand{\lequalsref}[1]{\overset{\eqref{#1}}{\lequals}}
\newcommand{\gequalsref}[1]{\overset{\eqref{#1}}{\lequals}}

\newcommand{\gtreqlesss}{{\;\;\gtreqless\;\;}}

\newcommand{\gtreqlesssref}[1]{\overset{\eqref{#1}}{\gtreqlesss}}

\newcommand{\LOCCw}{\;\underset{\text{LOCC}}{\longrightarrow}\;}

\newcommand{\dspiff}{\;\;\Longleftrightarrow\;\;}
\newcommand{\dspif}{\;\;\Longleftarrow\;\;}
\newcommand{\dspthen}{\;\;\Longrightarrow\;\;}

\newcommand{\dspdef}{\;\;\overset{\text{def.}}{\Longleftrightarrow}\;\;}

\newcommand{\dispt}[1]{\;\;\text{#1}\;\;}

\newcommand{\set}[1]{\{ #1 \}}
\newcommand{\bigset}[1]{\bigl\{ #1 \bigr\}}
\newcommand{\Bigset}[1]{\Bigl\{ #1 \Bigr\}}

\newcommand{\sset}[2]{\set{ #1 \;\vert\; #2 }}
\newcommand{\bigsset}[2]{\bigset{ #1 \;\big\vert\; #2 }}
\newcommand{\Bigsset}[2]{\Bigset{ #1 \;\Big\vert\; #2 }}

\newcommand{\mset}[1]{\set{#1}}

\newcommand{\smset}[2]{\mset{ #1 \;\vert\; #2 }}

\providecommand{\abs}[1]{{\lvert#1\rvert}}
\providecommand{\bigabs}[1]{{\big\lvert#1\big\rvert}}

\providecommand{\norm}[1]{{\lVert#1\rVert}}

\newcommand{\pinv}[1]{\hat{#1}}

\begin{document}
\title{Alternatives of entanglement depth and metrological entanglement criteria}
\author{Szilárd Szalay}
\email{szalay.szilard@wigner.hu}
\affiliation{HUN-REN Wigner Research Centre for Physics, P.O. Box 49, H-1525 Budapest, Hungary}
\affiliation{Department of Theoretical Physics, University of the Basque Country UPV/EHU,
P.O. Box 644, E-48080 Bilbao, Spain}
\affiliation{EHU Quantum Center, University of the Basque Country UPV/EHU,
Barrio Sarriena s/n, E-48940 Leioa, Biscay, Spain}
\author{Géza Tóth}
\email{toth@alumni.nd.edu}
\affiliation{Department of Theoretical Physics, University of the Basque Country UPV/EHU,
P.O. Box 644, E-48080 Bilbao, Spain}
\affiliation{EHU Quantum Center, University of the Basque Country UPV/EHU,
Barrio Sarriena s/n, E-48940 Leioa, Biscay, Spain}
\affiliation{IKERBASQUE, Basque Foundation for Science, E-48009 Bilbao, Spain}
\affiliation{Donostia International Physics Center DIPC,
Paseo Manuel de Lardizabal 4, E-20018 San Sebasti\'an, Spain}
\affiliation{HUN-REN Wigner Research Centre for Physics, P.O. Box 49, H-1525 Budapest, Hungary}

\date{April 17, 2025}

\begin{abstract}
We work out the general theory of 
one-pa\-rameter families of partial entanglement properties
and the resulting entanglement depth-like quantities.
Special cases of these are
the depth of partitionability,
the depth of producibility (or simply entanglement depth)
and the depth of stretchability,
which are based on one-parameter families of partial entanglement properties known earlier.
We also construct some further physically meaningful properties,
for instance the squareability, the toughness, the degree of freedom, and also several ones of entropic motivation.
Metrological multipartite entanglement criteria with the quantum Fisher information 
fit naturally into this framework. 
Here we formulate these for the depth of squareability, which therefore turns out to be the natural choice,
leading to stronger bounds than the usual entanglement depth. 
 Namely, the quantum Fisher information turns out to provide a lower bound not only on the maximal size of entangled subsystems,
 but also on the average size of entangled subsystems for a random choice of elementary subsystems.
We also formulate criteria with convex quantities for both cases, which are much stronger than the original ones.
In particular, the quantum Fisher information puts a lower bound on the average size of entangled subsystems.
We also argue that one-parameter partial entanglement properties, which carry entropic meaning,
are more suitable for the purpose of defining metrological bounds.
\end{abstract}

\maketitle{}

\newpage
\tableofcontents

\section{Introduction}
\label{sec:Intro}

One of the possible ways of 
characterizing the multipartite entanglement in an $n$-partite quantum system is \emph{producibility},
given by the size of the largest entangled subsystem~\cite{Sorensen-2001,Guhne-2005}.
A \emph{pure state} is $k$-producible, 
if the largest entangled subsystem is of size at most $k$,
that is,
if it can be given by a state vector of the form
\begin{equation}
\cket{\psi_1} \otimes \cket{\psi_2} \otimes \cdots \otimes \cket{\psi_m},
\end{equation}
where $\cket{\psi_i}$ are state vectors of the disjoint subsystems of size at most $k$~\cite{Sorensen-2001,Guhne-2005}.
A \emph{mixed state} $\rho$ is $k$-producible if it can be given as a mixture of pure $k$-producible states.
Note that a $k$-producible state is $k'$ producible for all $k'\geq k$, 
so the minimal value of $k$ for which the state is $k$-producible is the characteristic property of the state,
called the \emph{entanglement depth}, $D(\rho)$.
The entanglement depth can be detected by collective measurements~\cite{Sorensen-2001,Lucke-2014,Vitagliano-2017,Vitagliano-2018} or by correlation measurements~\cite{Guhne-2005}. If the density matrix is known in small systems, semidefinite programming can also be used~\cite{Jungnitsch-2011}. There is also a general theory that bounds the entanglement depth based on the expectation value of witness operators~\cite{Sun-2024}.

There have been many groundbreaking experiments
putting a lower bound on the entanglement depth of a quantum system,
aiming to produce larger and larger entanglement depth,
reaching the thousands~\cite{Esteve-2008,Gross-2010,Leroux-2010,Lucke-2014,Hosten-2016,McConnell-2015,Haas-2014,Zou-2018,Xin-2023}. Number-resolving detectors have recently been developed that will further increase the entanglement depth obtained in these systems~\cite{Hetzel-2023,Quensen-2025}.
At this point, an important question arises.
A system of $100$ particles of entanglement depth $D=20$ can be realized in various ways.
The two extreme cases are
when all $20$-particle groups are fully entangled,
given by the state vector
\begin{subequations}
\begin{equation}
\label{eq:state1}
\Bigl[\frac{1}{\sqrt 2}\bigl(\cket{0}^{\otimes 20}+\cket{1}^{\otimes 20}\bigr)\Bigr] ^{\otimes 5},
\end{equation}
and
when there is a single $20$-particle group that is fully entangled 
while the rest of the particles are fully separable,
given by the state vector
\begin{equation}
\label{eq:state2}
\Bigl[\frac{1}{\sqrt 2}\bigl(\cket{0}^{\otimes 20}+\cket{1}^{\otimes 20}\bigr)\Bigr]  \otimes \cket{0}^{\otimes 80}.
\end{equation}
\end{subequations}
Clearly, we would like to distinguish these two cases
and also \emph{many but not too many} further cases in-between,
by generalizing the concept of entanglement depth.

To handle such questions, we define
\emph{one-parameter partial entanglement properties},
which we call \emph{$f$-entanglement},
given by \emph{generator functions} $f$ (Section~\ref{sec:PSprops}).
The values of the generator functions, usually denoted with $k$, `parametrize' the classification.
We also introduce different entanglement measures, quantifying these properties,
the \emph{$(k,f)$-entanglement of formation} and
the \emph{relative entropy of $(k,f)$-entanglement} (Appendix~\ref{app:PSmeas}),
and
the \emph{$f$-entanglement depth} and
the \emph{$f$-entanglement depth of formation} (Section~\ref{sec:Depthmeas}).
The point here is that among the exponentially growing number of possibilities of partial entanglement~\cite{Szalay-2015b,Szalay-2017},
we single out some expressive ones,
which can be \emph{ordered} consecutively,
and the place in this ordering reflects the strength of that kind of entanglement.

For example, we already have the \emph{partitionability}, \emph{producibility} and \emph{stretchability of entanglement},
defined by the number of subsystems separable from each other, 
the size of the largest entangled subsystem,
and the difference of these two, respectively~\cite{Szalay-2019}.
Now we introduce among others the \emph{squareability of entanglement}, the \emph{entanglement toughness}
or the \emph{entanglement degree of freedom},
defined by the sum of squares of the sizes of entangled subsystems,
the size of the smallest entangled subsystem,
and the effective number of elementary subsystems needed for the description of the state, respectively (Section~\ref{sec:1param}).
All of these (except producibility) clearly distinguish the cases~\eqref{eq:state1} and~\eqref{eq:state2},
and they give rather different characterizations of the states.

A fundamental relation between quantum metrology and entanglement
is the metrological entanglement criterion~\cite{Pezze-2009}
\begin{equation}
\label{eq:teaser.FQ1prod}
F_\text{Q}(\rho,J^\text{z})/n \leq 1
\end{equation}
for all fully separable states,
where $F_\text{Q}$ is the \emph{quantum Fisher information},
characterizing metrological precision, hence playing a central role in quantum metrology~\cite{Pezze-2014,Toth-2014,Pezze-2018}.
$F_\text{Q}$ can be bounded from below based on measurements on the quantum state~\cite{Lucke-2011,Krischek-2011,Ockeloen-2013,Strobel-2014,Muessel-2014,Muessel-2015,Pezze-2016,Kruse-2016,Barontini-2015,Bohnet-2016}.
The criterion above gives a sufficient condition of entanglement,
that is, if the bound~\eqref{eq:teaser.FQ1prod} is violated then there must be some entanglement in the system.
Later this turned out to be a special case of the
metrological entanglement criterion~\cite{Hyllus-2012,Toth-2012,Pezze-2014,Toth-2014,Pezze-2018} 
\begin{subequations}
\label{eq:teaser}
\begin{equation}
\label{eq:teaser.FQD}
F_\text{Q}(\rho,J^\text{z})/n \leq D(\rho),
\end{equation}
which is now an inequality between two quantities, one from quantum entanglement theory, the other from quantum metrology, and it
gives a more detailed characterization of entanglement.
Here $D$ is the (producibility) \emph{entanglement depth}, as before.
A generalization of the bound~\eqref{eq:teaser.FQD} gives bounds on the possible values of
combinations of the depth of producibility and of partitionability,
and also of the depth of stretchability~\cite{Ren-2021},
by which the respective aspects of multipartite entanglement~\cite{Ren-2021} could be verified in experimental data~\cite{Ockeloen-2013,Strobel-2014,Muessel-2014,Muessel-2015,Pezze-2016,Kruse-2016,Barontini-2015,Bohnet-2016,Leroux-2010,Hosten-2016,Cox-2016,Bohnet-2014,Leibfried-2003,Sackett-2000,Monz-2011,Leibfried-2004,Leibfried-2005}.
The meaning of the bound~\eqref{eq:teaser.FQD} is that a low level of multipartite entanglement in the system restricts the metrological precision;
or, from the opposite point of view, a high precision indicates the presence of strong multipartite entanglement.

The bound~\eqref{eq:teaser.FQD} can be tight in some situations,
however, there are some relevant cases when it is very far from being tight.
For instance, let us consider a pure state of entanglement depth $D$
mixed with uncorrelated noise orthogonal to it.
For such a possibly weakly entangled state the quantum Fisher information is much smaller than $nD$.
However, the quantum Fisher information is the convex roof extension of the variance~\cite{Toth-2013,Yu-2013},
by which we derive the convex bound
\begin{equation}
\label{eq:teaser.FQDoF}
F_\text{Q}(\rho,J^\text{z})/n \leq D^\text{oF}(\rho)\leq D(\rho),
\end{equation}
\end{subequations}
which is \emph{much} stronger than the original bound~\eqref{eq:teaser.FQD}.
Here $D^\text{oF}(\rho)$ is the \emph{entanglement depth of formation},
which is defined as the \emph{convex roof extension} of the entanglement depth.
It is enlightening to formulate this in parallel to a similar formulation of the entanglement depth $D(\rho)$.
For pure states, both are just the entanglement depth, being the size of the largest entangled subsystem.
For mixed states, for any pure decomposition,
we form the \emph{maximum}, or the \emph{average} of the depth in the decomposition,
then $D(\rho)$ and $D^\text{oF}(\rho)$ are the minimum of these with respect to the decompositions in the two cases, respectively.
The bound~\eqref{eq:teaser.FQDoF} tells us that
not only the entanglement depth $D(\rho)$ is bounded from below, as in~\eqref{eq:teaser.FQD},
but also the \emph{average entanglement depth} for every possible decomposition $D^\text{oF}(\rho)$.
That is, if there are pure states of entanglement depth smaller than $F_\text{Q}/n$ in the mixture,
then there also has to be pure states of entanglement depth larger than $F_\text{Q}/n$,
with weight enough to compensate the lower depth states.

These metrological bounds fit naturally into the framework of one-parameter partial entanglement properties,
and the bounds~\eqref{eq:teaser} are also particular cases of our general results (Section~\ref{sec:Metro}).
First, we show in general that for states of any given partial separability property,
the direct bound on the quantum Fisher information is given simply by the squareability of that property (Section~\ref{sec:Metro.PS}).
This leads to an identity if that property itself is the squareability,
suggesting that entanglement squareability is the natural multipartite entanglement property
from the point of view, or for the purposes of quantum metrology.
For $f$-entanglement,
for all states,
the bound on the quantum Fisher information leads to a function of the $f$-entanglement depth,
which turns out to be the depth of an induced one-parameter property (Sections~\ref{sec:Metro.1param}-\ref{sec:Metro.1paramxmpl}).
This, again leads to an identity if that property itself is the squareability,
then the upper bound turns out to be simply the \emph{squareability entanglement depth} $D_\text{sq}(\rho)$,
which turns out to be stronger than~\eqref{eq:teaser.FQD},
\begin{subequations}
\label{eq:teaser2}
\begin{equation}
\label{eq:teaser2.FQDsq}
F_\text{Q}(\rho,J^\text{z}) \leq D_\text{sq}(\rho).
\end{equation}
Again, exploiting the key result 
that the quantum Fisher information is the convex roof extension of the variance~\cite{Toth-2013,Yu-2013},
the original bounds can be strengthened to convex bounds,
being the convex roof extensions of the original bounds (Sections~\ref{sec:Metro.1paramoF}-\ref{sec:Metro.1paramoFxmpl}).
For the case of squareability, the upper bound is the squareability depth of formation,
which turns out to be stronger than~\eqref{eq:teaser2.FQDsq},
\begin{equation}
\label{eq:teaser2.FQDsqoF}
F_\text{Q}(\rho,J^\text{z}) \leq D_\text{sq}^\text{oF}(\rho) \leq D_\text{sq}(\rho).
\end{equation}
\end{subequations}
From the general construction, applied to the case of the usual producibility and entanglement depth,
the bounds~\eqref{eq:teaser} in terms of usual depths
turn out to be weaker than the bounds~\eqref{eq:teaser2} in terms of squareability depths,
and we have
\begin{equation}
\label{eq:hipersuper}
\begin{array}{ccccc}
 & & nD^\text{oF}(\rho) & \leq & nD(\rho) \\[6pt]
 & & \rotatebox[origin=c]{90}{$\leq$} & & \rotatebox[origin=c]{90}{$\leq$} \\[6pt]
F_\text{Q}(\rho,J^\text{z}) & \leq & D^\text{oF}_\text{sq}(\rho) & \leq & D_\text{sq}(\rho) 
\end{array}
\end{equation}
altogether.
So the strongest bound is given in terms of the
squareability entanglement depth of formation $D^\text{oF}_\text{sq}(\rho)$,
which turns out to be $n$ times the \emph{average size of the entangled subsystems} (ASES),
which is the average size of the subsystem a randomly chosen particle belongs to,
where the particle is selected based on a uniform distribution 
(Section~\ref{sec:Metro.meaning}).

Beyond these, we also consider the question
which particular one-parameter partial entanglement properties are useful
for the formulation of the aforementioned metrological bounds (Section~\ref{sec:Metro.usefulness}),
or, which particular one-parameter partial entanglement properties can be detected by these metrological criteria.
This is motivated by the observation that, for instance,
entanglement toughness does not lead to meaningful metrological bounds (other than for biseparability),
while many other one-parameter properties do.
We formulate this question precisely, and argue for that
the \emph{dominance-monotonicity} is,
although not being necessary neither sufficient,
favorable to impose on the one-parameter property
to be metrologically useful (Section~\ref{sec:Metro.dommon}).
The dominance-monotone properties possess entropic meaning,
expressing the \emph{mixedness of the sizes of subsystems} separable from one another.

\section{Permutation invariant partial separability and one-parameter entanglement  properties}
\label{sec:PSprops}

In this section we quickly recall the structure 
of the permutation invariant partial entanglement properties~\cite{Szalay-2019};
then we work out the general description of the 
one-parameter entanglement properties, which we call $f$-entanglement.
Note that we use a simplified notation,
dropping the `-sep' from the subscript, 
$\mathcal{D}_{\pinv{\upsilon}}\equiv\mathcal{D}_{\pinv{\upsilon}\text{-sep}}$
used in the general case~\cite{Szalay-2019}
to distinguish correlation and entanglement.
This is because we do not consider correlations here~\cite{Szalay-2017},
although the construction could be carried out also for multipartite correlations.

\subsection{State spaces and classes}
\label{sec:PSprops.gen}

Let us have $n\geq2$ elementary subsystems,
and let $X\subseteq \set{1,2,\dots,n}$ and $\abs{X}$ denote an arbitrary \emph{subsystem} and its \emph{size}.
For the quantum theoretical description of each \emph{elementary subsystem} $l\in\set{1,2,\dots,n}$,
let us have the 
Hilbert space $\mathcal{H}_l$
of uniform dimension
$2\leq d:=\dim(\mathcal{H}_l)<\infty$,
then we have $\mathcal{H}_X=\bigotimes_{l\in X}\mathcal{H}_l$ for any subsystem $X$,
of dimension $\dim\bigl(\mathcal{H}_X\bigr)=d^{\abs{X}}$.
For the whole system, we use the notation 
$\mathcal{H}=\bigotimes_{l=1}^n\mathcal{H}_l$ for the Hilbert space, 
$\mathcal{P}=\sset{\proj{\psi}}{\cket{\psi}\in\mathcal{H},\norm{\psi}=1}$ for the space of pure states, and
$\mathcal{D}=\Conv(\mathcal{P})$ for the whole space of states (pure and mixed together),
which is the convex hull of pure states,
and we have $\mathcal{P}=\Extr(\mathcal{D})$, the set of extremal points of the state space $\mathcal{D}$.

\textit{On the first level}, we may form mixtures of states 
which are separable with respect to splits consisting of subsystems of (possibly different) fixed sizes.
Such properties are labeled by \emph{integer partitions} of $n$~\cite{Andrews-1984,Stanley-2012},
$\pinv{\xi}=\mset{x_1,x_2,\dots,x_{\abs{\pinv{\xi}}}}$,
which are multisets, containing the $x\in\pinv{\xi}$ parts $x\in \field{N}$, such that $\sum_{x\in\pinv{\xi}}x=n$.
$\abs{\pinv{\xi}}$ denotes the number of parts in $\pinv{\xi}$.
The set of the integer partitions is denoted with $\pinv{P}_\text{I}$.
(A multiset is a set allowing multiple instances of its elements.
It is the structure~\cite{Szalay-2019} naturally arising here:
the number of subsystem sizes varies, and the order of those is not relevant in this description, so $m$-tuples should not be used;
on the other hand, multiple subsystems of the same size arise, so sets could not be used either.
Note that we do not distinguish in the writing of sets and multisets,
sets of integers are always understood as multisets.)

The integer partition $\pinv{\upsilon}$ is called \emph{finer than or equal to} $\pinv{\xi}$,
denoted as $\pinv{\upsilon}\finereq\pinv{\xi}$,
if $\pinv{\xi}$ can be obtained as partial summations of $\pinv{\upsilon}$.
This is a proper partial order, called \emph{refinement}, 
derived from the refinement of set partitions~\cite{Szalay-2019}.
(For more details, see Appendix~\ref{app:1param.Gen},
for illustrations, see Figure~\ref{fig:PpI23456PPS}.)

A pure state is $\pinv{\xi}$-separable, 
if it is the projector to the product of state vectors of subsystems of sizes $x\in\pinv{\xi}$.
The space of those states is
\begin{subequations}
\label{eq:DPsepIp}
\begin{equation}
\label{eq:PsepIp}
\mathcal{P}_{\pinv{\xi}} := 
\Bigsset{\proj{\psi}\in\mathcal{P} }
{ \cket{\psi} = \bigotimes_{j=1}^\abs{\pinv{\xi}}\cket{\psi_{X_j}} },
\end{equation}
where $\cket{\psi_{X_j}}\in \mathcal{H}_{X_j}$ and $X_j$-s are disjoint subsystems of size $\abs{X_j}=x_j\in\pinv{\xi}$.
A general state is \emph{$\pinv{\xi}$-separable}, 
if it is the mixture of pure $\pinv{\xi}$-separable states.
The space of those states~\cite{Dur-1999,Dur-2000a,Acin-2001,Seevinck-2001,Seevinck-2008,Szalay-2012,Szalay-2015b,Szalay-2019} is
\begin{align}
\label{eq:DsepIp}
\mathcal{D}_{\pinv{\xi}} &:= \Conv \bigl(\mathcal{P}_{\pinv{\xi}}\bigr),
\intertext{and we have~\cite{Szalay-2019}}
\label{eq:PsepIpExtr}
\mathcal{P}_{\pinv{\xi}} &\phantom{:}= \Extr \bigl(\mathcal{D}_{\pinv{\xi}}\bigr).
\end{align}
\end{subequations}
The state space $\mathcal{D}_{\pinv{\xi}}$ is closed under LOCC.
Note that if a state is $\pinv{\upsilon}$-separable,
then it is also $\pinv{\xi}$-separable for all coarser $\pinv{\xi}$,
that is,
\begin{subequations}
\label{eq:oisomIp}
\begin{align}
\label{eq:oisomPIp}
\pinv{\upsilon}\finereq\pinv{\xi} &\dspiff \mathcal{P}_{\pinv{\upsilon}}\subseteq\mathcal{P}_{\pinv{\xi}},\\
\label{eq:oisomDIp}
\pinv{\upsilon}\finereq\pinv{\xi} &\dspiff \mathcal{D}_{\pinv{\upsilon}}\subseteq\mathcal{D}_{\pinv{\xi}},
\end{align}
\end{subequations}
so the refinement in $\pinv{P}_\text{I}$ encodes the inclusion hierarchy of the state spaces~\cite{Szalay-2019,Szalay-2015b}.

\textit{For example,} the pure states
given by state vectors~\eqref{eq:state1} and~\eqref{eq:state2}
are $\mset{20,20,20,20,20}$-separable, and the latter is additionally $\mset{20,1,1,\dots,1}$-separable.
On the other hand, the projectors onto the state vectors
$\frac{1}{\sqrt 2}\bigl(\cket{0}^{\otimes 100}+\cket{1}^{\otimes 100}\bigr)$
 and 
$\cket{0}^{\otimes 100}$
are $\top$-separable, and the latter is additionally $\bot$-separable,
given by the coarsest and the finest integer partitions 
\begin{subequations}
\label{eq:topbot}
\begin{align}
\label{eq:topbot.t}
\top&=\mset{n}\in\pinv{P}_\text{I},\\
\label{eq:topbot.b}
\bot&=\mset{1,1,\dots,1}\in\pinv{P}_\text{I},
\end{align}
\end{subequations}
labeling \emph{trivial separability} and \emph{full separability}, respectively.
Note also that these permutation invariant properties do not distinguish among,
e.g., pure states given by the state vectors
$\frac{1}{\sqrt 2}\bigl(\cket{000}+\cket{011}\bigr)$,
$\frac{1}{\sqrt 2}\bigl(\cket{000}+\cket{101}\bigr)$ and
$\frac{1}{\sqrt 2}\bigl(\cket{000}+\cket{110}\bigr)$,
all of these are $\mset{2,1}$-separable.

\textit{On the second level}, we would like to form mixtures of states 
of different, possibly incompatible $\pinv{\xi}$-separability in a general way.
This is a natural need, since properties like producibility cannot be given by a single $\pinv{\xi}$ in general.
For example, in the case of $n=4$,
both $\mset{3,1}$-separable and $\mset{2,2}$-separable states are needed to form $3$-producible states (see later),
while $\mset{3,1}\nfinereq\mset{2,2}$ and $\mset{3,1}\ncoarsereq\mset{2,2}$.
Exploiting~\eqref{eq:oisomIp}, 
such properties are labeled by \emph{down-sets} (also called order-ideals)~\cite{Szalay-2019,Davey-2002} of integer partitions,
which are sets
$\pinv{\vs{\xi}}=\set{\pinv{\xi}_1,\pinv{\xi}_2,\dots,\pinv{\xi}_{\abs{\pinv{\vs{\xi}}}}}$ closed downwards,
that is, if $\pinv{\xi}\in\pinv{\vs{\xi}}$ and $\pinv{\upsilon}\finereq\pinv{\xi}$, then $\pinv{\upsilon}\in\pinv{\vs{\xi}}$.
For example, in the case of $n=4$,
the down-set labeling $3$-producibility is $\set{\set{3,1},\set{2,2},\set{2,1,1},\set{1,1,1,1}}$
(see Figure~\ref{fig:PpI23456PPS}),
on the other hand, $\set{\set{3,1},\set{2,2}}$ is not a down-set,
since it is not closed downwards, it does not contain, e.g., $\set{2,1,1}\finereq\set{2,2}$.
The set of the integer partition down-sets is denoted with $\pinv{P}_\text{II}$.

The down-set $\pinv{\vs{\upsilon}}$ is called \emph{finer than or equal to} $\pinv{\vs{\xi}}$,
denoted as $\pinv{\vs{\upsilon}}\finereq \pinv{\vs{\xi}}$, if and only if $\pinv{\vs{\upsilon}}\subseteq \pinv{\vs{\xi}}$.
We call this partial order \emph{refinement} too~\cite{Szalay-2019}.
(For illustrations, see Figure~\ref{fig:PpI23456PPS}.)

A pure state is $\pinv{\vs{\xi}}$-separable,
if it is $\pinv{\xi}$-separable for a $\pinv{\xi}\in\pinv{\vs{\xi}}$.
The space of those states~\cite{Acin-2001,Seevinck-2008,Szalay-2015b,Szalay-2019} is
\begin{subequations}
\label{eq:DPsepIIp}
\begin{equation}
\label{eq:PsepIIp}
\mathcal{P}_{\pinv{\vs{\xi}}} := \bigcup_{\pinv{\xi}\in\pinv{\vs{\xi}}}\mathcal{P}_{\pinv{\xi}}.
\end{equation}
A general state is \emph{$\pinv{\vs{\xi}}$-separable}, 
if it is the mixture of pure $\pinv{\vs{\xi}}$-separable states,
that is,
being the mixture of $\pinv{\xi}$-separable states for (usually different) partitions $\pinv{\xi}\in\pinv{\vs{\xi}}$.
The space of those states is
\begin{align}
\label{eq:DsepIIp}
\mathcal{D}_{\pinv{\vs{\xi}}} &:=
\Conv\bigl(\mathcal{P}_{\pinv{\vs{\xi}}}\bigr)
=\Conv\Bigl(\bigcup_{\pinv{\xi}\in\pinv{\vs{\xi}}}\mathcal{D}_{\pinv{\xi}}\Bigr),
\intertext{and we have~\cite{Szalay-2019}}
\label{eq:PsepIIpExtr}
\mathcal{P}_{\pinv{\vs{\xi}}} &\phantom{:}= \Extr \bigl(\mathcal{D}_{\pinv{\vs{\xi}}}\bigr)
= \bigcup_{\pinv{\xi}\in\pinv{\vs{\xi}}}\Extr\bigl(\mathcal{D}_{\pinv{\xi}}\bigr).
\end{align}
\end{subequations}
The state space $\mathcal{D}_{\pinv{\vs{\xi}}}$ is closed under LOCC.
Note that if a state is $\pinv{\vs{\upsilon}}$-separable, 
then it is also $\pinv{\vs{\xi}}$-separable for all coarser $\pinv{\vs{\xi}}$,
that is,
\begin{subequations}
\label{eq:oisomIIp}
\begin{align}
\label{eq:oisomPIIp}
\pinv{\vs{\upsilon}}\finereq\pinv{\vs{\xi}} &\dspiff \mathcal{P}_{\pinv{\vs{\upsilon}}}\subseteq\mathcal{P}_{\pinv{\vs{\xi}}},\\
\label{eq:oisomDIIp}
\pinv{\vs{\upsilon}}\finereq\pinv{\vs{\xi}} &\dspiff \mathcal{D}_{\pinv{\vs{\upsilon}}}\subseteq\mathcal{D}_{\pinv{\vs{\xi}}},
\end{align}
\end{subequations}
so the refinement in $\pinv{P}_\text{II}$ encodes the inclusion hierarchy of the state spaces~\cite{Szalay-2019,Szalay-2015b}.

\emph{For example,}
the (principal) down-set $\set{\bot}$ and $\downset\set{\top}\equiv \pinv{P}_\text{I}$ 
label \emph{full separability} and \emph{trivial separability} again.
(Here $\downset\set{\dots}$ denotes the down-closure of the set $\set{\dots}$,
it contains all the elements finer than or equal to the elements of $\set{\dots}$,
so it is always a down-set.
For example, $\downset\set{\mset{2,2,1},\mset{3,1,1}}=\set{\mset{2,2,1},\mset{3,1,1},\mset{2,1,1,1},\mset{1,1,1,1,1}}$, see also Figure~\ref{fig:PpI23456PPS}.)
All the other partial separability properties lie between these two,
for example,
the down-set containing all the nontrivial partitions 
$\downset\sset{\mset{m,n-m}}{m=1,2,\dots \lfloor n/2\rfloor}\equiv \pinv{P}_\text{I}\setminus\set{\top}$
labels \emph{biseparability}.
(For further examples, see Section~\ref{sec:PSprops.1paramxpl}
and Figure~\ref{fig:PpI23456PPS}.)
In the subsequent sections we will consider
one-parameter families of partial separability properties,
which fit also in this second level.

\textit{On the third level}, we would like to describe states
having some partial separability properties,
and not having some other ones;
leading to a disjoint covering of the whole state space $\mathcal{D}$,
i.e., \emph{classification} of states~\cite{Szalay-2015b,Szalay-2018,Szalay-2019}.
First of all, to be able to formulate also coarser classifications,
we select the partial separability properties $\pinv{P}_\text{II*}\subseteq \pinv{P}_\text{II}$
with respect to which the classification is carried out.
Exploiting~\eqref{eq:oisomIIp},
such class properties are labeled by \emph{up-sets} (also called order-filters)~\cite{Szalay-2019,Davey-2002} of integer partition down-sets,
which are sets
$\pinv{\vvs{\xi}}=\set{\pinv{\vs{\xi}}_1,\pinv{\vs{\xi}}_2,\dots,\pinv{\vs{\xi}}_{\abs{\pinv{\vvs{\xi}}}}}\subseteq\pinv{P}_\text{II*}$ closed upwards,
that is, if $\pinv{\vs{\xi}}\in\pinv{\vvs{\xi}}$ 
and $\pinv{\vs{\upsilon}}\in\pinv{P}_\text{II*}$ such that $\pinv{\vs{\upsilon}}\coarsereq\pinv{\vs{\xi}}$, 
then $\pinv{\vs{\upsilon}}\in\pinv{\vvs{\xi}}$.
The set of the up-sets of integer partition down-sets is denoted with $\pinv{P}_\text{III*}$.

The up-set $\pinv{\vvs{\upsilon}}$ is called \emph{coarser than or equal to} $\pinv{\vvs{\xi}}$,
denoted as $\pinv{\vvs{\upsilon}}\finereq \pinv{\vvs{\xi}}$, if and only if $\pinv{\vvs{\upsilon}}\subseteq \pinv{\vvs{\xi}}$.
We call this \emph{refinement} too~\cite{Szalay-2019}.

A general state is \emph{strictly $\pinv{\vvs{\xi}}$-separable}, if it is
separable with respect to all $\pinv{\vs{\xi}} \in\pinv{\vvs{\xi}}$,
and entangled with respect to all $\pinv{\vs{\xi}}\in\cmpl{\pinv{\vvs{\xi}}}= \pinv{P}_\text{II*} \setminus \pinv{\vvs{\xi}}$.
The \emph{class} of those states is
\begin{equation}
\label{eq:CsepIIIp}
\mathcal{C}_{\pinv{\vvs{\xi}}} :=
 \Bigl(\bigcap_{\pinv{\vs{\xi}}'\in\cmpl{\pinv{\vvs{\xi}}}} \cmpl{\mathcal{D}_{\pinv{\vs{\xi}}'}}\Bigr) \cap 
 \Bigl(\bigcap_{\pinv{\vs{\xi}} \in      \pinv{\vvs{\xi}} }       \mathcal{D}_{\pinv{\vs{\xi}} }\Bigr)
=\Bigl(\bigcap_{\pinv{\vs{\xi}} \in      \pinv{\vvs{\xi}} }       \mathcal{D}_{\pinv{\vs{\xi}} }\Bigr)
 \setminus\Bigl(\bigcup_{\pinv{\vs{\xi}}'\in\cmpl{\pinv{\vvs{\xi}}}} {\mathcal{D}_{\pinv{\vs{\xi}}'}}\Bigr).
\end{equation}
(This gives the labeling scheme for the possible intersections of the state spaces $\mathcal{D}_{\pinv{\vs{\xi}}}$,
see Appendix A of~\cite{Szalay-2018}.)
Because of the LOCC-closedness of the state spaces $\mathcal{D}_{\pinv{\vs{\xi}}}$,
the refinement gives necessary condition for LOCC convertibility~\cite{Szalay-2015b,Szalay-2019}
as
\begin{equation}
\label{eq:ohomomCIIIp}
\mathcal{C}_{\pinv{\vvs{\upsilon}}} \LOCCw \mathcal{C}_{\pinv{\vvs{\xi}}}
\dspthen \pinv{\vvs{\upsilon}}\finereq\pinv{\vvs{\xi}},
\end{equation}
where the left-hand side is a shorthand notation for that
there exist an LOCC map $\Lambda:\mathcal{D}\to\mathcal{D}$ and a state $\rho \in \mathcal{C}_{\pinv{\vvs{\upsilon}}}$ 
such that $\Lambda(\rho)\in \mathcal{C}_{\pinv{\vvs{\xi}}}$.
That is, LOCC map can bring to a class only if it is `finer-than-or-equally-fine' with respect to partial separability.

\emph{Examples} are given in~\cite{Szalay-2019} for the general case,
and will be shown in Sections~\ref{sec:PSprops.1paramxpl} and~\ref{sec:PSprops.1param} for the one-parameter case.
(We will also use the notation $\upset\set{\dots}$, which is the up-closure of the set $\set{\dots}$,
it contains all the elements coarser than or equal to the elements of $\set{\dots}$,
so it is always an up-set.)

\subsection{Examples of \texorpdfstring{$f$}{f}-entanglement properties}
\label{sec:PSprops.1paramxpl}

\begin{figure*}\centering
\includegraphics{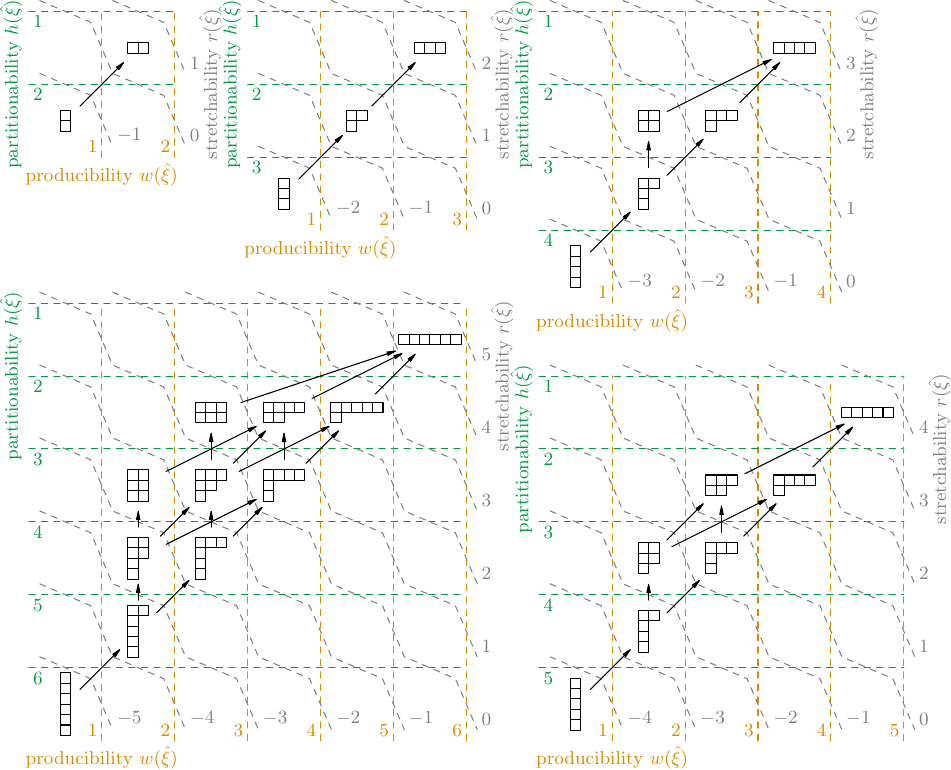}
\caption{The down-sets~\eqref{eq:vxik} corresponding to partitionability, producibility and stretchability,
illustrated on the posets $\pinv{P}_\text{I}$ for $n=2,3,4,5,6$.
(The integer partitions $\pinv{\xi}\in\pinv{P}_\text{I}$ are represented by their Young diagrams,
and the refinement order $\finereq$ is denoted by \emph{consecutive} arrows,
which are pointing from the finer partition to the coarser one.
The refinement is a partial order, that is, not every pair of partitions are connected by arrows.
The integer partitions describing trivial separability~\eqref{eq:topbot.t} and full separability~\eqref{eq:topbot.b}
are the $\top$ and $\bot$ elements of the poset $\pinv{P}_\text{I}$, drawn in the upper right and lower left corners of these plots, respectively.
The three kinds of down-sets~\eqref{eq:vxik} contain the Young diagrams below the green, to the left from the yellow,
and below the gray dashed lines.
These are down-sets, that is, the borders of any of them are crossed by arrows in only one direction.
These three kinds of down-sets~\eqref{eq:vxik} form chains~\eqref{eq:vximonk} for the indexing $k$,
that is, the dashed lines (borders) of the same color do not cross one another.)}
\label{fig:PpI23456PPS}
\end{figure*}

Here we recall 
three notable one-parameter families of properties~\cite{Szalay-2015b,Szalay-2017,Szalay-2019},
the \emph{$k$-partitionability} (for $k\in[1,n]_\field{Z}$),
the \emph{$k$-producibility} (for $k\in[1,n]_\field{Z}$)
and the \emph{$k$-stretchability} (for $k\in[-(n-1),n-1]_\field{Z}\setminus\set{\pm(n-2)}$),
serving as examples for both the general construction above,
and also for the one-parameter families of properties in the subsequent sections.
These properties are defined by the down-sets of integer partitions,
where the number of parts is at least $k$,
\begin{subequations}
\label{eq:vxik}
\begin{align}
\label{eq:vxik.part}
\pinv{\vs{\xi}}_{k\text{-part}} &:= \bigsset{\pinv{\xi}\in \pinv{P}_\text{I}}{\abs{\pinv{\xi}}\geq k},
\intertext{the size of the largest part is at most $k$,}
\label{eq:vxik.prod}
\pinv{\vs{\xi}}_{k\text{-prod}} &:= \bigsset{\pinv{\xi}\in \pinv{P}_\text{I}}{\max(\pinv{\xi})\leq k},
\intertext{and the difference of these is at most $k$,}
\label{eq:vxik.str}
\pinv{\vs{\xi}}_{k\text{-str}}  &:= \bigsset{\pinv{\xi}\in \pinv{P}_\text{I}}{\max(\pinv{\xi})-\abs{\pinv{\xi}}\leq k},
\end{align}
\end{subequations}
respectively.
These are all down-sets
(this is illustrative to derive directly,
although it also follows from the monotonicity of the generator functions~\eqref{eq:hwr} in the general construction later),
and these form chains, that is,
\begin{subequations}
\label{eq:vximonk}
\begin{align}
\label{eq:vximonk.part}
k\geq k' &\dspiff \pinv{\vs{\xi}}_{k\text{-part}} \finereq \pinv{\vs{\xi}}_{k'\text{-part}},\\
\label{eq:vximonk.prod}
k\leq k' &\dspiff \pinv{\vs{\xi}}_{k\text{-prod}} \finereq \pinv{\vs{\xi}}_{k'\text{-prod}},\\
\label{eq:vximonk.str}
k\leq k' &\dspiff \pinv{\vs{\xi}}_{k\text{-str}}  \finereq \pinv{\vs{\xi}}_{k'\text{-str}}.
\end{align}
\end{subequations}
(For illustrations, see Figure~\ref{fig:PpI23456PPS},
which also explains why the values $\pm(n-2)$ are missing from the possible values of $k$
in the case of stretchability:
all the partitions differs from $\top$ and $\bot$
both in the number of parts \emph{and} the size of the largest part.)
These lead to the 
\emph{$k$-partitionably separable} (also called $k$-partitionable, or $k$-separable 
\cite{Acin-2001,Seevinck-2008,Szalay-2015b,Szalay-2017,Szalay-2019}) states, 
which can be mixed by using states which can be separated into at least $k$ subsystems;
\emph{$k$-producibly separable} (also called $k$-producible 
\cite{Seevinck-2001,Guhne-2005,Guhne-2006,Toth-2010,Szalay-2015b,Szalay-2017,Szalay-2019}) states,
which can be mixed by using entanglement inside subsystems of size at most $k$;
and \emph{$k$-stretchably separable} states~\cite{Szalay-2019},
which can be mixed by using entanglement of Dyson-rank at most $k$;
the spaces of which are
\begin{subequations}
\label{eq:DsepIIpk}
\begin{align}
\label{eq:DsepIIpk.part}
\mathcal{D}_{k\text{-part}} &:= \mathcal{D}_{\pinv{\vs{\xi}}_{k\text{-part}}},\\
\label{eq:DsepIIpk.prod}
\mathcal{D}_{k\text{-prod}} &:= \mathcal{D}_{\pinv{\vs{\xi}}_{k\text{-prod}}},\\
\label{eq:DsepIIpk.str}
\mathcal{D}_{k\text{-str}}  &:= \mathcal{D}_{\pinv{\vs{\xi}}_{k\text{-str}}},
\end{align}
\end{subequations}
closed under LOCC,
forming nested subsets
\begin{subequations}
\label{eq:DsepIIpmonk}
\begin{align}
\label{eq:DsepIIpmonk.part}
k\geq k' &\dspiff \mathcal{D}_{k\text{-part}} \subseteq \mathcal{D}_{k'\text{-part}}, \\
\label{eq:DsepIIpmonk.prod}
k\leq k' &\dspiff \mathcal{D}_{k\text{-prod}} \subseteq \mathcal{D}_{k'\text{-prod}}, \\
\label{eq:DsepIIpmonk.str}
k\leq k' &\dspiff \mathcal{D}_{k\text{-str}}  \subseteq \mathcal{D}_{k'\text{-str}},
\end{align}
\end{subequations}
by~\eqref{eq:oisomDIIp} and~\eqref{eq:vximonk}.
We have 
$\mathcal{D}_{n\text{-part}}=\mathcal{D}_{1\text{-prod}}=\mathcal{D}_{-(n-1)\text{-str}}$, containing the fully separable states,
$\mathcal{D}_{2\text{-part}}=\mathcal{D}_{(n-1)\text{-prod}}=\mathcal{D}_{(n-3)\text{-str}}$, containing the biseparable states,
$\mathcal{D}_{1\text{-part}}=\mathcal{D}_{n\text{-prod}}=\mathcal{D}_{(n-1)\text{-str}}=\mathcal{D}$, containing all the states.
There are some further coincidences among the partitionability, producibility and stretchability properties,
as can be read off from Figure~\ref{fig:PpI23456PPS},
but these three one-parameter properties are different for the most values of $k$.

The strict partitionability, producibility and stretchability properties are
given by the up-sets
\begin{subequations}
\label{eq:vvxik}
\begin{align}
\label{eq:vvxik.part}
\pinv{\vvs{\xi}}_{k\text{-part}} &:= \upset\bigset{\pinv{\vs{\xi}}_{k\text{-part}}}
= \bigsset{\pinv{\vs{\xi}}_{k'\text{-part}}}{k\geq k'},\\
\label{eq:vvxik.prod}
\pinv{\vvs{\xi}}_{k\text{-prod}} &:= \upset\bigset{\pinv{\vs{\xi}}_{k\text{-prod}}}
= \bigsset{\pinv{\vs{\xi}}_{k'\text{-prod}}}{k\leq k'},\\
\label{eq:vvxik.str}
\pinv{\vvs{\xi}}_{k\text{-str}}  &:= \upset\bigset{\pinv{\vs{\xi}}_{k\text{-str}}}
= \bigsset{\pinv{\vs{\xi}}_{k'\text{-str}}}{k\leq k'},
\end{align}
\end{subequations}
by~\eqref{eq:vximonk},
forming chains,
\begin{subequations}
\label{eq:vvximonk}
\begin{align}
\label{eq:vvximonk.part}
k\leq k' &\dspiff \pinv{\vvs{\xi}}_{k\text{-part}} \finereq \pinv{\vvs{\xi}}_{k'\text{-part}},\\
\label{eq:vvximonk.prod}
k\geq k' &\dspiff \pinv{\vvs{\xi}}_{k\text{-prod}} \finereq \pinv{\vvs{\xi}}_{k'\text{-prod}},\\
\label{eq:vvximonk.str}
k\geq k' &\dspiff \pinv{\vvs{\xi}}_{k\text{-str}}  \finereq \pinv{\vvs{\xi}}_{k'\text{-str}}.
\end{align}
\end{subequations}
These lead to
the \emph{strictly $k$-partitionably separable states},
the \emph{strictly $k$-producibly separable states}
and the \emph{strictly $k$-stretchably separable states}~\cite{Szalay-2019},
the classes~\eqref{eq:CsepIIIp} of which are
\begin{subequations}
\label{eq:CsepIIIpk}
\begin{align}
\begin{split}
&\mathcal{C}_{k\text{-part}} :=
\mathcal{C}_{\pinv{\vvs{\xi}}_{k\text{-part}}} \\
&\quad = \begin{cases}
\mathcal{D}_{n\text{-part}}& \text{for $k=n$},\\
\mathcal{D}_{k\text{-part}}\setminus\mathcal{D}_{(k+1)\text{-part}}& \text{else},
\end{cases}
\end{split}\\
\begin{split}
&\mathcal{C}_{k\text{-prod}} :=
\mathcal{C}_{\pinv{\vvs{\xi}}_{k\text{-prod}}} \\
&\quad = \begin{cases}
\mathcal{D}_{1\text{-prod}}& \text{for $k=1$},\\
\mathcal{D}_{k\text{-prod}}\setminus\mathcal{D}_{(k-1)\text{-prod}}& \text{else},
\end{cases}
\end{split}\\
\begin{split}
&\mathcal{C}_{k\text{-str}} :=
\mathcal{C}_{\pinv{\vvs{\xi}}_{k\text{-str}}} \\
&\quad = \begin{cases}
\mathcal{D}_{-(n-1)\text{-str}}& \text{for $k=-(n-1)$},\\
\mathcal{D}_{-(n-3)\text{-str}}\setminus\mathcal{D}_{-(n-1)\text{-str}}& \text{for $k=-(n-3)$},\\
\mathcal{D}_{ (n-1)\text{-str}}\setminus\mathcal{D}_{ (n-3)\text{-str}}& \text{for $k= (n-1)$},\\
\mathcal{D}_{k\text{-str}}\setminus\mathcal{D}_{(k-1)\text{-str}}& \text{else},
\end{cases}
\end{split}
\end{align}
\end{subequations}
having a simple structure with respect to LOCC,
\begin{subequations}
\label{eq:CsepIIIpmonk}
\begin{align}
\mathcal{C}_{k\text{-part}} \LOCCw \mathcal{C}_{k'\text{-part}} &\dspthen k \leq k',\\
\mathcal{C}_{k\text{-prod}} \LOCCw \mathcal{C}_{k'\text{-prod}} &\dspthen k \geq k',\\
\mathcal{C}_{k\text{-str}}  \LOCCw \mathcal{C}_{k'\text{-str}}  &\dspthen k \geq k',
\end{align}
\end{subequations}
by~\eqref{eq:vvximonk} and~\eqref{eq:ohomomCIIIp}.
The elements of the class $\mathcal{C}_{k\text{-prod}}$,
the strictly $k$-producibly separable states
($k$-producibly separable 
but not $(k-1)$-producibly separable)
are also called states of \emph{entanglement depth} $k$~\cite{Sorensen-2001,Lucke-2014,Chen-2016}.
We have 
$\mathcal{C}_{n\text{-part}}=\mathcal{C}_{1\text{-prod}}=\mathcal{C}_{-(n-1)\text{-str}}$, containing the fully separable states,
$\mathcal{C}_{2\text{-part}}=\mathcal{C}_{(n-1)\text{-prod}}=\mathcal{C}_{(n-3)\text{-str}}$, containing the strictly biseparable states,
$\mathcal{C}_{1\text{-part}}=\mathcal{C}_{n\text{-prod}}=\mathcal{C}_{(n-1)\text{-str}}$, containing the genuinely multipartite entangled states.
There are some further coincidences among the strict partitionability, producibility and stretchability properties,
as can be read off from Figure~\ref{fig:PpI23456PPS},
but these three strict one-parameter properties are different for the most values of $k$.

\subsection{\texorpdfstring{$f$}{f}-entanglement properties in general}
\label{sec:PSprops.1param}

\textit{In general}, one-parameter partial separability properties, like 
partitionability, producibility and stretchability,
can be given by sub- or super-level sets of monotone functions over the integer partitions.
Let a \emph{generator function} 
be a function assigning a real number to all the integer partitions
\begin{subequations}
\label{eq:genf}
\begin{equation}
\label{eq:genf.f}
f:\pinv{P}_\text{I}\longrightarrow \field{R},
\end{equation}
where $\pinv{P}_\text{I}$ is the set of integer partitions.
The function $f$ is either increasing or decreasing monotone,
\begin{equation}
\label{eq:genf.mon}
\pinv{\upsilon}\finereq\pinv{\xi} \dspthen f(\pinv{\upsilon})\lesseqgtr f(\pinv{\xi}).
\end{equation}
\end{subequations}
Although $f$ is decreasing if and only if $-f$ is increasing,
we keep using both kinds of generator functions,
since both arise naturally in the construction.
For example, for partitionability, producibility and stretchability~\eqref{eq:vxik},
the generator functions are 
\begin{subequations}
\label{eq:hwr}
\begin{align}
\label{eq:hwr.h}
h(\pinv{\xi})&:=\abs{\pinv{\xi}},\\
\label{eq:hwr.w}
w(\pinv{\xi})&:=\max(\pinv{\xi}),\\ 
\label{eq:hwr.r}
r(\pinv{\xi})&:=\max(\pinv{\xi})-\abs{\pinv{\xi}},
\end{align}
\end{subequations}
respectively,
the \emph{height} (decreasing), \emph{width} (increasing) and \emph{(Dyson-) rank} (increasing)
of the Young diagram representing the partition~\cite{Szalay-2019}.
In the following, like in~\eqref{eq:genf.mon},
the upper and lower relation signs are always understood
for the increasing and decreasing cases, respectively.
The values of the generator functions, usually denoted with $k\in f(\pinv{P}_\text{I})=\sset{f(\pinv{\xi})}{\pinv{\xi}\in\pinv{P}_\text{I}}$,
will then `parametrize' the one-parameter classification.
Note that these values are not integers in general, contrary to our main examples~\eqref{eq:hwr}.

Note that, if a generator function $f:\pinv{P}_\text{I}\to\field{R}$ 
is composed with a monotone function $g:\field{R}\to\field{R}$,
the resulting function $g\circ f$ is also a generator function~\eqref{eq:genf},
the monotonicity of which is flipped if $g$ is decreasing.

Because of the monotonicity~\eqref{eq:genf.mon},
the extremal values of $f$ are $f(\bot)$ and $f(\top)$,
that is, 
\begin{equation}
\label{eq:frange}
f(\bot)\lesseqgtr f(\pinv{\xi})\lesseqgtr f(\top).
\end{equation}
Note that, as also in our main examples~\eqref{eq:hwr},
the function $f$ is not injective in general, that is,
$f$ may take the same value for different partitions,
and $\abs{f(\pinv{P}_\text{I})}\leq\abs{\pinv{P}_\text{I}}$.
For later use, we extend $f$ to $\pinv{P}_\text{II}$ as
\begin{subequations}
\begin{equation}
\label{eq:fII}
f(\pinv{\vs{\xi}}):=\begin{cases}
\max_{\pinv{\xi}\in\pinv{\vs{\xi}}} f(\pinv{\xi})& \text{if $f$ is increasing},\\
\min_{\pinv{\xi}\in\pinv{\vs{\xi}}} f(\pinv{\xi})& \text{if $f$ is decreasing}.
\end{cases}
\end{equation}
This is then monotone for the refinement of the second level of the construction,
increasing/decreasing if the original $f$ is increasing/decreasing for the refinement of the first level~\eqref{eq:genf.mon},
\begin{equation}
\label{eq:genf.monII}
\pinv{\vs{\upsilon}}\finereq\pinv{\vs{\xi}} \dspthen f(\pinv{\vs{\upsilon}})\lesseqgtr f(\pinv{\vs{\xi}}),
\end{equation}
\end{subequations}
which is easy to check by definition~\eqref{eq:fII}.
This is an extension in the sense that $f(\downset\set{\pinv{\xi}}) = f(\pinv{\xi})$,
by~\eqref{eq:genf}.

Also for later use, let us have the notation for the level sets (inverse image)
and sub/super-level sets
of the generator function $f:\pinv{P}_\text{I}\to\field{R}$
for $k\in f(\pinv{P}_\text{I})$,
\begin{subequations}
\label{eq:fset}
\begin{align}
\label{eq:fset.level}
f^{-1}(k) &:= \bigsset{\pinv{\xi}\in\pinv{P}_\text{I}}{f(\pinv{\xi})=k},\\
\label{eq:fset.sslevel}
f^\lesseqgtr(k) &:= \bigsset{\pinv{\xi}\in\pinv{P}_\text{I}}{f(\pinv{\xi})\lesseqgtr k}.
\end{align}
\end{subequations}
For any generator function $f$,
for any given $k\in f(\pinv{P}_\text{I})$,
the monotonicity~\eqref{eq:genf.mon} leads to that
\begin{equation}
\label{eq:vxir}
\pinv{\vs{\xi}}_{k,f} := f^\lesseqgtr(k) \equiv \bigsset{\pinv{\xi}\in\pinv{P}_\text{I}}{f(\pinv{\xi})\lesseqgtr k}
\end{equation}
is a nonempty down-set, $\pinv{\vs{\xi}}_{k,f}\in\pinv{P}_\text{II}$,
describing a \emph{one-parameter partial separability property},
which we call \emph{$(k,f)$-separability}.
These form chains in $\pinv{P}_\text{II}$,
\begin{equation}
\label{eq:vxirchain}
k\lesseqgtr k' \dspiff \pinv{\vs{\xi}}_{k,f} \finereq \pinv{\vs{\xi}}_{k',f},
\end{equation}
by the definition~\eqref{eq:vxir}.
Let us denote these chains as 
$\pinv{P}_{\text{II},f}:=\bigsset{\pinv{\vs{\xi}}_{k,f}}{k\in f(\pinv{P}_\text{I})}\subseteq\pinv{P}_\text{II}$.
We also have
\begin{equation}
\label{eq:fIIkf}
f(\pinv{\vs{\xi}}_{k,f}) = k
\end{equation}
simply by~\eqref{eq:fII} and~\eqref{eq:vxir}.

\begin{figure}\centering
\includegraphics{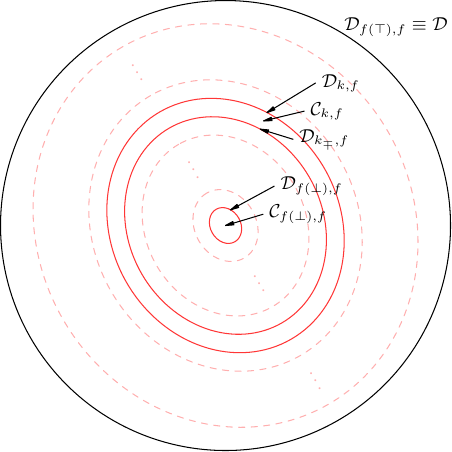}
\caption{The nested state spaces $\mathcal{D}_{k,f}$ given in~\eqref{eq:DsepIIpf}
and disjoint classes $\mathcal{C}_{k,f}$ given in~\eqref{eq:CsepIIIpf} as differences of the state spaces.
The numbering $k\in f(\pinv{P}_\text{I})$ increases/decreases outwards
in the case of increasing/decreasing generator function~\eqref{eq:genf}.
LOCC maps cannot bring outwards~\eqref{eq:CsepIIIpfmonk}.
Different generator functions lead to different such onion like structures,
intersecting nontrivially in general.}
\label{fig:nested}
\end{figure}

In this general one-parameter case,
for the generator function $f$ and the value $k\in f(\pinv{P}_\text{I})$,
let the \emph{$(k,f)$-separable states}, or
`states of $f$-separability $k$',
be those given by the nonempty down-set $\pinv{\vs{\xi}}_{k,f}\in\pinv{P}_{\text{II},f}\subseteq\pinv{P}_\text{II}$
in the general construction~\eqref{eq:DPsepIIp}.
These are the mixtures of pure $\pinv{\xi}$-separable states for which $f(\pinv{\xi})\lesseqgtr k$,
the space of which is 
\begin{subequations}
\label{eq:DPsepIIpf}
\begin{align}
\label{eq:DsepIIpf}
\mathcal{D}_{k,f} &:=
\mathcal{D}_{ \pinv{\vs{\xi}}_{k,f}} 
\equiv \Conv \bigl(\mathcal{P}_{k,f}\bigr)
\equiv \Conv \bigl(\mathcal{P}_{ \pinv{\vs{\xi}}_{k,f}}\bigr),
\intertext{and we have}
\label{eq:PsepIIpfExtr}
\mathcal{P}_{k,f} &:=
\mathcal{P}_{ \pinv{\vs{\xi}}_{k,f}} 
\equiv \Extr \bigl(\mathcal{D}_{k,f}\bigr)
\equiv \Extr \bigl(\mathcal{D}_{ \pinv{\vs{\xi}}_{k,f}}\bigr),
\end{align}
where
\begin{equation}
\label{eq:PsepIIpf}
\mathcal{P}_{k,f} 
\equiv \bigcup_{\substack{\pinv{\xi}\in\pinv{P}_\text{I}:\\ f(\pinv{\xi})\lesseqgtr k}} \mathcal{P}_{\pinv{\xi}}
\equiv \bigcup_{\substack{\pinv{\vs{\xi}}\in\pinv{P}_\text{II}:\\ f(\pinv{\vs{\xi}})\lesseqgtr k}} \mathcal{P}_{\pinv{\vs{\xi}}}
\end{equation}
\end{subequations}
by~\eqref{eq:DPsepIIp} and~\eqref{eq:fII}.
The state space $\mathcal{D}_{k,f}$ is
closed under LOCC,
and both form nested subsets,
\begin{subequations}
\begin{align}
\label{eq:PsepIIpfmonk}
k\lesseqgtr k'  &\dspiff  \mathcal{P}_{k,f} \subseteq \mathcal{P}_{k',f},\\
\label{eq:DsepIIpfmonk}
k\lesseqgtr k'  &\dspiff  \mathcal{D}_{k,f} \subseteq \mathcal{D}_{k',f},
\end{align}
\end{subequations}
by~\eqref{eq:oisomIIp} and~\eqref{eq:vxirchain}.
(For illustrations, see Figure~\ref{fig:nested}.)
Note that the whole state spaces arise for the value $k=f(\top)$, that is,
$\mathcal{P}_{f(\top),f}=\mathcal{P}_{\downset\set{\top}}=\mathcal{P}$
and
$\mathcal{D}_{f(\top),f}=\mathcal{D}_{\downset\set{\top}}=\mathcal{D}$.

The strict $f$-separability properties
can be constructed by the nonempty up-sets of $\pinv{P}_{\text{II},f}$~\cite{Szalay-2019}.
Since $\pinv{P}_{\text{II},f}$ is a chain, we have that,
for the values $k\in f(\pinv{P}_\text{I})$,
the \emph{strict $f$-separability properties} are given by the up-sets
\begin{equation}
\label{eq:vvxir}
\pinv{\vvs{\xi}}_{k,f} := \upset\bigset{\pinv{\vs{\xi}}_{k,f}}
= \bigsset{\pinv{\vs{\xi}}_{k',f}}{k'\in f(\pinv{P}_\text{I}), k\lesseqgtr k'}
\end{equation}
by~\eqref{eq:vxirchain}, forming chains again,
\begin{equation}
\label{eq:vvxirchain}
k\gtreqless k' \dspiff \pinv{\vvs{\xi}}_{k,f}  \finereq \pinv{\vvs{\xi}}_{k',f}.
\end{equation}
Let us denote these chains as 
$\pinv{P}_{\text{III},f}:=\bigsset{\pinv{\vvs{\xi}}_{k,f}}{k\in f(\pinv{P}_\text{I})}$.
These lead to the
\emph{strictly $(k,f)$-separable states}, or
`states of $f$-separability strictly $k$',
the class~\eqref{eq:CsepIIIp} of which is
\begin{subequations}
\label{eq:CsepIIIpf}
\begin{equation}
\mathcal{C}_{k,f} := 
\mathcal{C}_{\pinv{\vvs{\xi}}_{k,f}} = \begin{cases}
\mathcal{D}_{f(\bot),f}& \text{for $k=f(\bot)$},\\
\mathcal{D}_{k,f} \setminus \mathcal{D}_{k_\mp,f}& \text{else},
\end{cases}
\end{equation}
where 
$k_\mp$ denotes the values previous/next to $k$ among the possible values of $f$
(understood for increasing and decreasing $f$, respectively),
that is,
\begin{align}
k_-&:=\max\bigsset{k'\in f(\pinv{P}_\text{I})}{k'< k},\\
k_+&:=\min\bigsset{k'\in f(\pinv{P}_\text{I})}{k'> k}.
\end{align}
\end{subequations}
For example, for producibility~\eqref{eq:hwr.w}, we have $k_- = k-1$,
however, 
for more general generator functions $f$, the recipe for $k\mapsto k_\mp$ is more involved.
(Indeed, if $k\neq f(\bot)$, the class~\eqref{eq:CsepIIIp} is
$\mathcal{C}_{\pinv{\vvs{\xi}}_{k,f}}= \bigcap_{k\gtrless k'} \cmpl{\mathcal{D}_{k',f}} \cap \bigcap_{k\lesseqgtr k'} \mathcal{D}_{k',f }
=\cmpl{\mathcal{D}_{k_\mp,f}}\cap\mathcal{D}_{k,f}$
by~\eqref{eq:vvxir} and~\eqref{eq:DsepIIpfmonk}.
For illustrations, see Figure~\ref{fig:nested}.)
From the classes~\eqref{eq:CsepIIIpf} we can build up the state spaces~\eqref{eq:DsepIIpf} simply as
\begin{equation}
\label{eq:CsepIIIpfDIIp}
\mathcal{D}_{k,f} = \bigcup_{k'\in f(\pinv{P}_\text{I}), k' \lesseqgtr k}\mathcal{C}_{k',f}.
\end{equation}
(For illustrations, see Figure~\ref{fig:nested}.)
The classes have a simple structure with respect to LOCC,
\begin{equation}
\label{eq:CsepIIIpfmonk}
\mathcal{C}_{k,f} \LOCCw \mathcal{C}_{k',f} \dspthen k \gtreqless k',
\end{equation}
by~\eqref{eq:ohomomCIIIp} and~\eqref{eq:vvxirchain}.

For partitionability, producibility and stretchability,
we get back the state spaces~\eqref{eq:DsepIIpk}-\eqref{eq:DsepIIpmonk}
and classes~\eqref{eq:CsepIIIpk}-\eqref{eq:CsepIIIpmonk}
by the height, width and rank~\eqref{eq:hwr} as generator functions.

\subsection{Remarks}
\label{sec:PSprops.remarks}

Here we list some remarks 
on the one-parameter classification scheme,
one paragraph each.

Note that,
since no subsystem has a distinguished role,
the one-parameter classification presented in this section
was defined in the framework of permutation invariant 
partial separability (partial entanglement)~\cite{Szalay-2019},
being a motivated particular case of the general theory of partial separability~\cite{Szalay-2015b},
demonstrated also experimentally~\cite{Garcia-Perez-2023},
worked out in parallel with partial correlation~\cite{Szalay-2017,Szalay-2018,Szalay-2019}.
Note that the whole construction could have been carried out also in the general setting,
when not only the sizes of the subsystems matter
(if, for some reason, entanglement with some particular subsystems are of higher value as a resource);
and also for partial correlations.

Note that
the point in using $f$-entanglement is that
the whole structure of partial entanglement classification
($\pinv{P}_\text{III}$ in the permutation invariant case~\cite{Szalay-2019}, or $P_\text{III}$ in general~\cite{Szalay-2015b})
is getting too involved rapidly with the increasing number of subsystems.
Using generator functions~\eqref{eq:genf} by their sub/super-level sets~\eqref{eq:vxir}
provides a coarsening of the structure in a motivated way,
and leads to a simple, chain like classification~\eqref{eq:DsepIIpfmonk},~\eqref{eq:CsepIIIpfmonk}.
The number of classes is greatly reduced to 
$\abs{\pinv{P}_{\text{III},f}}=\abs{f(\pinv{P}_\text{I})}\ll\abs{\pinv{P}_\text{III}}$,
the number of the values of the generator function $f$.
An extreme case is when
the generator function $f$ is injective,
that is, it takes different values for all partitions $\pinv{\xi}\in\pinv{P}_\text{I}$,
then it leads to a total order on $\pinv{P}_\text{I}$.
This might be an advantage if the goal is to be able to order the `value' of different kinds of Level I partial separability,
although we end up with $\abs{\pinv{P}_{\text{III},f}}=\abs{f(\pinv{P}_\text{I})}=\abs{\pinv{P}_\text{I}}$ different classes,
which is the number of integer partitions of $n$, known to increase rapidly~\cite{oeisA000041}.
When the generator function $f$ is not injective,
that is, it may take the same value for different partitions,
then $\abs{\pinv{P}_{\text{III},f}}=\abs{f(\pinv{P}_\text{I})}<\abs{\pinv{P}_\text{I}}$.
The width, height and rank~\eqref{eq:hwr} were examples for this case,
for which 
$\abs{w(\pinv{P}_\text{I})}=\abs{h(\pinv{P}_\text{I})}=n$
and $\abs{r(\pinv{P}_\text{I})}=2n-3$ if $n\geq3$.
(Further examples are shown in Section~\ref{sec:1param}.)
The other extreme case is when
the generator function $f$ is constant,
that is, it takes the same $k_0=f(\pinv{\xi})$ value for all partitions $\pinv{\xi}\in\pinv{P}_\text{I}$.
Then the only state space is the only class, $\mathcal{D}_{k_0,f}=\mathcal{C}_{k_0,f}\equiv\mathcal{D}$,
the whole state space.
Such generator functions are hence useless.

Note that there are different generator functions which lead to the same classification, with different $k$ values.
An obvious example is taking $cf$ instead of $f$ with a nonzero real constant $c$,
then the $f$-entanglement classification is the same as the $(cf)$-entanglement classification.
More generally, any $g\circ f$ leads to the same classification as $f$ if $g$ is strictly monotone,
and a possibly coarser classification if $g$ is monotone but not strictly.
(If $g$ is not monotone then $g\circ f$ is not necessarily a generator function.)
These symmetries and coarsenings of the one-parameter properties are elaborated in Appendix~\ref{app:trafg},
since these play an important role in the depths of the one-parameter properties, and then in the metrological multipartite entanglement criteria,
formulated in the subsequent sections.

Note that
the conjugation $\pconj$, being the reflection of the Young diagram with respect to its `diagonal'~\cite{Stanley-2012},
is a natural involution on the $\pinv{P}_\text{I}$ set of integer partitions.
The height and width generator functions of producibility and partitionability~\eqref{eq:hwr} are related by conjugation,
$h=w\circ\pconj$, $h\circ\pconj=w$,
however, the classification itself is not invariant for conjugation (as can already be seen for $n=4$ in Figure~\ref{fig:PpI23456PPS}).
This is because refinement is not transformed well by conjugation~\cite{Szalay-2019},
so the refinement-monotonicity~\eqref{eq:genf} of a generator function is not preserved, neither flipped.
In other words, for a generator function $f$, the transformed function $f\circ\pconj$ is not a generator function in general.
We will turn back to this issue in Section~\ref{sec:1param.remarks}.

Note that 
in this work we use only the convex hulls of the Level~I notions 
($\xi$-separability or partition-separability) of partial separability~\cite{Szalay-2015b,Szalay-2019},
however, the finest description of partial separability
can be given by the whole lattice structure generated by the intersection and convex hull of union of $\xi$-separability.
Significant results were achieved recently in this direction
when, by the aid of multipartite entanglement of X-states and GHZ-diagonal states~\cite{Han-2016a,Han-2016b,Han-2017a},
not only partial separability criteria were formulated~\cite{Han-2017b,Chen-2017,Ha-2018,Han-2018,Han-2021a},
and the convex structure and related Bell inequalities were investigated~\cite{Han-2019,Han-2021b},
but also the nondistributivity~\cite{Han-2020}
and the infinite cardinality~\cite{Ha-2022} of this lattice were demonstrated.

\section{\texorpdfstring{$f$}{f}-entanglement depth}
\label{sec:Depthmeas}

In this section we work out the general description of the
discrete valued, `entanglement depth like' quantities, the \emph{$f$-entanglement depth},
and also its convex/concave variant, the \emph{$f$-entanglement depth of formation},
naturally characterizing one-parameter entanglement properties of quantum states, given by the generator function $f$.
We note that the appropriate generalization of the entanglement of formation and the relative entropy of entanglement
can also be used for the characterization of the one-parameter entanglement properties of quantum states~\cite{Szalay-2019},
which is recalled in Appendix~\ref{app:PSmeas} for the sake of completeness.

\subsection{Examples of \texorpdfstring{$f$}{f}-entanglement depth}
\label{sec:Depthmeas.Depthfxmpl}

For the three notable one-parameter families of properties, 
partitionability, producibility and stretchability (Section~\ref{sec:PSprops.1paramxpl}),
we also have
the \emph{partitionability entanglement depth},
the \emph{producibility entanglement depth} (or simply \emph{entanglement depth}~\cite{Sorensen-2001,Lucke-2014}),
and the \emph{stretchability entanglement depth},
\begin{subequations}
\label{eq:Depth}
\begin{align}
\label{eq:Depth.part}
D_\text{part}(\rho) &:= 
\max\bigsset{k}{\rho\in \mathcal{D}_{k\text{-part}}},\\
\label{eq:Depth.prod}
D_\text{prod}(\rho) &:= 
\min\bigsset{k}{\rho\in \mathcal{D}_{k\text{-prod}}} \equiv D(\rho),\\
\label{eq:Depth.str}
D_\text{str}(\rho)  &:= 
\min\bigsset{k}{\rho\in \mathcal{D}_{k\text{-str}}}.
\end{align}
\end{subequations}
These express how deep the state is located in the onion like classification defined by the partitionability, producibility and stretchability (see Figure~\ref{fig:nested}).
These are the bounds of the possible $k$ values for which 
$\rho\in\mathcal{D}_{k\text{-part}}$,
$\rho\in\mathcal{D}_{k\text{-prod}}$ and
$\rho\in\mathcal{D}_{k\text{-str}}$, respectively~\eqref{eq:DsepIIpk},~\eqref{eq:DsepIIpfmonk};
or the values $k$ for which
$\rho\in\mathcal{C}_{k\text{-part}}$,
$\rho\in\mathcal{C}_{k\text{-prod}}$ and
$\rho\in\mathcal{C}_{k\text{-str}}$, respectively~\eqref{eq:CsepIIIpk},~\eqref{eq:CsepIIIpf}.
These take the integer values in 
$[1,n]_\field{Z}$ in the first two cases,
and in  $[-(n-1),n-1]_\field{Z}\setminus\set{\pm(n-2)}$ in the third;
$n$, $1$ and $-(n-1)$ respectively for fully separable states;
$2$, $n-1$ and $n-3$ respectively for strictly biseparable states; and
$1$, $n$ and $n-1$ respectively for genuinely multipartite entangled states.
Because of the LOCC-closedness of the state spaces 
$\mathcal{D}_{\pinv{\vs{\xi}}}$,
these are LOCC-monotones,
increasing in the first case and decreasing in the second and third.
This is shown in general in the following subsection.

\subsection{\texorpdfstring{$f$}{f}-entanglement depth in general}
\label{sec:Depthmeas.Depthf}

\textit{In general}, for the one-parameter properties~\eqref{eq:vxir},
defined by the generator function $f$ over the permutation invariant properties~\eqref{eq:genf},
we have the \emph{$f$-entanglement depth} (or depth of $f$-entanglement)
\begin{equation}
\label{eq:Depthf}
D_f(\rho) := 
\begin{cases} 
\min\bigsset{k\in f(\pinv{P}_\text{I})}{\rho\in \mathcal{D}_{k,f} }, \\
\max\bigsset{k\in f(\pinv{P}_\text{I})}{\rho\in \mathcal{D}_{k,f} },
\end{cases}
\end{equation}
for increasing or decreasing $f$, respectively.
This expresses how deep the state is located in the onion like classification defined by the one-parameter property given by the generator function $f$ (see Figure~\ref{fig:nested}).
The range of $D_f$ is the discrete range $f(\pinv{P}_\text{I})\subset\field{R}$ of $f$, 
and
\begin{subequations}
\begin{equation}
\label{eq:DepthfC}
D_f(\rho) = k \dspiff \rho\in \mathcal{C}_{k,f},
\end{equation}
by~\eqref{eq:CsepIIIpf},
or, for the state spaces
\begin{equation}
\label{eq:DepthfD}
D_f(\rho) \lesseqgtr k \dspiff \rho\in \mathcal{D}_{k,f}
\end{equation}
for all $k\in f(\pinv{P}_\text{I})$
by~\eqref{eq:CsepIIIpfDIIp}.
\end{subequations}
In other words,
the $f$-entanglement depth is that $k$ value,
for which $\rho\in\mathcal{C}_{k,f}$,
that is,
the \emph{$f$-separability classes}~\eqref{eq:CsepIIIpf} are just the
level sets (inverse images) of the $f$-entanglement depth~\eqref{eq:Depthf},
\begin{subequations}
\begin{equation}
\label{eq:DepthfClevel}
\mathcal{C}_{k,f} = D_f^{-1}(k) \equiv\bigsset{\rho\in\mathcal{D}}{D_f(\rho)=k}
\end{equation}
for all $k\in f(\pinv{P}_\text{I})$,
and \emph{$f$-separability spaces}~\eqref{eq:DsepIIpf} are just the
sub/super-level sets of the $f$-entanglement depth~\eqref{eq:Depthf},
\begin{equation}
\label{eq:DepthfDlevel}
\mathcal{D}_{k,f} = D_f^\lesseqgtr(k) \equiv \bigsset{\rho\in\mathcal{D}}{D_f(\rho) \lesseqgtr k}
\end{equation}
\end{subequations}
for all $k\in f(\pinv{P}_\text{I})$.
(For illustrations, see Figure~\ref{fig:nested}.)
Note that,
although the $f$-entanglement depth~\eqref{eq:Depthf} is neither convex nor concave
(it is a step-function),
it is still nonincreasing/nondecreasing monotone with respect to LOCC~\cite{Bennett-1996a,Bennett-1996b,Vidal-2000,Horodecki-2001,Horodecki-2009}
in the two cases of increasing/decreasing $f$, respectively,
simply by~\eqref{eq:CsepIIIpfmonk} and~\eqref{eq:DepthfC}.
This makes it a proper entanglement measure~\cite{Horodecki-2009}.
It is moreover nonincreasing/nondecreasing on average with respect to selective LOCC~\cite{Bennett-1996a,Bennett-1996b,Vidal-2000,Horodecki-2001,Horodecki-2009}
in the two cases of increasing/decreasing $f$, respectively.
That is, $\sum_j q_j D_f( \rho'_j )\lesseqgtr D_f(\rho)$,
where $\rho'_j=\Lambda_j(\rho)/q_j$ and $q_j=\Tr(\Lambda_j(\rho))\neq0$
for selective LOCC map $\Lambda=\sum_j\Lambda_j$ with outcome maps $\Lambda_j$.
This holds even for selective separable operations 
(where the Kraus operators of $\Lambda_j(\rho)=\sum_i K_{j,i} \rho K_{j,i}^\dagger$ are elementary tensors
$K_{j,i}=\bigotimes_{l=1}^n K_{j,i,l}$,
which is an operation class larger than LOCC~\cite{Bennett-1999}),
because
$\mathcal{D}_{f,k}$ is closed under such maps $\rho\mapsto\rho'_j$,
so $D_f(\rho'_j)\lesseqgtr D_f(\rho)$ by~\eqref{eq:DepthfD},
leading to $\sum_j q_j D_f( \rho'_j )\lesseqgtr D_f(\rho)$.

An alternative way of formalizing the $f$-entanglement depth,
given from the point of view of the general classification $\pinv{P}_\text{II}$, is
\begin{equation}
\label{eq:Depthf2}
D_f(\rho) = 
\begin{cases} 
\min\limits_{\pinv{\vs{\xi}}\in\pinv{P}_\text{II}}\bigsset{f(\pinv{\vs{\xi}})}{\rho\in \mathcal{D}_{\pinv{\vs{\xi}}} }, \\
\max\limits_{\pinv{\vs{\xi}}\in\pinv{P}_\text{II}}\bigsset{f(\pinv{\vs{\xi}})}{\rho\in \mathcal{D}_{\pinv{\vs{\xi}}} },
\end{cases}
\end{equation}
for increasing or decreasing $f$, respectively.
(For the proof, see Appendix~\ref{app:Depthfs.Depthf2}.)
From this, it is also easy to see that
if the generator functions $f_1,f_2:\pinv{P}_\text{I}\to\field{R}$ are both increasing or both decreasing,
then
\begin{equation}
\label{eq:Depthfmon}
f_1 \leq f_2 \dspthen D_{f_1} \leq D_{f_2}. 
\end{equation}
(Indeed, if $f_1(\pinv{\xi}) \leq f_2(\pinv{\xi})$ for all $\pinv{\xi}\in\pinv{P}_\text{I}$,
then $f_1(\pinv{\vs{\xi}}) \leq f_2(\pinv{\vs{\xi}})$ for all $\pinv{\vs{\xi}}\in\pinv{P}_\text{II}$ by~\eqref{eq:fII},
then 
    $\min_{\pinv{\vs{\xi}}} \bigsset{f_1(\pinv{\vs{\xi}})}{\rho\in \mathcal{D}_{\pinv{\vs{\xi}}} } 
\leq \min_{\pinv{\vs{\xi}}} \bigsset{f_2(\pinv{\vs{\xi}})}{\rho\in \mathcal{D}_{\pinv{\vs{\xi}}} }$,
and the same holds for $\max$,
then the form~\eqref{eq:Depthf2} for $D_f$ leads to the claim.)
This is remarkable, 
since the one-parameter classifications given by the generator functions $f_1$ and $f_2$ may be (actually, they usually are) incompatible,
that is, may lead to classes intersecting nontrivially.

Note that the $f$-entanglement depth~\eqref{eq:Depthf}
can also be formulated in terms of optimization over pure decompositions.
Having the restriction of $D_f$ in~\eqref{eq:Depthf} to the pure states,
we have
\begin{equation}
\label{eq:Depthfdec}
D_f(\rho) = \begin{cases} 
\min\limits_{\set{(p_j,\pi_j)}\decomp \rho}
 \max\limits_j D_f(\pi_j), \\
\max\limits_{\set{(p_j,\pi_j)}\decomp \rho}
 \min\limits_j D_f(\pi_j),
\end{cases}
\end{equation}
for increasing or decreasing $f$, respectively.
(For the proof, see Appendix~\ref{app:Depthfs.Depthfdec}.
The first minimization/maximization is taken over all the $\rho=\sum_{j=1}^m p_j\pi_j$ pure convex decompositions of $\rho$,
that is, for all $m\in\field{N}$, for all $j=1,2,\dots,m$, $\pi_j\in\mathcal{P}$, $p_j>0$, $\sum_{j=1}^m p_j=1$
for which we use the shorthand notation $\decomp$ above.
This is a discrete function, the minimum or maximum are taken for $m\leq \dim(\mathcal{D})+1= d^{2n}$ due to Carathéodory's theorem.)
The meaning of the formulation~\eqref{eq:Depthfdec} of the $f$-entanglement depth is that
the state $\rho$ can optimally be mixed by the use of pure states of $f$-entanglement at most/least $D_f(\rho)$.

Note that if a generator function $f:\pinv{P}_\text{I}\to\field{R}$ 
is composed with a \emph{monotone} function $g:\field{R}\to\field{R}$,
then the $f$-entanglement depth~\eqref{eq:Depthf} is transformed as
\begin{equation}
\label{eq:gmeasuresfD}
D_{g\circ f}  = g\circ D_f. 
\end{equation}
(For the proof, see Appendix~\ref{app:trafg.fDepth}.)
Note that if $g$ is decreasing then the LOCC monotonicity of $D_{g\circ f}$ is the opposite as that of $D_f$.
Then, from bounds in generator functions,
we may also have more involved bounds for the depths 
than the simple case~\eqref{eq:Depthfmon}.
That is, if the generator functions $f_1,f_2:\pinv{P}_\text{I}\to\field{R}$
and the monotone functions $g_1,g_2:\field{R}\to\field{R}$
are such that $g_1\circ f_1$ and $g_2\circ f_2$ are both increasing or both decreasing,
then we have
\begin{equation}
\label{eq:gmeasuresfDmon}
g_1\circ f_1 \leq g_2\circ f_2 \dspthen
g_1\circ D_{f_1} \leq g_2\circ D_{f_2}
\end{equation}
by~\eqref{eq:Depthfmon} and~\eqref{eq:gmeasuresfD}.

For partitionability, producibility and stretchability,
we get back the depths~\eqref{eq:Depth}
by the height, width and rank~\eqref{eq:hwr} as generator functions.
Note also the alternative formulations~\eqref{eq:Depthf2} and~\eqref{eq:Depthfdec} of those.
Also, an immediate consequence of~\eqref{eq:Depthfmon} for these is that,
since $r<w$~\eqref{eq:hwr},
the (producibility) entanglement depth~\eqref{eq:Depth.prod} is an upper bound of the stretchability entanglement depth~\eqref{eq:Depth.str},
$D_\text{str}<D_\text{prod}\equiv D$.
This is of course a rather loose bound,
however, by the aid of~\eqref{eq:gmeasuresfDmon},
we may have stronger, even strict bounds in some cases.
For this, we recall the bounds among the generator functions $h$, $w$ and $r$~\eqref{eq:hwr},
\begin{subequations}
\label{eq:hwrbounds}
\begin{align}
n/w                                 &\leq  h  \leq  n+1-w, \\
n/h                                 &\leq  w  \leq  n+1-h,\\
n/h-h                               &\leq  r  \leq  n+1-2h,\\
-(n+1)+2w                           &\leq  r  \leq  w-n/w,\\
\frac12\bigl(\sqrt{r^2+4n}-r\bigr)  &\leq  h  \leq  \frac12(n+1-r),\\
\frac12\bigl(\sqrt{r^2+4n}+r\bigr)  &\leq  w  \leq  \frac12(n+1+r),
\end{align}
\end{subequations}
see (58) in~\cite{Szalay-2019}.
These lead to
\begin{subequations}
\label{eq:Dbounds}
\begin{align}
n/D_\text{prod}                        &\leq  D_\text{part}  \leq  n+1-D_\text{prod}, \\
n/D_\text{part}                        &\leq  D_\text{prod}  \leq  n+1-D_\text{part},\\
n/D_\text{part}-D_\text{part}          &\leq  D_\text{str}   \leq  n+1-2D_\text{part},\\
-(n+1)+2D_\text{prod}                  &\leq  D_\text{str}   \leq  D_\text{prod}-n/D_\text{prod},\\
\sqrt{D_\text{str}^2+4n}-D_\text{str}  &\leq 2D_\text{part}  \leq  n+1-D_\text{str},\\
\sqrt{D_\text{str}^2+4n}+D_\text{str}  &\leq 2D_\text{prod}  \leq  n+1+D_\text{str},
\end{align}
\end{subequations}
where the inequalities in the fourth and sixth rows were obtained by applying increasing $g$ functions, 
the remaining inequalities were obtained by using decreasing ones for the application of~\eqref{eq:gmeasuresfDmon}.

\subsection{Examples of \texorpdfstring{$f$}{f}-entanglement depth of formation}
\label{sec:Depthmeas.DepthOFfxmpl}

The $f$-entanglement depths~\eqref{eq:Depthf} are discrete valued, step functions over the state space,
which leads to some inconvenient consequences.
For instance, for the usual (producibility) entanglement depth~\eqref{eq:Depth.prod},
mixing even an infinitely small amount of $n$-qubit GHZ state $\cket{\psi_\text{GHZ}}=(\cket{00\dots0}+\cket{11\dots1})/\sqrt{2}$,
having maximal entanglement depth $D\bigl(\proj{\psi_\text{GHZ}}\bigr)=n$,
to a separable state $\cket{\psi_\text{sep}}:=\cket{10\dots0}$ orthogonal to it,
having minimal entanglement depth $D\bigl(\proj{\psi_\text{sep}}\bigr)=1$,
that is,
$\rho_\epsilon:=(1-\epsilon)\proj{\psi_\text{sep}} + \epsilon\proj{\psi_\text{GHZ}}$,
we get maximal entanglement depth $D(\rho_\epsilon)=n$ for all $\epsilon>0$.
(This is because every pure convex decomposition of such a state contains pure states given by state vectors
of the form $a\cket{\psi_\text{sep}}+b\cket{\psi_\text{GHZ}}$ 
(see Schrödinger's mixture theorem~\cite{Schrodinger-1936,Gisin-1989,Hughston-1993}),
being separable if and only if $b=0$.)

To avoid such situations, we may formulate the continuous variant of the $f$-entanglement depths
by the convex or concave roof construction,
expressing the average $f$-entanglement depth of the optimal decomposition.
For the partitionability, producibility and stretchability entanglement depths~\eqref{eq:Depth}
these are
the \emph{partitionability entanglement depth of formation},
the \emph{producibility entanglement depth of formation} (or simply \emph{entanglement depth of formation}),
and the \emph{stretchability entanglement depth of formation},
\begin{subequations}
\label{eq:DepthOF}
\begin{align}
\label{eq:DepthOF.part}
D^\text{oF}_\text{part}(\rho) &:=
\max_{\set{(p_j,\pi_j)}\decomp \rho}
 \sum_j p_j D_\text{part}(\pi_j),\\
\label{eq:DepthOF.prod}
D^\text{oF}_\text{prod}(\rho) &:=
\min_{\set{(p_j,\pi_j)}\decomp \rho}
 \sum_j p_j D_\text{prod}(\pi_j) \equiv D^\text{oF}(\rho) ,\\
\label{eq:DepthOF.str}
D^\text{oF}_\text{str}(\rho) &:=
\min_{\set{(p_j,\pi_j)}\decomp \rho}
 \sum_j p_j D_\text{str}(\pi_j).
\end{align}
\end{subequations}
These take the values in the continuous ranges $[1,n]_\field{R}$ and $[-(n-1),n-1]_\field{R}$ in the first two and the third cases, respectively.

For the example above, for $\epsilon>0$, we have the much more expressive
$1<D^\text{oF}(\rho_\epsilon)\leq (1-\epsilon)1 +\epsilon n$ upper bound on the $f$-entanglement depth of formation,
reflecting the physical situation much better.

\subsection{\texorpdfstring{$f$}{f}-entanglement depth of formation in general}
\label{sec:Depthmeas.DepthOFf}

\textit{In general}, for the one-parameter properties~\eqref{eq:vxir},
defined by the generator function $f$ over the permutation invariant properties~\eqref{eq:genf},
we have the \emph{$f$-entanglement depth of formation}
\begin{equation}
\label{eq:DepthOFf}
D^\text{oF}_f(\rho) := \begin{cases} 
\min\limits_{\set{(p_j,\pi_j)}\decomp \rho}
 \sum_j p_j D_f(\pi_j),\\
\max\limits_{\set{(p_j,\pi_j)}\decomp \rho}
 \sum_j p_j D_f(\pi_j),
\end{cases}
\end{equation}
for increasing or decreasing $f$, respectively.
This is the \emph{convex/concave roof extension}~\cite{Uhlmann-1998,Uhlmann-2010}
of the $f$-entanglement depth~\eqref{eq:Depthf}.
(For the proof, see Appendix~\ref{app:Depthfs.DepthOFf}.)
The meaning of the formula~\eqref{eq:DepthOFf} of the $f$-entanglement depth of formation is that
the state $\rho$ can optimally be mixed by the use of pure states of \emph{average} $f$-entanglement $D^\text{oF}_f(\rho)$.
These take the values in the continuous ranges $\bigl[\min f(\pinv{P}_\text{I}),\max f(\pinv{P}_\text{I})\bigr]_\field{R}$.
These are entanglement monotones
(convex/concave 
and nonincreasing/nondecreasing on average with respect to selective LOCC
in the two cases of increasing/decreasing $f$, respectively).
Indeed, this follows by the convex roof construction from
$\sum_j q_j D_f( \pi'_j )\lesseqgtr D_f(\pi)$,
where $\pi'_j=\Lambda_j(\pi)/q_j$ and $q_j=\Tr(\Lambda_j(\pi))\neq0$
for selective LOCC map $\Lambda=\sum_j\Lambda_j$ with outcome maps $\Lambda_j$ of Kraus rank $1$
\cite{Vidal-2000,Horodecki-2001,Szalay-2015b}.
This holds even for selective separable operations 
because
$\mathcal{D}_{f,k}\cap\mathcal{P}=\mathcal{P}_{f,k}$ is closed under such maps $\pi\mapsto\pi'_j$,
so $D_f(\pi'_j)\lesseqgtr D_f(\pi)$ by~\eqref{eq:DepthfD},
leading to $\sum_j q_j D_f( \pi'_j )\lesseqgtr D_f(\pi)$.

From~\eqref{eq:Depthfdec} it easily follows that
\begin{equation}
\label{eq:DepthfoFDepthf}
D^\text{oF}_f(\rho) \lesseqgtr D_f(\rho).
\end{equation}
(Indeed, 
$\min_j\set{D_f(\pi_j)}\leq\sum_j p_j D_f(\pi_j)\leq\max_j\set{D_f(\pi_j)}$
for all particular decompositions,
leading to the bounds in the minimization or maximization of these functions
with respect to the decompositions in~\eqref{eq:Depthfdec} and~\eqref{eq:DepthOFf}.)
From this, we readily have
\begin{subequations}
\begin{equation}
\label{eq:DepthfoFD}
D^\text{oF}_f(\rho) \lesseqgtr k \dspif \rho\in \mathcal{D}_{k,f}
\end{equation}
for all $k\in f(\pinv{P}_\text{I})$
by~\eqref{eq:DepthfD}.
Here, contrary to~\eqref{eq:DepthfD}, we do not have necessary and sufficient condition,
which is not a disadvantage, quite the contrary:
the motivation for the introduction of entanglement depth of formation in Section~\ref{sec:Depthmeas.DepthOFfxmpl}
was just the manifestation of this.
On the other hand, $f(\bot) \lesseqgtr D^\text{oF}_f(\rho)$,
so we have equivalence for the finest member of the $f$ based classification,
\begin{equation}
D^\text{oF}_f(\rho) = f(\bot) \dspiff \rho\in \mathcal{D}_{f(\bot),f} \equiv \mathcal{C}_{f(\bot),f}
\end{equation}
\end{subequations}
by~\eqref{eq:CsepIIIpf}.

It is also easy to see that
if the generator functions $f_1,f_2:\pinv{P}_\text{I}\to\field{R}$ are both increasing or both decreasing,
then
\begin{equation}
\label{eq:DepthfoFmon}
f_1 \leq f_2 \dspthen D^\text{oF}_{f_1}(\rho) \leq D^\text{oF}_{f_2}(\rho).
\end{equation}
(Indeed, this is the consequence of~\eqref{eq:Depthfmon},
and the monotonicity of the convex/concave roof extension,
if $D_{f_1}(\rho) \leq D_{f_2}(\rho)$, then $D^\text{oF}_{f_1}(\rho) \leq D^\text{oF}_{f_2}(\rho)$.)

Note that for the $f$-entanglement depth of formation,
we do not have such a strong property as~\eqref{eq:gmeasuresfD} for the $f$-entanglement depth.
The convex and concave roof construction used in the definition of the $f$-entanglement depth of formation
restricts the possibilities to convex and concave $g$ functions, respectively.
If a generator function $f:\pinv{P}_\text{I}\to\field{R}$
is composed with a \emph{convex} or \emph{concave} \emph{monotone} function $g:\field{R}\to\field{R}$,
then for the $f$-entanglement depth of formation~\eqref{eq:DepthOFf} we have
\begin{subequations}
\label{eq:gmeasuresfDoF}
\begin{align}
\label{eq:gmeasuresfDoF.conv}
D^\text{oF}_{g\circ f} \geq g\circ D^\text{oF}_f,
\intertext{or}
\label{eq:gmeasuresfDoF.conc}
D^\text{oF}_{g\circ f} \leq g\circ D^\text{oF}_f,
\end{align}
for the cases of convex or concave $g$ functions, respectively.
If $g$ is affine,
that is, of the form $g(u)=au+b$,
then the $f$-entanglement depth of formation~\eqref{eq:DepthOFf} is transformed as
\begin{equation}
\label{eq:gmeasuresfDoF.affine}
D^\text{oF}_{g\circ f} = g\circ D^\text{oF}_f,
\end{equation}
\end{subequations}
by~\eqref{eq:gmeasuresfD} and the definition~\eqref{eq:DepthOFf}.
(For the proof, see Appendix~\ref{app:trafg.fDepthoF}.)
Then, from bounds in generator functions,
we may also have more complex, and possibly stronger bounds for the depths of formations
than the simple case~\eqref{eq:DepthfoFmon}.
That is, if the generator functions $f_1,f_2:\pinv{P}_\text{I}\to\field{R}$ 
and the monotone functions $g_1,g_2:\field{R}\to\field{R}$
are such that $g_1\circ f_1$ and $g_2\circ f_2$ are both increasing or both decreasing
\emph{and} $g_1$ is convex, $g_2$ is concave,
then we have 
\begin{equation}
\label{eq:gmeasuresfDoFmon}
g_1\circ f_1 \leq g_2\circ f_2 \dspthen
g_1\circ D^\text{oF}_{f_1} \leq g^\text{oF}_2\circ D_{f_2}
\end{equation}
by~\eqref{eq:DepthfoFmon} and~\eqref{eq:gmeasuresfDoF}.

For partitionability, producibility and stretchability,
we get back the depths of formations~\eqref{eq:DepthOF}
by the height, width and rank~\eqref{eq:hwr} as generator functions.
As in~\eqref{eq:Dbounds}, we also have 
\begin{subequations}
\label{eq:DoFbounds}
\begin{align}
n/D^\text{oF}_\text{prod}                                  \leq\; & D^\text{oF}_\text{part}  \leq  n+1-D^\text{oF}_\text{prod}, \\
n/D^\text{oF}_\text{part}                                  \leq\; & D^\text{oF}_\text{prod}  \leq  n+1-D^\text{oF}_\text{part},\\
n/D^\text{oF}_\text{part}-D^\text{oF}_\text{part}          \leq\; & D^\text{oF}_\text{str}   \leq  n+1-2D^\text{oF}_\text{part},\\
-(n+1)+2D^\text{oF}_\text{prod}                            \leq\; & D^\text{oF}_\text{str}   \leq  D^\text{oF}_\text{prod}-n/D^\text{oF}_\text{prod},\\
\sqrt{(D^\text{oF}_\text{str})^2+4n}-D^\text{oF}_\text{str}  &\leq 2D^\text{oF}_\text{part}  \leq  n+1-D^\text{oF}_\text{str},\\
\sqrt{(D^\text{oF}_\text{str})^2+4n}+D^\text{oF}_\text{str}  &\leq 2D^\text{oF}_\text{prod}  \leq  n+1+D^\text{oF}_\text{str},
\end{align}
\end{subequations}
based on the relations recalled in~\eqref{eq:hwrbounds},
where the inequalities in the fourth and sixth rows came by applying increasing $g$ functions,    
the remaining inequalities by decreasing ones, 
which were 
convex on the left-hand side,
identity in the middle,
and concave on the right-hand side in all rows
for the application of~\eqref{eq:gmeasuresfDoFmon}.

\subsection{Remarks}
\label{sec:Depthmeas.remarks}

Here we list some remarks 
on the $f$-entanglement depth and $f$-entanglement depth of formation of one-parameter properties,
one paragraph each.

Note that,
again, since no subsystem has a distinguished role,
the entanglement depth and entanglement depth of formation presented here
was defined in the framework of permutation invariant 
partial separability~\cite{Szalay-2019},
being a motivated particular case of the general theory of partial separability~\cite{Szalay-2015b},
demonstrated also experimentally~\cite{Garcia-Perez-2023},
worked out in parallel with partial correlation~\cite{Szalay-2017,Szalay-2018,Szalay-2019}.
Note that the whole construction could have been carried out also in the general setting,
when not only the sizes of the subsystems matter;
and also
the $f$-correlation depth can be defined analogously for partial correlations,
but not its convex/concave roof extension,
the analogous definition for $f$-correlation depth of formation would lead to the $f$-entanglement depth of formation.

Note that
we considered both increasing and decreasing generator functions~\eqref{eq:genf}
for the definition of one-parameter properties~\eqref{eq:vxir} in Section~\ref{sec:PSprops},
since both arise naturally in well motivated situations~\eqref{eq:hwr}.
The resulting $f$-entanglement depth~\eqref{eq:Depthf} 
and $f$-entanglement depth of formation~\eqref{eq:DepthOFf}
are then decreasing/increasing monotones with respect to LOCC for the case of increasing/decreasing generator functions.
Having increasing LOCC monotones for measuring entanglement is not a real problem,
the important point is the monotonicity itself.
Although $-1$ times an increasing monotone is a decreasing one,
such quantity would be less expressive,
so instead of doing this,
we prefer to call the increasing LOCC monotones separability measures/monotones instead of entanglement measures/monotones.

Note that
the LOCC monotonicity of the $f$-entanglement depth
could be rather difficult to prove for general $f$ from scratch,
however, it was easy to show thanks to the 
whole construction of the (permutation invariant) partial separability~\cite{Szalay-2015b,Szalay-2019},
recalled in Section~\ref{sec:PSprops}.

Note that
the characterizations
\eqref{eq:Depthfdec} of $f$-entanglement depth and 
\eqref{eq:DepthOFf} of $f$-entanglement depth of formation
by convex decompositions
resemble the case of the Schmidt number~\cite{Terhal-2000,Sanpera-2001} (sometimes also called Schmidt rank)
and the (convex roof extended) Schmidt rank (we would call it Schmidt rank of formation)
measuring \emph{bipartite} entanglement.
These are based on the Schmidt rank of bipartite pure states,
which is also a lower semicontinuous function,
so the LOCC monotonicity can be shown analogously to the case of $f$-entanglement depth and $f$-entanglement depth of formation
(see Appendix~\ref{app:Depthfs.DepthOFf}).

Note that 
the $f$-entanglement depth
signals the presence or absence of given
one-parameter partial entanglement properties $\pinv{\vs{\xi}}_{k,f}\in\pinv{P}_{\text{II},f}$~\eqref{eq:DepthfC},
which is a rather coarse description.
In case of the presence of a given partial entanglement property $\pinv{\vs{\xi}}_{k,f}$,
it is still a question how strong it is.
This can be characterized by the 
$(k,f)$-entanglement of formation $E^\text{oF}_{k,f}$
or the
relative entropy of $(k,f)$-entanglement $E^\text{R}_{k,f}$~\eqref{eq:rfEoFR},
being proper entanglement measures.
(See Appendix~\ref{app:PSmeas} for the application of the general construction~\cite{Szalay-2019}
to the one-parameter case.)
By the faithfulness of these,
we readily have
\begin{subequations}
\begin{align}
E^\text{oF}_{k,f}(\rho) = E^\text{R}_{k,f}(\rho) = 0,
&\dspiff
D_f(\rho) \lesseqgtr k,\\
E^\text{oF}_{k,f}(\rho) = E^\text{R}_{k,f}(\rho) = 0,
&\dspthen
D_f^\text{oF}(\rho) \lesseqgtr k,
\end{align}
\end{subequations}
for 
the $f$-entanglement depth and
the $f$-entanglement depth of formation
by~\eqref{eq:DepthfD} and~\eqref{eq:DepthfoFD}.
To find further, more general bounds among these quantities represents an important research direction.

\section{Some notable cases of \texorpdfstring{$f$}{f}-en\-tan\-gle\-ment}
\label{sec:1param}

For the one-parameter properties
partitionability, producibility and stretchability,
the generator functions were 
the height, width and rank~\eqref{eq:hwr} of the Young diagram representing the partition-type of the system~\cite{Szalay-2019}.
In this section we consider some further, motivated generator functions~\eqref{eq:genf},
which lead to some meaningful one-parameter properties.
The proof of some general properties of some of these functions are recalled in
Appendices~\ref{app:powerf.Genfqlim} and~\ref{app:powerf.Genfqmon} for the convenience of the reader.
The proof of the monotonicity~\eqref{eq:genf.mon} of the generator functions are given in Appendix~\ref{app:1param.Mon}.
The observations \eqref{eq:gmeasuresfD} and~\eqref{eq:gmeasuresfDoF}
give us a considerable freedom in the construction.

As illustration, we also plot the generator functions $f(\pinv{\xi})$ against the height $h(\pinv{\xi})$,
being the natural gradation of the poset $\pinv{P}_\text{I}$,
flipped, so that those plots resemble the height vs.~width plot of Figure~\ref{fig:PpI23456PPS}.
Note that multiple partitions $\pinv{\xi}$ may fall on each vertex in those figures in general.
The down-sets $\pinv{\vs{\xi}}_{k,f}$ are to the left/right of vertical lines (not drawn)
for the increasing/decreasing cases, respectively.

\subsection{Max and min generator functions, toughness}
\label{sec:1param.maxmin}

The producibility was given by the width~\eqref{eq:hwr.w} of the partition,
which is the size of the largest part.
We can smoothen this concept by taking into account the first $m$ largest parts.
We get generator function also by taking into account the first $m$ smallest parts.

\begin{figure}\centering
\includegraphics{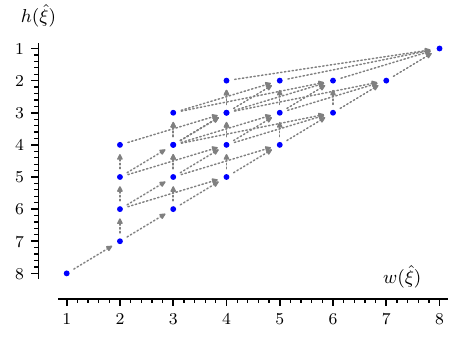}
\includegraphics{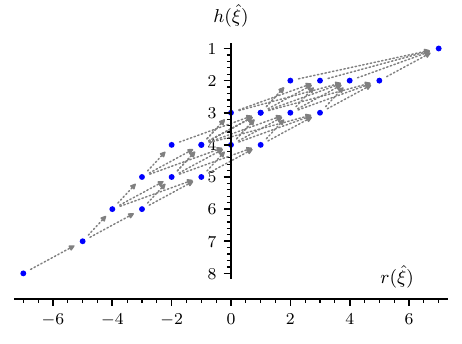}
\includegraphics{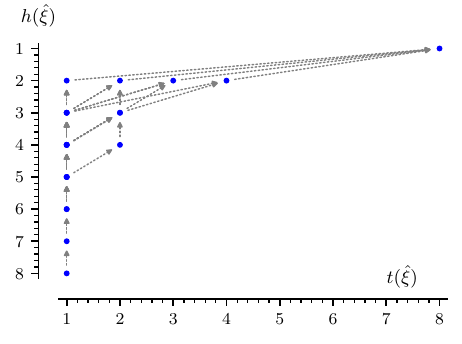}
\caption{The height~\eqref{eq:hwr.h}
vs.~width~\eqref{eq:hwr.w}, rank~\eqref{eq:hwr.r} and toughness~\eqref{eq:t}
plots of $\pinv{P}_\text{I}$ for $n=8$.
See also figure~\ref{fig:PpI23456PPS} for the width for $n\leq6$.}
\label{fig:fPp.wrt}
\end{figure}

\begin{subequations}
\label{eq:extr}
For $m=1,2,\dots,n$, let us have the sum of the largest $m$ parts as
\begin{equation}
\label{eq:extr.wm}
w_m(\pinv{\xi}) :=
\max_{\pinv{\xi}'\subseteq\pinv{\xi},\abs{\pinv{\xi}'}\leq m} \sum_{x\in\pinv{\xi}'} x,
\end{equation}
which is an increasing generator function, see~\eqref{eq:mon.wm}.
For $m=1,2,\dots,n$, let us have the sum of the smallest $m$ parts as
\begin{equation}
\label{eq:extr.tm}
t_m(\pinv{\xi}) :=
\min_{\pinv{\xi}'\subseteq\pinv{\xi},\abs{\pinv{\xi}'}\geq m} \sum_{x\in\pinv{\xi}'} x,
\end{equation}
which is an increasing generator function, see~\eqref{eq:mon.tm}.
\end{subequations}
(For illustration, see Figure~\ref{fig:fPp.wrt}.)

Clearly, we have $w_1=\max=w$, the width~\eqref{eq:hwr.w},
and $w_m\leq w_{m'}$ if $m<m'$.
We also have $w_m(\bot) = m$ for the finest partition,
and $w_m(\pinv{\xi})=w_m(\top) = n$ for all $m\geq h(\pinv{\xi})$,
so the range of $w_m$ is between these values~\eqref{eq:frange}.
For a given producibility, states with
$w_m$-depth steeper with $m$ are more entangled,
as the largest $m$ parts of the corresponding partitions are larger.
Let us define
\begin{equation}
\label{eq:t}
t(\pinv{\xi}) := \min(\pinv{\xi}),
\end{equation}
which is an increasing generator function, see~\eqref{eq:mon.tm},
which we call the \emph{toughness} of the partition $\pinv{\xi}$,
or of the Young diagram,
by which we have $t_1=\min=t$.
Also, $t_m\leq t_{m'}$ if $m<m'$.
We also have $t_m(\bot) = m$ for the finest partition,
and $t_m(\pinv{\xi})=t_m(\top) = n$ for all $m\geq h(\pinv{\xi})$,
so the range of $t_m$ is between these values~\eqref{eq:frange}.
Note that $t$ takes the values $1,2,\dots,\lfloor n/2 \rfloor,n$,
and $t(\top)=n$ uniquely.
For a given partitionability, states with
$t_m$-depth steeper with $m$ are more entangled,
as the smallest $m$ parts of the corresponding partitions are larger.
Clearly, $t_m\leq w_m$, and 
$w_m(\pinv{\xi})+t_{h(\pinv{\xi})-m}(\pinv{\xi}) = n$
if $m<h(\pinv{\xi})$.

For these functions we have 
the $(k,w_m)$- and $(k,t_m)$-entanglement of formation 
and the relative entropy of $(k,w_m)$- and $(k,t_m)$-entanglement~\eqref{eq:rfEoFR}
in the usual way~\cite{Szalay-2019}.
We also have the corresponding depths,
the $w_m$- and $t_m$-entanglement depth~\eqref{eq:Depthf}
and $w_m$- and $t_m$-entanglement depth of formation~\eqref{eq:DepthOFf}.
For example, for the function $w_2$, if the depth of $w_2$-entanglement is $D_{w_2}(\rho)=k$, then,
to mix it, there is a need for pure entanglement
where the two largest entangled subsystems together are of size at least $k$.
For the toughness $t$, the \emph{(entanglement) toughness depth}, or \emph{depth of (entanglement) toughness}
$D_\text{tgh}(\rho):=D_t(\rho)$ may be of particular interest.
For example, if $D_\text{tgh}(\rho)=k$,
then, to mix $\rho$, there is a need for pure entanglement
where the smallest entangled subsystem is of size at least $k$.
In particular, if $D_\text{tgh}(\rho)=2$,
then there is a need for pure entanglement
where there are no elementary subsystems separable from the rest of the system.
On the other hand, 
$D_\text{tgh}(\rho)=n$ for genuinely multipartite entangled states $\rho\in\mathcal{C}_{n\text{-tgh}}$,
and $D_\text{tgh}(\rho)\leq \lfloor n/2 \rfloor$ for biseparable states $\rho\in\mathcal{D}_{\lfloor n/2 \rfloor\text{-tgh}}$.
Note that, e.g., fully separable states cannot be identified by toughness.

\subsection{Power sums and power means}
\label{sec:1param.pw}

\begin{figure}\centering
\includegraphics{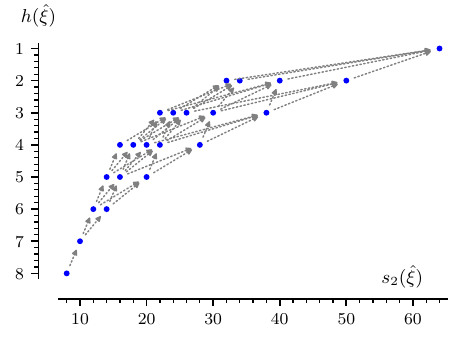}
\includegraphics{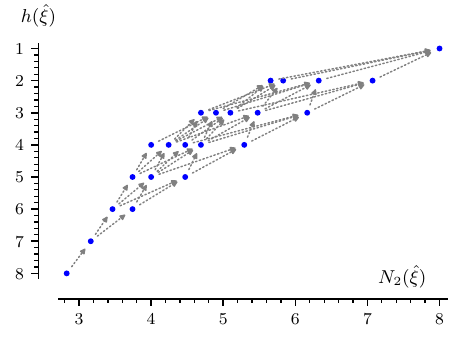}
\includegraphics{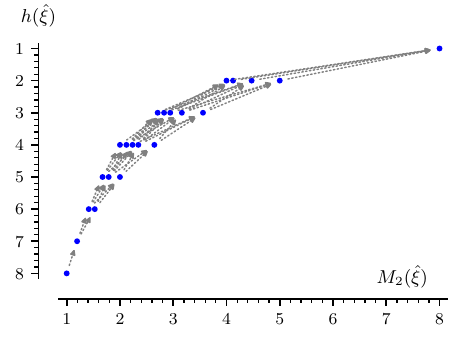}
\caption{The height~\eqref{eq:hwr.h}
vs.~power-sum~\eqref{eq:sq}, $q$-sum \eqref{eq:Nq} and $q$-mean~\eqref{eq:Mq}
plots of $\pinv{P}_\text{I}$ for $q=2$, $n=8$.}
\label{fig:fPp.s2N2M2}
\end{figure}

We may experiment with other characterizations of partitions as well.

\begin{subequations}
For $q\in\field{R}$, let us have
the \emph{power sum}
\begin{equation}
\label{eq:sq}
s_q(\pinv{\xi}) := \sum_{x\in\pinv{\xi}} x^q,
\end{equation}
which is an increasing/decreasing generator function for $1 \lesseqgtr q$, see~\eqref{eq:mon.sq}.
For $q\in\field{R}$, $q\neq0$, let us have
the \emph{$q$-sum} ($q$-norm for $q\geq1$)
\begin{equation}
\label{eq:Nq}
N_q(\pinv{\xi}) := \bigl(s_q(\pinv{\xi}) \bigr)^{1/q}, 
\end{equation}
which is an increasing generator function for $q<0$ and $1\leq q$ and decreasing generator function for $0< q\leq 1$, see~\eqref{eq:mon.Nq}.
For $q\in\field{R}$, let us have
the \emph{$q$-mean}
\begin{equation}
\label{eq:Mq}
M_q(\pinv{\xi}) := \Bigl(\frac{s_q(\pinv{\xi})}{\abs{\pinv{\xi}}}\Bigr)^{1/q},
\end{equation}
which is an increasing generator function for $1\leq q$, see~\eqref{eq:mon.Mq}.
\end{subequations}
(For illustration, see Figure~\ref{fig:fPp.s2N2M2}.)

The power sum has the limits
$s_{q\to\infty}(\pinv{\xi}\neq\bot)=\infty$ and $s_{q\to\infty}(\bot)=n$, see~\eqref{eq:qlim.spinf}, and
$s_{q\to-\infty}$ gives the number of size-one subsystems, see~\eqref{eq:qlim.sninf}.
For $q=0$ and $1$, it is
$s_0(\pinv{\xi})=\abs{\pinv{\xi}}=h(\pinv{\xi})$, leading to partitionability~\eqref{eq:hwr.h},
and $s_1(\pinv{\xi})=n$, being constant, not leading to a meaningful property.
For all $q$ parameter values,
$s_q(\bot)=n$ and $s_q(\top)=n^q$ for the finest and coarsest partitions,
so the range of $s_q$ is between these values~\eqref{eq:frange}.

The $q$-sum has the limits 
$N_{q\to\infty}=\max$, see~\eqref{eq:qlim.Npinf}, and
$N_{q\to-\infty}=\min$, see~\eqref{eq:qlim.Nninf},
leading to the $m=1$ case (toughness~\eqref{eq:t} and producibility~\eqref{eq:hwr.w})
of the $t_m$ and $w_m$ properties~\eqref{eq:extr}.
For $q=1$, it is
$N_1(\pinv{\xi})=n$, being constant, not leading to a meaningful property,
and $N_q$ does not have limit as $q\to0$, see~\eqref{eq:qlim.N0p}-\eqref{eq:qlim.N0m},
so we cannot properly interpolate between $t$ (toughness) and $w$ (producibility) by that. 
For all $q$ parameter values,
$N_q(\bot)=n^{1/q}$ and $N_q(\top)=n$ for the finest and coarsest partitions,
so the range of $N_q$ is between these values~\eqref{eq:frange}.

The $q$-mean covers more, it has the limits 
$M_{q\to\infty}=\max$, see~\eqref{eq:qlim.Mpinf}, and
$M_{q\to-\infty}=\min$, see~\eqref{eq:qlim.Mninf}, again,
while being continuous at $q=0$ by its limit,
the geometric mean
$M_0(\pinv{\xi})=(\prod_{x\in\pinv{\xi}}x)^{1/\abs{\pinv{\xi}}}$, see~\eqref{eq:qlim.M0}.
Also, while $N_1(\pinv{\xi})=n$, being constant, does not lead to a meaningful property,
$M_1(\pinv{\xi})=n/\abs{\pinv{\xi}}$ leads to reciprocal-partitionability.
So $M_q$ covers toughness, reciprocal-partitionability and producibility,
however, it does not interpolate among these, since it fails to be a generator function for $q<1$.
Such interpolation would be nice to have,
and it will indeed be possible by the use of Rényi generator functions,
see Section~\ref{sec:1param.entr}.
For all $q$ parameter values,
$M_q(\bot)=1$ and $M_q(\top)=n^{1/q}$ for the finest and coarsest partitions,
so the range of $M_q$ is between these values~\eqref{eq:frange}.

It holds in general for these functions that
for higher $q$, the larger parts get more emphasis.
This diverges for $s_q$, but not for $N_q$ and $M_q$,
where the smaller parts are more suppressed then.
Note that, for a given $q\neq0$ parameter, $s_q$ and $N_q$ are connected by the strictly monotone function $g(u)=u^{1/q}$,
so they lead to the same classification by 
\eqref{eq:gprops.strict} and~\eqref{eq:gstates.strictD}-\eqref{eq:gstates.strictC},
although the interpolation properties are different,
and the resulting depths have different values and meanings,
see~\eqref{eq:gmeasuresfD}.

For these functions we have
the $(k,s_q)$-, $(k,N_q)$- and $(k,M_q)$-entanglement of formation 
and the relative entropy of $(k,s_q)$-, $(k,N_q)$- and $(k,M_q)$-entanglement~\eqref{eq:rfEoFR}
in the usual way~\cite{Szalay-2019}.
We also have the corresponding depths,
the $s_q$-, $N_q$- and $M_q$-entanglement depth~\eqref{eq:Depthf}
and $s_q$-, $N_q$- and $M_q$-entanglement depth of formation~\eqref{eq:DepthOFf}.
For example, if the depth of $M_2$-entanglement of a state $\rho$ is $D_{M_2}(\rho)=k$, then,
to mix it, there is a need for pure entanglement
where the quadratic mean $M_2$ of the sizes of entangled subsystems is at least $k$.

\subsection{Probabilistic functions}
\label{sec:1param.prob}

The power based generator functions in the previous section
may gain some motivation in probabilistic scenarios.

If we consider an arbitrary partition $\xi=\set{X_1,X_2,\dots,X_{\abs{\xi}}}$ of the whole system
(that is, $X_i\subseteq\set{1,2,\dots,n}$ nonempty and disjoint)
of type $\pinv{\xi}$
(that is, $\mset{\abs{X_1},\abs{X_2},\dots,\abs{X_{\abs{\xi}}}}=\mset{x_1,x_2,\dots,x_{\abs{\xi}}}=\pinv{\xi}$),
then the probability of getting a particular subsystem $X_i$
by picking an elementary subsystem randomly with equal probabilities $1/n$
is $\abs{X_i}/n=x_i/n$.
Later we use the notation $\pinv{\xi}/n=\smset{x/n}{x\in\pinv{\xi}}$
for these probabilities.
The $q$-th (raw) moment of the subsystem size is then
\begin{subequations}
\label{eq:xipmom1}
\begin{equation}
\label{eq:xipmom1.raw}
\sum_{x\in\pinv{\xi}}\frac{x}{n} x^q
=\frac{s_{q+1}(\pinv{\xi})}{n},
\end{equation}
which is basically the $(q+1)$-th power sum~\eqref{eq:sq} of $\pinv{\xi}$.
We note that the $q$-th central moment of the subsystem size
(meaningful only for $2\leq q\in\field{N}$) is
\begin{equation}
\label{eq:xipmom1.central}
\sum_{x\in\pinv{\xi}} \frac{x}{n}\Bigl(x - \sum_{x'\in\pinv{\xi}} \frac{x'}{n}x'\Bigr)^q
= \sum_{x\in\pinv{\xi}} \frac{x}{n}\Bigl(x-\frac{s_2(\pinv{\xi})}{n}\Bigr)^q,
\end{equation}
\end{subequations}
however, it is not monotone~\eqref{eq:genf}, so not a proper generator function
(this can already be seen for the $n=3$, $q=2$ case).

We may consider a different situation as well,
when we pick a particular subsystem $X\in\xi$ with equal probability $1/\abs{\xi}=1/\abs{\pinv{\xi}}$.
The $q$-th (raw) moment of the subsystem size is then
\begin{subequations}
\label{eq:xipmom2}
\begin{equation}
\label{eq:xipmom2.raw}
\sum_{x\in\pinv{\xi}}\frac{1}{\abs{\pinv{\xi}}} x^q
=\frac{s_q(\pinv{\xi})}{\abs{\pinv{\xi}}},
\end{equation}
which is basically the $q$-th power of the $q$-mean~\eqref{eq:Mq} of $\pinv{\xi}$,
increasing for $q\geq1$.
For $q=1$, it gives back the average size $M_1$.
We note that the $q$-th central moment of the subsystem size
(meaningful only for $2\leq q\in\field{N}$) is
\begin{equation}
\label{eq:xipmom2.central}
\sum_{x\in\pinv{\xi}} \frac{1}{\abs{\pinv{\xi}}}\Bigl(x - \sum_{x'\in\pinv{\xi}} \frac{1}{\abs{\pinv{\xi}}}x'\Bigr)^q
= \sum_{x\in\pinv{\xi}} \frac{1}{\abs{\pinv{\xi}}} \Bigl(x-\frac{n}{\abs{\pinv{\xi}}}\Bigr)^q,
\end{equation}
\end{subequations}
however, it is not monotone~\eqref{eq:genf}, so not a proper generator function
(this can already be seen for the $n=3$, $q=2$ case).

\subsection{Entropy based generator functions}
\label{sec:1param.entr}

\begin{figure*}\centering
\includegraphics{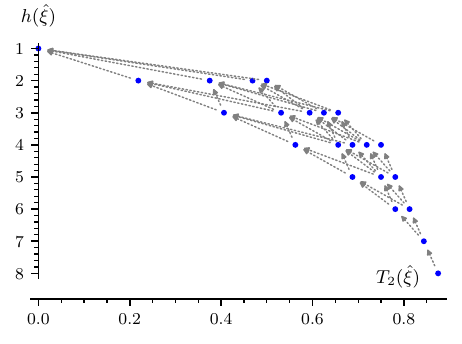}
\includegraphics{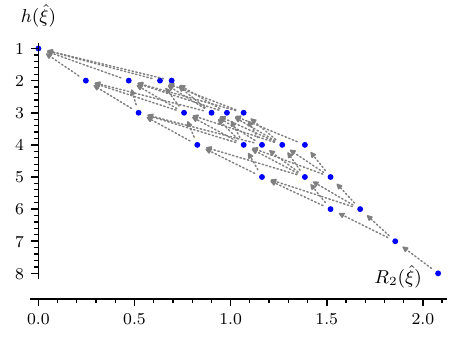}
\includegraphics{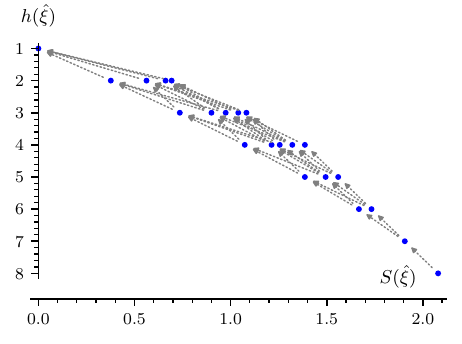}
\includegraphics{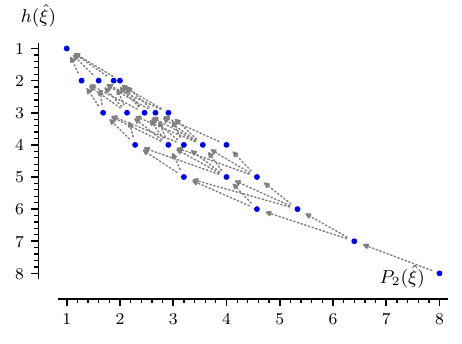}
\caption{The height~\eqref{eq:hwr.h} 
vs.~Tsallis function~\eqref{eq:Tq}, Rényi function~\eqref{eq:Rq}, Shannon function~\eqref{eq:S} and $P_q$ function~\eqref{eq:Pq}
plots of $\pinv{P}_\text{I}$ for $q=2$, $n=8$.}
\label{fig:fPp.T2R2SP2}
\end{figure*}

We may also use entropy based generator functions
through the probabilities $\pinv{\xi}/n=\smset{x/n}{x\in\pinv{\xi}}$,
which allows us to interpolate between toughness, partitionability and producibility.
Note that we do not call these entropies in general,
since, on the one hand, they act on $\pinv{P}_\text{I}$,
on the other hand, larger ranges of parameter $q$ are allowed here.
(For some elaboration on this, see Section~\ref{sec:1param.remarks} and Appendix~\ref{app:1param}.)

\begin{subequations}
\label{eq:qentr}
For $q\in\field{R}$, let us have
the \emph{Tsallis function}
\begin{equation}
\label{eq:Tq}
T_q(\pinv{\xi}) = \frac1{1-q}\Bigl(\sum_{x\in\pinv{\xi}}\bigl(\frac{x}{n}\bigr)^q -1 \Bigr)
= \frac{s_q(\pinv{\xi})/n^q-1}{1-q},
\end{equation}
which is a decreasing generator function, see~\eqref{eq:mon.Tq}.
For $q\in\field{R}$, let us have
the \emph{Rényi function}
\begin{equation} 
\label{eq:Rq}
R_q(\pinv{\xi}) = \frac1{1-q} \ln\Bigl( \sum_{x\in\pinv{\xi}} \bigl(\frac{x}{n}\bigr)^q\Bigr)
= \frac{\ln(s_q(\pinv{\xi})) - \ln (n^q)}{1-q},
\end{equation}
which is a decreasing generator function, see~\eqref{eq:mon.Rq}.
Let us also have
the \emph{Shannon function}
\begin{equation}
\label{eq:S} 
S(\pinv{\xi}) = -\sum_{x\in\pinv{\xi}} \frac{x}{n}\ln\bigl(\frac{x}{n}\bigr)
= \ln(n) - \frac1n\sum_{x\in\pinv{\xi}} x\ln(x),
\end{equation}
which is a decreasing generator function, see~\eqref{eq:mon.S}.
\end{subequations}
(For illustration, see Figure~\ref{fig:fPp.T2R2SP2}.)

Note that for the $0\leq q$ case 
these are the Rényi, Tsallis and Shannon entropies of
the normalized partitions $\pinv{\xi}/n=\smset{x/n}{x\in\pinv{\xi}}$.
Normalization is needed to avoid the discontinuity at $q=1$,
however, the resulting quantities are insensitive to the size $n$ of the system
(contrary to the max and min and power based generator functions in Sections~\ref{sec:1param.maxmin} and~\ref{sec:1param.pw}).
This could be considered as a disadvantage,
if the goal was to compare systems of different sizes,
however, this is not the case in the classification of partial separability,
since in the definition of the properties~\eqref{eq:vxir}
the value of the function does not matter. 

The Tsallis generator function has the limits
$T_{q\to+\infty}=0$, see~\eqref{eq:qlim.Tpinf}, and
$T_{q\to-\infty}(\top)=0$, while
$T_{q\to-\infty}(\pinv{\xi}\neq\top)=\infty$, see~\eqref{eq:qlim.Tninf}.
For $q=0$ and $1$, it is
$T_0(\pinv{\xi})=\abs{\pinv{\xi}}-1$, see~\eqref{eq:qlim.T0}, leading to a partitionability-like quantity,
and $T_1=S$, see~\eqref{eq:qlim.T1}, the Shannon generator function.
For all $q$ parameter values,
$T_q(\bot)=\frac{n^{1-q}-1}{1-q}=\ln_{(q)}(n)$ and $T_q(\top)=0$ for the extremal partitions,
so the range of $T_q$ is between these values~\eqref{eq:frange}.

The Rényi generator function has the limits
$R_{q\to+\infty}(\pinv{\xi})=\ln(n)-\ln(\max(\pinv{\xi}))$, see~\eqref{eq:qlim.Rpinf}, leading to a logarithmically flipped producibility, and
$R_{q\to-\infty}(\pinv{\xi})=\ln(n)-\ln(\min(\pinv{\xi}))$, see~\eqref{eq:qlim.Rninf}, leading to a logarithmically flipped toughness.
For $q=0$ and $1$, it is
$R_0(\pinv{\xi})= \ln(\abs{\pinv{\xi}})$, leading to logarithmic partitionability~\eqref{eq:qlim.R0}
and $R_1=S$, the Shannon generator function~\eqref{eq:qlim.R1}.
For all $q$ parameter values,
$R_q(\bot)=S(\bot)=\ln(n)$ and $R_q(\top)=S(\top)=0$ for the extremal partitions,
so the range of $R_q$ and $S$ is between these values~\eqref{eq:frange}.

We have then that $R_q$ for $q\in[-\infty,\infty]$
nicely interpolates between the negative logarithmic toughness, the logarithmic partitionability and the negative logarithmic producibility.
Taking the exponential of the Rényi generator function,
we have
\begin{equation}
\label{eq:Pq}
P_q(\pinv{\xi}) =
\ee^{R_q(\pinv{\xi})}
= \Bigl(\sum_{x\in\pinv{\xi}} \bigl(\frac{x}{n}\bigr)^q\Bigr)^{1/(1-q)}
= s_q\Bigl(\frac{\pinv{\xi}}{n}\Bigr)^{1/(1-q)},
\end{equation}
which is a decreasing generator function~\eqref{eq:mon.Pq}
(continuous at $q=1$),
interpolating between the reciprocal-toughness~\eqref{eq:qlim.Pninf}, 
the partitionability~\eqref{eq:qlim.P0}
and the reciprocal-producibility~\eqref{eq:qlim.Ppinf}.
Note that $P_q=\ee^{R_q} = \ee_{(q)}^{T_q}$
with the $q$-deformed exponential $e_{(q)}(x)=(1+(1-q)x)^{1/(1-q)}$,
in accordance with that $T_q$ is given by the $q$-deformed logarithm $\ln_{(q)}(x)=(x^{1-q}-1)/(1-q)$.
For all $q$ parameter values,
$P_q(\bot)=n$ and $P_q(\top)=1$ for the extremal partitions,
so the range of $P_q$ is between these values~\eqref{eq:frange}.

As before, for high $q$, the larger parts get more emphasis by $R_q$ and $P_q$,
in the sense that the smaller parts are more suppressed,
since these entropy based generator functions are bounded in $q$. This does not hold for $T_q$.
Note that, for a given $q\neq1$ parameter, $T_q$, $R_q$ and $P_q$ are connected with $s_q$ 
by the strictly monotone functions $g(u)=\frac{1}{1-q}(u/n^q-1)$, $g(u)=\frac{1}{1-q}\ln(u/n^q)$ and $g(u)=(u/n^q)^{1/(1-q)}$,
so they lead to the same classification by 
\eqref{eq:gprops.strict} and~\eqref{eq:gstates.strictD}-\eqref{eq:gstates.strictC},
although the interpolation properties are different,
and the resulting depths have different values and meanings,
see~\eqref{eq:gmeasuresfD}.

For $0\leq q$ the Rényi and Tsallis (and Shannon) functions
are indeed the respective entropies of the subsystem size distribution $\pinv{\xi}/n=\smset{x/n}{x\in\pinv{\xi}}$.
So in this case we call them \emph{entropic generator functions} (contrary to the more general `entropy based').
We will turn back to entropic (dominance-monotone) properties later 
in Sections~\ref{sec:1param.remarks} and~\ref{sec:Metro.dommon} and Appendix~\ref{app:1param},
since they play an important role in the usefulness of one-parameter properties in formulating metrological bounds,
here we just mention the most important point.
The entropic one-parameter properties
(and also those which are monotone functions of them,
such as $s_q$~\eqref{eq:sq}, $N_q$~\eqref{eq:Nq}, $P_q$~\eqref{eq:Pq} for $0\leq q$)
can also be endowed with entropic motivation then:
they characterize the mixedness of the subsystem sizes.
The more pure the subsystem size distribution, the more entangled the state is.

For these functions we have
the $(k,T_q)$-, $(k,R_q)$-, $(k,S)$- and $(k,P_q)$-entanglement of formation
and the relative entropy of $(k,T_q)$-, $(k,R_q)$-, $(k,S)$- and $(k,P_q)$-entanglement~\eqref{eq:rfEoFR}
in the usual way~\cite{Szalay-2019}.
We also have the corresponding depths,
the $T_q$-, $R_q$-, $S$- and $P_q$-entanglement depth~\eqref{eq:Depthf}
and $T_q$-, $R_q$-, $S$- and $P_q$-entanglement depth of formation~\eqref{eq:DepthOFf}.
For example, if the depth of Shannon-entanglement of a state $\rho$ is $D_S(\rho)=k$, then,
to mix it, there is a need for pure entanglement
where the Shannon function of the sizes of entangled subsystems is at most $k$.

Let us emphasize that the one-parameter multipartite entanglement properties
given by entropy based generator functions~\eqref{eq:qentr} 
are about the subsystem size distribution in a partition of a multipartite system,
which we could call `Rényi property' or `Tsallis property of multipartite entanglement'.
The resulting depths and other entanglement measures should not be confused with,
e.g., Rényi or Tsallis entanglement entropies~\cite{Vidal-2000,Horodecki-2001},
which are bipartite entanglement measures,
the Rényi or Tsallis entropies of the Schmidt coefficients for pure states,
and the convex roof extensions of these for general states.

\subsection{\texorpdfstring{The $q=2$ case: squareability}{The q=2 case: squareability}}
\label{sec:1param.2}

The degree $2$ case of the above, power based generator functions
are of particular importance, as we will see in Section~\ref{sec:Metro}.
Let us collect these here.
First, let us have the \emph{squareability}
\begin{equation}
\label{eq:sq2}
s_2(\pinv{\xi})=\sum_{x\in\pinv{\xi}} x^2
\end{equation}
of the partition $\pinv{\xi}\in\pinv{P}_\text{I}$ by~\eqref{eq:sq}, 
by which we also have
$N_2 = s_2^{1/2}$ by~\eqref{eq:Nq},
$T_2 = 1-s_2/n^2$ by~\eqref{eq:Tq},
$R_2 = 2\ln(n)-\ln(s_2)$ by~\eqref{eq:Rq} and
$P_2 = n^2/s_2$ by~\eqref{eq:Pq}.
(For illustration, see Figures~\ref{fig:fPp.s2N2M2}, and~\ref{fig:fPp.T2R2SP2}.)
Note that $s_2:\pinv{P}_\text{I}\to\field{N}$ is not injective,
(neither $N_2$, $T_2$, $R_2$, $P_2$)
for example $s_2(\mset{2,2,2})=s_2(\mset{3,1,1,1})=12$.
Note also that $s_2(\bot)=n$ and $s_2(\top)=n^2$ \emph{uniquely},
and $s_2(\pinv{\xi}\finer\top)\leq(n-1)^2+1^2=n^2-2n+2$.

For the squareability $s_2$ we have
the space of $k$-squareable states
and class of strictly $k$-squareable states
\begin{subequations}
\begin{align}
\label{eq:Dsq2}
\mathcal{D}_{k\text{-sq}} &:= \mathcal{D}_{k,s_2}
\equiv \Conv \Biggl(  \bigcup_{\substack{\pinv{\xi}\in\pinv{P}_\text{I}:\\ \sum_{x\in\pinv{\xi}} x^2\leq k}} \mathcal{P}_{\pinv{\xi}}  \Biggr),\\
\mathcal{C}_{k\text{-sq}} &:= \mathcal{C}_{k,s_2}
\equiv \begin{cases}
\mathcal{D}_{n\text{-sq}}& \text{for $k=n$},\\
\mathcal{D}_{k\text{-sq}} \setminus \mathcal{D}_{k_-\text{-sq}}& \text{else},
\end{cases}
\end{align}
\end{subequations}
by~\eqref{eq:DsepIIpf} and~\eqref{eq:CsepIIIpf}.
The whole state space is $\mathcal{D}_{n^2\text{-sq}}\equiv\mathcal{D}$,
the space of biseparable states is $\mathcal{D}_{(n^2-2n+2)\text{-sq}}$,
the class of genuinely multipartite entangled states is $\mathcal{C}_{n^2\text{-sq}}=\mathcal{D}_{n^2\text{-sq}}\setminus\mathcal{D}_{(n^2-2n+2)\text{-sq}}$,
and the space of fully separable states is $\mathcal{D}_{n\text{-sq}}\equiv\mathcal{C}_{n\text{-sq}}$.

\begin{figure*}\centering
\includegraphics{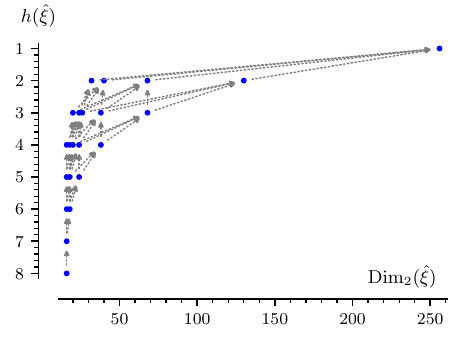}
\includegraphics{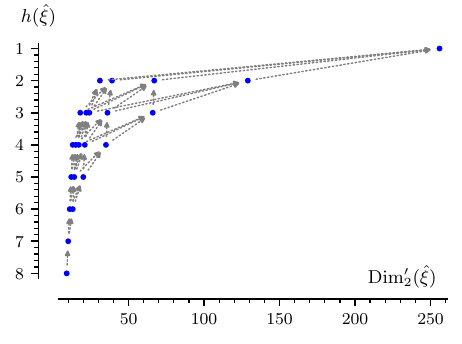}
\includegraphics{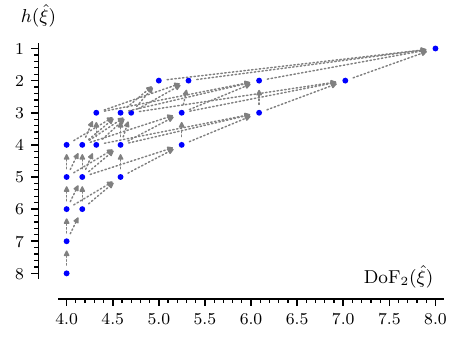}
\includegraphics{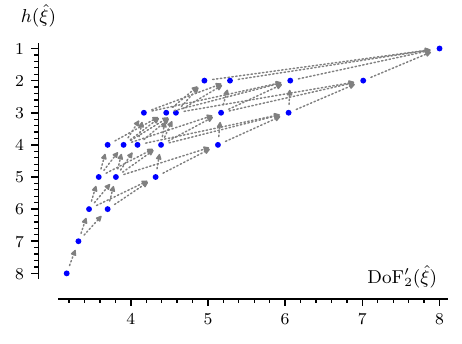}
\caption{The height~\eqref{eq:hwr.h} 
vs.~entanglement dimension~\eqref{eq:Dimd}, projective entanglement dimension~\eqref{eq:Dimpd},
entanglement degree of freedom~\eqref{eq:DoFd} and projective entanglement degree of freedom~\eqref{eq:DoFpd}
plots of $\pinv{P}_\text{I}$ for qubits $d=2$, $n=8$.}
\label{fig:fPp.Dim2Dimp2DoF2DoFp2}
\end{figure*}

For the squareability $s_2$ we have
the squareability-entanglement of formation 
and the relative entropy of squareability-entanglement~\eqref{eq:rfEoFR}
in the usual way~\cite{Szalay-2019}.
We also have the corresponding depths,
the (entanglement) squareability depth 
and the (entanglement) squareability depth of formation
\begin{subequations}
\label{eq:sq2ent}
\begin{align}
\label{eq:sq2ent.Depth}
\begin{split}
D_\text{sq} := D_{s_2}
&\equiv \min\bigsset{k\in s_2(\pinv{P}_\text{I})}{\rho\in \mathcal{D}_{k\text{-sq}} }\\
&\equiv \min\limits_{\set{(p_j,\pi_j)}\decomp \rho}  \max\limits_j D_\text{sq}(\pi_j),
\end{split} \\
\label{eq:sq2ent.DepthoF}
D^\text{oF}_\text{sq} := D^\text{oF}_{s_2}
&\equiv \min\limits_{\set{(p_j,\pi_j)}\decomp \rho} \sum_j p_j D_\text{sq}(\pi_j),
\end{align}
\end{subequations}
by~\eqref{eq:Depthf},~\eqref{eq:Depthfdec} and~\eqref{eq:DepthOFf}.
For example, if the squareability depth of a state $\rho$ is $D_\text{sq}(\rho)=k$, then,
to mix it, there is a need for pure entanglement
where the squareability of the sizes of entangled subsystems is at least $k$.
On the other hand, 
$D_\text{sq}(\rho)=n^2$ for genuinely multipartite entangled states $\rho\in\mathcal{C}_{n^2\text{-sq}}$,
$D_\text{sq}(\rho)\leq n^2-2n+2$ for biseparable states $\rho\in\mathcal{D}_{(n^2-2n+2)\text{-sq}}$,
and $D_\text{sq}(\rho)=n$ for fully separable states $\rho\in\mathcal{C}_{n\text{-sq}}\equiv\mathcal{D}_{n\text{-sq}}$.

\subsection{Entanglement dimension and entanglement degree of freedom}
\label{sec:1param.DOF}

After the variants of power functions,
a radically different way of characterization can be given by exponentials.

For $d=\dim(\mathcal{H}_i)\geq2$, let us have the \emph{entanglement dimension}
(or \emph{effective dimension})
\begin{subequations} 
\label{eq:Dim}
\begin{equation} 
\label{eq:Dimd}
\Dim_d(\pinv{\xi}) := \sum_{x\in\pinv{\xi}} d^x,
\end{equation}
which is an increasing generator function, see~\eqref{eq:mon.Dimd}.
It expresses the dimension of the hypothetical Hilbert space,
which is needed
to describe the components $\bigset{\cket{\psi_X}\in\mathcal{H}_X}_{X\in\xi}$
of a $\pinv{\xi}$-separable (unnormalized) vector
$\cket{\psi_\xi}=\bigotimes_{X\in\xi}\cket{\psi_X}$ one by one, where $\pinv{\xi}=\sset{\abs{X}}{X\in\xi}$.
If normalizations and complex phases are also taken into account,
then we have the \emph{projective entanglement dimension}
(or \emph{effective projective dimension})
\begin{equation} 
\label{eq:Dimpd}
\Dim'_d(\pinv{\xi}) := \sum_{x\in\pinv{\xi}} (d^x-1) +1 
= \Dim_d(\pinv{\xi})-\abs{\pinv{\xi}} +1,
\end{equation}
\end{subequations}
which is an increasing generator function, see~\eqref{eq:mon.Dimpd}.
It expresses the dimension of the hypothetical Hilbert space,
the parameters of the pure states of which are needed
to describe a $\pinv{\xi}$-separable pure state $\proj{\psi_\xi}$.
For all $d\geq2$ dimensions,
$\Dim_d(\bot) = nd$ and $\Dim_d(\top) = d^n$, as well as
$\Dim'_d(\bot) = n(d-1)+1$ and $\Dim'_d(\top) = d^n$,
for the extremal partitions,
so the ranges of $\Dim_d$ and $\Dim'_d$ are between these values~\eqref{eq:frange}, respectively.
(For illustration, see Figures~\ref{fig:fPp.Dim2Dimp2DoF2DoFp2}.)

For $d=\dim(\mathcal{H}_i)$, let us have the \emph{entanglement degree of freedom}
(or \emph{effective size})
\begin{subequations} 
\label{eq:DoF}
\begin{equation}
\label{eq:DoFd}
\DoF_d(\pinv{\xi}) := \log_d\Bigl(\sum_{x\in\pinv{\xi}} d^x\Bigr)
=\log_d\bigl( \Dim_d(\pinv{\xi}) \bigr),
\end{equation}
which is an increasing generator function, see~\eqref{eq:mon.DoFd}.
It expresses the number of hypothetical $d$-dimensional Hilbert spaces,
the composite system of which are needed
to describe the components of a $\pinv{\xi}$-separable (unnormalized) vector one by one.
If normalizations and complex phases are also taken into account,
then we have the \emph{projective entanglement degree of freedom}
(or \emph{effective projective size})
\begin{equation}
\label{eq:DoFpd}
\DoF'_d(\pinv{\xi}) := \log_d\bigl( \Dim'_d(\pinv{\xi}) \bigr)
=\log_d\bigl( \Dim_d(\pinv{\xi})-\abs{\pinv{\xi}} +1 \bigr),
\end{equation}
\end{subequations}
which is an increasing generator function, see~\eqref{eq:mon.DoFpd}.
It expresses the number of hypothetical qu\textit{d}its,
the parameters of the pure states of the composite system of which are needed
to describe a $\pinv{\xi}$-separable pure state $\proj{\psi_\xi}$.
For all $d\geq2$ dimensions,
$\DoF_d(\bot) = \log_d(n)+1$ and $\DoF_d(\top) = n$, as well as
$\DoF'_d(\bot) = \log_d(n(d-1)+1)$ and $\DoF'_d(\top) = n$,
for the extremal partitions,
so the ranges of $\DoF_d$ and $\DoF'_d$ are between these values~\eqref{eq:frange}, respectively.
(For illustration, see Figures~\ref{fig:fPp.Dim2Dimp2DoF2DoFp2}.)

Note that,
contrary to the generator functions introduced earlier,
$\Dim$, $\Dim'$, $\DoF$ and $\DoF'$ are dimension-sensitive,
which may be an advantage when entanglement is compared among systems
with different $d=\dim(\mathcal{H}_l)$ dimension of the Hilbert space of elementary subsystems.

Note that, for a given dimension $d$, the functions $\DoF_d$ and $\DoF'_d$ are connected with $\Dim_d$ and $\Dim'_d$ 
by the strictly monotone functions $g(u)=\ln_d(u)$,
so they lead to the same classification by 
\eqref{eq:gprops.strict} and~\eqref{eq:gstates.strictD}-\eqref{eq:gstates.strictC},
although the resulting depths have different values and meanings,
see~\eqref{eq:gmeasuresfD}.

For these functions we have
the $(k,\Dim)$-, $(k,\Dim')$-, $(k,\DoF)$- and $(k,\DoF')$-entanglement of formation 
and the relative entropy of $(k,\Dim)$-, $(k,\Dim')$-, $(k,\DoF)$- and $(k,\DoF')$-entanglement~\eqref{eq:rfEoFR}
in the usual way~\cite{Szalay-2019}.
We also have the corresponding depths,
the $\Dim$-, $\Dim'$-, $\DoF$- and $\DoF'$-entanglement depth~\eqref{eq:Depthf}
and $\Dim$-, $\Dim'$-, $\DoF$- and $\DoF'$-entanglement depth of formation~\eqref{eq:DepthOFf}.
For example, if the depth of $\DoF$-entanglement of a state $\rho$ is $D_{\DoF_d}(\rho)=k$ then,
to mix it, there is a need for pure entanglement
where the degree of freedom is at least $k$.

\subsection{Remarks}
\label{sec:1param.remarks}

Here we list some remarks 
on the generator functions defining one-parameter partial entanglement properties,
one paragraph each.

Note that
for partitionability, producibility, stretchability and toughness,
the generator functions were 
the height, width, rank~\eqref{eq:hwr} and toughness~\eqref{eq:t}
of the Young diagram representing the partition~\cite{Szalay-2019}.
In some sense, these characterize `global properties' of the partitions $\pinv{\xi}\in\pinv{P}_\text{I}$,
describing the multipartite entanglement properties.
We also defined functions of the form
$f(\pinv{\xi}) = \sum_{x\in\pinv{\xi}} f_0(x)$,
such as the power sum~\eqref{eq:sq}, the squareability~\eqref{eq:sq2},
and the entanglement dimension~\eqref{eq:Dim}
(or monotone functions thereof, such as entropy based generator functions~\eqref{eq:qentr},~\eqref{eq:Pq}
and the entanglement degree of freedom~\eqref{eq:DoF})
characterizing `local properties' in some sense.
It is quite remarkable that
the power based local properties can reach the global properties as limits,
thanks to the continuity in the parameter $q$ 
(see Appendix~\ref{app:powerf.Genfqlim}).

Recall that if a generator function $f$ is injective then
it leads to a total order on $\pinv{P}_\text{I}$ (see Section~\ref{sec:PSprops.remarks}).
For the power-sum $s_q$~\eqref{eq:sq} for $q\in\field{N}$, injectivity means that the
nonlinear Diophantine equation 
$\sum_{x\in\pinv{\xi}}x^q=\sum_{y\in\pinv{\upsilon}}y^q$,
for $\pinv{\xi},\pinv{\upsilon}\in\pinv{P}_\text{I}$
has no nontrivial ($\pinv{\xi}\neq\pinv{\upsilon}$) solution.
Although several of these equations have nontrivial solutions
(for example, $2^2+2^2+2^2=3^2+1^2+1^2+1^2=12$ for $n=6$, $q=2$, or
$2^3+2^3+2^3+2^3=3^3+1^3+1^3+1^3+1^3+1^3=32$ for $n=8$, $q=3$),
these  seem to be rather sporadic.
On the other hand, for $q\not\in\field{N}$ the equation becomes transcendental, and we expect the injectivity of $s_q$.
For the entanglement dimension $f=\Dim_d$~\eqref{eq:Dimd} for $2\leq q\in\field{N}$, we again have the
exponential Diophantine equation 
$\sum_{x\in\pinv{\xi}}d^x=\sum_{y\in\pinv{\upsilon}}d^y$,
for $\pinv{\xi},\pinv{\upsilon}\in\pinv{P}_\text{I}$,
which, again, has some sporadic nontrivial solutions
(for example, $2^1+2^1+2^1+2^1=2^2+2^1+2^1=2^2+2^2=8$ for $n=4$, $d=2$, or
$3^2+3^2+3^2+3^2+3^2+3^2=3^3+3^1+3^1+3^1+3^1+3^1+3^1+3^1+3^1+3^1=54$ for $n=12$, $d=3$).
For the case $d\not\in\field{N}$, we expect the injectivity of $\Dim_d$,
although the meaning of $\Dim_d$ is not established for noninteger $d$,
which is just a parameter then.
These remarks hold also for generator functions which are strictly monotone functions of $s_q$ or $\Dim_d$, by~\eqref{eq:gprops.strict}.

Note that
in Section~\ref{sec:1param.entr} we defined some generator functions resembling entropies for $0\leq q$,
and our only concern was the refinement-monotonicity~\eqref{eq:genf}.
For those who are familiar with the theory of majorization
and its deep connection to entropies of probability distributions~\cite{Bengtsson-2006,Marshall-2010,Sagawa-2022},
it is an immediate question,
how the refinement $\finereq$ in the poset $\pinv{P}_\text{I}$ is related to the majorization.
In the discrete case of integer partitions, majorization is called dominance order~\cite{Brylawski-1973,Stanley-2012}, and
it turns out that if $\pinv{\xi}$ is coarser than $\pinv{\upsilon}$,
then $\pinv{\xi}$ dominates $\pinv{\upsilon}$,
but the reverse implication does not hold.
(For the proof, and for the characterization of the orders used, see Appendix~\ref{app:1param.Gen}.)
Consequently, any dominance-monotone function can be used as a generator function for one-parameter properties.
Recall that in the continuous case, 
Schur-concavity (decreasing majorization-monotonicity) is the defining property of entropies,
these are then monotones with respect to the mixing of discrete probability distributions~\cite{Bengtsson-2006}.
Therefore any entropy of the \emph{normalized} integer partitions $\pinv{\xi}/n$ 
(or dominance-monotone function of the integer partition $\pinv{\xi}$)
is a proper generator function, which we also call entropy.
The generator functions given in Section~\ref{sec:1param.entr}
are indeed entropies for $0\leq q$,
and generator functions but not entropies for $q<0$.
(For the refinement-monotonicity and dominance-monotonicity of the generator functions,
see Appendices~\ref{app:1param.Mon} and~\ref{app:1param.dmon}, respectively.)
We will come back to dominance-monotonicity later,
since it plays an important role in the construction of metrological multipartite entanglement criteria in Section~\ref{sec:Metro}.

Note that the $q$-dependent generator functions $s_q$, $N_q$
$R_q$, $P_q$ and $T_q$
are strictly monotone, convex/concave/affine functions of one another for fixed $q$.
As was already mentioned in Section~\ref{sec:PSprops.remarks},
this means that they lead to the same classification~\eqref{eq:gstates.strictD}-\eqref{eq:gstates.strictC}.
(See Appendix~\ref{app:trafg} for more details.)
Beyond this, the main point here is that
the strictly monotone connection among these functions
lends $s_q$ and all power based generator functions an entropic character
for the parameters $0\leq q$ (inside the defined range),
expressing the \emph{mixedness} of the size distribution.
This is noteworthy,
since they fail to have central moment character~\eqref{eq:xipmom1.central},~\eqref{eq:xipmom2.central},
expressing the \emph{spread} of the size distribution.

Note that,
for a generator function $f$, the transformed function $f\circ\pconj$ is not a generator function in general 
($\pconj$ is the conjugation of integer partitions, see Section~\ref{sec:PSprops.remarks},~\cite{Stanley-2012}).
The dominance order, contrary to the refinement, transforms well with conjugation~\cite{Brylawski-1973},
$\pinv{\upsilon}$ is dominated by $\pinv{\xi}$ if and only if $\pinv{\xi}^\pconj$ is dominated by $\pinv{\upsilon}^\pconj$.
($\pconj$ is an \emph{antiautomorphism} of the dominance order.)
Then dominance-monotonicity is just flipped by conjugation,
$f$ is an increasing dominance-monotone if and only if $f\circ\pconj$ is a decreasing one, and vice versa;
so, in particular, $f\circ\pconj$ is also a generator function~\eqref{eq:poIs} in this case.
For example, $w_m$~\eqref{eq:extr.wm} is an increasing dominance-monotone~\eqref{eq:dmon.wm},
then we may define $h_m:=w_m\circ\pconj$, which is a decreasing dominance-monotone, so decreasing generator function.
On the other hand, $t_m$~\eqref{eq:extr.tm} is a generator function~\eqref{eq:mon.tm}, which is not dominance-monotone
(a counterexample can be given for $n=4$ already for $m=1$, the toughness~\eqref{eq:t}, where the partitions
$\mset{2,1,1}$, $\mset{2,2}$ and $\mset{3,1}$ are more and more dominant~\eqref{eq:poIpdom.p},
but $t(\mset{2,1,1})=1$, $t(\mset{2,2})=2$ and $t(\mset{3,1})=1$),
and $t_m\circ\pconj$ is not even a generator function
(we have
$\mset{1,1,1,1}\finereq\mset{2,1,1}\finereq\mset{2,2}$,
but 
$t(\mset{1,1,1,1}^\pconj)=t(\mset{4})=4$,
$t(\mset{2,1,1}^\pconj)=t(\mset{3,1})=1$ and
$t(\mset{2,2}^\pconj)=t(\mset{2,2})=2$).

\section{Metrology and \texorpdfstring{$f$}{f}-entanglement}
\label{sec:Metro}

In this section 
we construct bounds on the \emph{quantum Fisher information} in collective spin-z measurement of qubits,
for general one-parameter multipartite entanglement properties, given in terms of their \emph{depths}.
We call these metrological multipartite entanglement (partial separability) criteria.
It turns out that this is given by the squareability of that property, 
suggesting that entanglement squareability is natural from the point of view of quantum metrology.
We also formulate \emph{stronger}, general \emph{convex roof type bounds},
which are particularly simple to use in the case when it is given in terms of 
the \emph{entanglement depth of formation} and \emph{entanglement squareability depth of formation}.
We also consider the usefulness of different one-parameter properties for the purpose of the formulation of metrological entanglement criteria,
and identify dominance-monotonicity to be an important property for that.

\subsection{Quantum Fisher information}
\label{sec:Metro.qFI}

The Cramér-Rao bound~\cite{Helstrom-1969,Holevo-2011,Pezze-2014,Toth-2014,Pezze-2018}
\begin{equation}
\label{eq:QCramerRao}
(\Delta\theta)^2\geq\frac1{m F_\text{Q}(\rho,A)}
\end{equation}
is of central importance in quantum metrology,
it gives a lower bound
on the precision of the estimation
of the phase shift $\theta$ generated by the self-adjoint operator $A$ as
$\rho\mapsto \ee^{-i\theta A}\rho \ee^{i\theta A}$,
in case of $m$ independent repetition.
This bound is given by the \emph{quantum Fisher information}~\cite{Braunstein-1994,Pezze-2014},
which can be written as
\begin{subequations}
\begin{equation}
\label{eq:Fisher}
F_\text{Q}(\rho,A) = 
2\sum_{\substack{i,j=1,\\\lambda_i+\lambda_j\neq0}}^d 
\frac{(\lambda_i-\lambda_j)^2}{\lambda_i+\lambda_j} \abs{\bra{\phi_i}A\cket{\phi_j}}^2,
\end{equation}
where $\rho=\sum_{i=1}^d\lambda_i\proj{\phi_i}$ is the eigendecomposition of the state.
For pure states $\rho=\proj{\phi}\in\mathcal{P}$, this reduces to (four times) the variance~\cite{Pezze-2014},
\begin{equation}
\label{eq:FisherPure}
\begin{split}
F_\text{Q}\bigl(\proj{\phi},A\bigr) &= 4\bigl(\bra{\phi}A^2\cket{\phi}-\bra{\phi}A\cket{\phi}^2\bigr)\\
&= 4 \Var\bigl(\proj{\phi},A\bigr),
\end{split}
\end{equation}
\end{subequations}
it is convex in the quantum state~\cite{Fujiwara-2001,Pezze-2014},
\begin{subequations}
\begin{equation}
\label{eq:FisherConv}
F_\text{Q}\Bigl(\sum_i p_i \rho_i, A\Bigr) \leq
\sum_i p_i F_\text{Q}(\rho_i, A), 
\end{equation}
moreover, it is the convex roof extension of (four times) the variance~\cite{Toth-2013,Yu-2013},
\begin{equation}
\label{eq:FisherConvRoof}
F_\text{Q}(\rho,A) = \min_{\set{(p_j,\pi_j)}\decomp \rho}
 \sum_j p_j 4\Var(\pi_j,A).
\end{equation}
These properties have also led to new types of uncertainty relations~\cite{Toth-2022,Chiew-2022}.
From these, it also follows that
\begin{equation}
\label{eq:FisherUpper}
F_\text{Q}(\rho,A) \leq 4\bigl(\Tr(\rho A^2) -\Tr(\rho A)^2\bigr)
 = 4 \Var(\rho,A),
\end{equation}
\end{subequations}
since the variance is concave.
(It is moreover the concave roof extension of itself~\cite{Toth-2013}.)

\subsection{Partial entanglement criteria}
\label{sec:Metro.PS}
 
For $k$-producibly separable states,
for the collective spin-z observable 
$J^\text{z}=\sum_{l=1}^n \frac{1}{2}\sigma^\text{z}_l\otimes \Id_{\cmpl{\set{l}}}$
of the $n$-qubit system
(where $\cmpl{\set{l}}:=\set{1,2,\dots,n}\setminus\set{l}$),
the quantum Fisher information obeys a $k$-dependent (attainable) upper-bound~\cite{Toth-2012,Hyllus-2012}.
Here and in the following subsection we recall and extend the proof 
of those bounds for $k$-producibly separable states~\cite{Toth-2012,Hyllus-2012}
to states of general permutation invariant entanglement properties~\eqref{eq:DsepIIp},
then to states of one-parameter entanglement properties~\eqref{eq:DsepIIpf}.

First we consider pure states only.
Let us have a permutation invariant partial separability property $\pinv{\xi}\in\pinv{P}_\text{I}$.
The (attainable) upper bound for the quantum Fisher information of the collective spin-z observable of $n$ qubits
$J^\text{z}$ for $\pinv{\xi}$-separable pure states $\pi\in\mathcal{P}_{\pinv{\xi}}$ is
\begin{equation}
\label{eq:genFisherUpperpure}
\max_{\pi\in\mathcal{P}_{\pinv{\xi}}} F_\text{Q}(\pi,J^\text{z})
= s_2(\pinv{\xi}).
\end{equation}
Indeed,
\begin{equation*}
\begin{split}
&\max_{\pi\in\mathcal{P}_{\pinv{\xi}}} 
    F_\text{Q}(\pi,J^\text{z}) 
\equalsref{eq:FisherPure}
  \max_{\pi\in\mathcal{P}_{\pinv{\xi}}}
    4 \Var(\pi, J^\text{z})\\
&\quad\equals
  \max_{\forall \xi \in s^{-1}(\pinv{\xi})}
  \max_{\forall \pi\in\mathcal{P}_\xi}
    4 \Var(\pi, J^\text{z})\\
&\quad\equalsref{eq:sumvar}
  \max_{\forall \xi \in s^{-1}(\pinv{\xi})}
  \max_{\forall \pi\in\mathcal{P}_\xi}
   \sum_{X\in\xi} 4 \Var(\pi_X, J_X^z)\\
&\quad\equals
  \max_{\forall \xi \in s^{-1}(\pinv{\xi})}
  \sum_{X\in\xi} 
  \max_{\pi_X\in\mathcal{P}_X}
  4 \Var(\pi_X, J_X^z)\\
&\quad\equalsref{eq:varJXboundqubit}
  \max_{\forall \xi \in s^{-1}(\pinv{\xi})}
  \sum_{X\in\xi} \abs{X}^2
\equals
  \sum_{x\in\pinv{\xi}} x^2 \equalsref{eq:sq2} s_2(\pinv{\xi}),
\end{split}
\end{equation*}
where 
the \emph{first equality} holds, because the quantum Fisher information is four times the variance for pure states~\eqref{eq:FisherPure};
the \emph{second equality} holds, using the definition $\mathcal{P}_\xi=\bigsset{\bigotimes_{X\in\xi}\pi_X}{\pi_X\in\mathcal{P}_X}$,
where $\xi=\set{X_1,X_2,\dots,X_{\abs{\xi}}}$ is a partition of the whole system into parts $X_j$ of sizes given by $\pinv{\xi}$,
that is, $s(\xi)=\bigsset{\abs{X}}{X\in\xi}=\pinv{\xi}$ (multiset),
so we have the $\mathcal{P}_{\pinv{\xi}}=\bigcup_{\xi\in s^{-1}(\pinv{\xi})}\mathcal{P}_\xi$ union (see~\cite{Szalay-2019} for more details);
the \emph{third equality} holds,
because the variance-squared of collective operators in product states 
is the sum of the variances of the local operators (see Appendix~\ref{app:MetroVar.sum});
the \emph{fourth equality} holds, because the maximization of the sum is taken over independent variables for each summand;
the \emph{fifth equality} holds, because the variance-squared of collective spin-z measurements is upper-bounded by $\abs{X}^2/4$,
and the bound can be attained (see Appendix~\ref{app:MetroVar.JzXbound});
the \emph{sixth equality} holds, because the function maximized is constant with respect to $\xi$
on the set $s^{-1}(\pinv{\xi})$,
and the \emph{the last equality} is just the definition~\eqref{eq:sq2}.

Now we turn to mixed states.
Let us have a permutation invariant partial separability property $\pinv{\vs{\xi}}\in\pinv{P}_\text{II}$.
The (attainable) upper bound for the quantum Fisher information of the collective spin-z observable of $n$ qubits
$J^\text{z}$
for $\pinv{\vs{\xi}}$-separable states $\rho\in\mathcal{D}_{\pinv{\vs{\xi}}}$ is
\begin{equation}
\label{eq:genFisherUpper}
\max_{\rho\in\mathcal{D}_{\pinv{\vs{\xi}}}} F_\text{Q}(\rho,J^\text{z})
= \max_{\pinv{\xi}\in\pinv{\vs{\xi}}} s_2(\pinv{\xi})
\equiv s_2(\pinv{\vs{\xi}}).
\end{equation}
Indeed,
\begin{equation*}
\begin{split}
&\max_{\rho\in\mathcal{D}_{\pinv{\vs{\xi}}}} 
    F_\text{Q}(\rho,J^\text{z}) 
\equalsref{eq:DPsepIIp}
  \max_{\pi\in\mathcal{P}_{\pinv{\vs{\xi}}}} 
    F_\text{Q}(\pi,J^\text{z})\\
&\quad\equalsref{eq:PsepIIp}
  \max_{\pinv{\xi}\in\pinv{\vs{\xi}}}
  \max_{\pi\in\mathcal{P}_{\pinv{\xi}}}
    F_\text{Q}(\pi,J^\text{z})
\equalsref{eq:genFisherUpperpure}
  \max_{\pinv{\xi}\in\pinv{\vs{\xi}}}
  s_2(\pinv{\xi})
 \equalsref{eq:fII} s_2(\pinv{\vs{\xi}}),
\end{split}
\end{equation*}
where the \emph{first equality} holds, because the quantum Fisher information is a convex function in the first variable~\eqref{eq:FisherConv},
and the maximization is taken over a convex set~\eqref{eq:DsepIIp},
so the maximum is attained on the extremal points~\eqref{eq:PsepIIpExtr};
the \emph{second equality} holds, because the maximum is taken over the union of state spaces~\eqref{eq:PsepIIp};
the \emph{third equality} holds, because of the result~\eqref{eq:genFisherUpperpure} for pure states;
the \emph{fourth equality} is just definition~\eqref{eq:fII}, noting that $s_2$ is an increasing generator function~\eqref{eq:mon.sq}. 

The bound~\eqref{eq:genFisherUpper} gives a necessary but not sufficient criterion of $\pinv{\vs{\xi}}$-separability.
If $\rho$ is $\pinv{\vs{\xi}}$-separable then $F_\text{Q}(\rho,J^\text{z})\leq s_2(\pinv{\vs{\xi}})$, or, contrapositively,
if $F_\text{Q}(\rho,J^\text{z})>s_2(\pinv{\vs{\xi}})$ then $\rho$ is $\pinv{\vs{\xi}}$-entangled.

From~\eqref{eq:genFisherUpper} we also have the side-result
that for all states $\rho\in\mathcal{D}$,
\begin{equation}
\label{eq:FisherBrutal}
F_\text{Q}(\rho,J^\text{z})
\leq \min_{\pinv{\vs{\xi}}\in\pinv{P}_\text{II}: \rho\in\mathcal{D}_{\pinv{\vs{\xi}}}} s_2(\pinv{\vs{\xi}}),
\end{equation}
since~\eqref{eq:genFisherUpper} holds for all $\pinv{\vs{\xi}}$
for which $\rho$ is $\pinv{\vs{\xi}}$-separable.

For general permutation invariant partial separability properties $\pinv{\vs{\xi}}\in\pinv{P}_\text{II}$,
the right-hand side of~\eqref{eq:genFisherUpper} cannot be evaluated to get a closed form.
We will see some examples in Section~\ref{sec:Metro.1paramxmpl} when it can.
To find the maximum of $s_2$~\eqref{eq:sq2} for the partitions $\pinv{\xi}\in\pinv{\vs{\xi}}$,
it is useful that $s_2$ increases if we move a subsystem from a smaller part to a larger one,
that is, for $x_i\geq x_j$, we have
\begin{equation}
\label{eq:s2useful}
\begin{split}
&s_2(\set{\dots,x_i,\dots,x_j,\dots})\\
&\quad< s_2(\set{\dots,x_i+1,\dots,x_j-1,\dots}),
\end{split}
\end{equation}
which is easy to check.
Note that
such step is usually not related by refinement in $\pinv{P}_\text{I}$~\eqref{eq:poIp.x},
which is, however, not a problem,
since~\eqref{eq:s2useful} is only used for the maximization~\eqref{eq:genFisherUpper} over the whole $\pinv{P}_\text{I}$,
not related to the classification structure.

\subsection{\texorpdfstring{$f$}{f}-entanglement criteria}
\label{sec:Metro.1param}

For the one-parameter partial separability properties $\pinv{\vs{\xi}}_{k,f}\in\pinv{P}_{\text{II},f}$, given in~\eqref{eq:vxir}
by the generator function $f$,
the (attainable) bound~\eqref{eq:genFisherUpper} takes the form
\begin{subequations}
\label{eq:FisherfBound}
\begin{equation}
\label{eq:Fisherkf}
\begin{split}
\max_{\rho\in\mathcal{D}_{k,f}} F_\text{Q}(\rho,J^\text{z})
&= \max_{\pinv{\xi}\in f^\lesseqgtr(k)} s_2(\pinv{\xi})\\
&\equiv s_2(\pinv{\vs{\xi}}_{k,f})
=:b_f(k)
\end{split}
\end{equation}
by~\eqref{eq:DsepIIpf}.
This gives a necessary but not sufficient criterion of $(k,f)$-separability.
If $\rho$ is $(k,f)$-separable then $F_\text{Q}(\rho,J^\text{z})\leq b_f(k)$, or, contrapositively,
if $F_\text{Q}(\rho,J^\text{z})>b_f(k)$ then $\rho$ is $(k,f)$-entangled.
On the other hand,
for all generator functions $f$, 
for all states $\rho\in\mathcal{D}$,
we have the bound in terms of the $f$-entanglement depth~\eqref{eq:Depthf} as
\begin{equation}
\label{eq:FisherfDepth}
\begin{split}
F_\text{Q}(\rho,J^\text{z})
&\leq 
\max_{\pinv{\xi}\in f^\lesseqgtr(D_f(\rho))} s_2(\pinv{\xi})\\
&\equiv s_2\bigl(\pinv{\vs{\xi}}_{D_f(\rho),f}\bigr) = b_f(D_f(\rho)) =: B_f(\rho).
\end{split}
\end{equation}
\end{subequations}
(Indeed, noting that $\rho\in\mathcal{D}_{D_f(\rho),f}$ by~\eqref{eq:Depthf} or~\eqref{eq:DepthfDlevel},
this is just
$F_\text{Q}(\rho,J^\text{z})\leq\max_{\rho'\in\mathcal{D}_{D_f(\rho),f}} F_\text{Q}(\rho',J^\text{z})$,
the right-hand side of which we already have in~\eqref{eq:Fisherkf}.)
This expresses the relation of two functions over the state space $\mathcal{D}$,
and this point of view will lead us to a completely new type of (convex) multipartite entanglement criteria later (see Section~\ref{sec:Metro.1paramoFxmpl}),
which are more sophisticated than the criteria of mere $(k,f)$-separability.

It follows by construction that the bound $b_f(k)$ is monotone,
that is, 
for a generator function $f$,
for all $k,k'\in f(\pinv{P}_\text{I})$, we have
\begin{equation}
\label{eq:monB.gen}
k \lesseqgtr k' \dspthen b_f(k) \leq b_f(k').
\end{equation}
(Indeed, for $k \lesseqgtr k'$ the maximization in~\eqref{eq:Fisherkf}
is taken over larger set, $f^\lesseqgtr(k)\subseteq f^\lesseqgtr(k')$, recalling~\eqref{eq:vxirchain}.)
Then for a generator function $f$, the composition $b_f\circ f$ is also a generator function~\eqref{eq:genf},
and the~\eqref{eq:FisherfDepth} form of the bound
is actually the depth of the induced $(b_f\circ f)$-entanglement,
\begin{equation}
\label{eq:BfDbff}
B_f(\rho)  
= b_f(D_f(\rho)) = D_{b_f\circ f}(\rho),
\end{equation}
by~\eqref{eq:gmeasuresfD},
that is, an entanglement measure itself,
and the bound~\eqref{eq:FisherfDepth} 
takes the form
\begin{equation}
\label{eq:FisherfDbff}
F_\text{Q}(\rho,J^\text{z}) \leq D_{b_f\circ f}(\rho)
\end{equation}
by the depth of the induced one-parameter property.

Since the left-hand side of~\eqref{eq:FisherfDepth} or~\eqref{eq:FisherfDbff} does not depend
on the particular one-parameter property given by the generator function $f$,
we also have that, for all states $\rho\in\mathcal{D}$,
\begin{equation}
\label{eq:Fisher1paramBrutal}
F_\text{Q}(\rho,J^\text{z})\leq 
\min_\text{$f$ as in~\eqref{eq:genf}} B_f(\rho)
= \min_\text{$f$ as in~\eqref{eq:genf}} D_{b_f\circ f}(\rho),
\end{equation}
where the minimization is taken for all the possible generator functions~\eqref{eq:genf}.
This is equivalent to the bound~\eqref{eq:FisherBrutal}, now formulated for one-parameter partial entanglement properties,
which can be seen by considering injective generator functions.

\subsection{Examples of \texorpdfstring{$f$}{f}-entanglement criteria}
\label{sec:Metro.1paramxmpl}

Now let us see the bounds~\eqref{eq:FisherfBound}
for the case of some particular one-parameter partial entanglement properties.

\begin{figure}\centering
\includegraphics{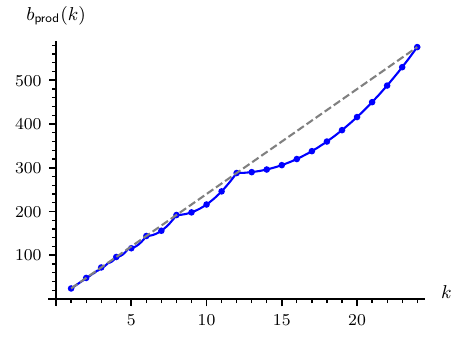}
\includegraphics{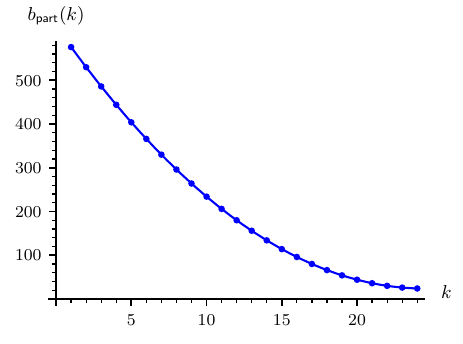}
\includegraphics{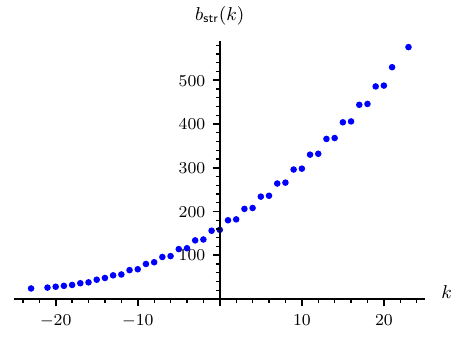}
\caption{The bound~\eqref{eq:Fisherkf} for $k$-producible~\eqref{eq:FisherBoundprod.k}, $k$-partitionable~\eqref{eq:FisherBoundpart.k} and $k$-stretchable~\eqref{eq:FisherBoundstr.k} states for $n=24$.
In the first case, the weaker bound~\eqref{eq:FisherBoundprod2.k} is also shown with dashed line.}
\label{fig:Bk.whr}
\end{figure}

For producibility,
$\pinv{\vs{\xi}}_{k\text{-prod}}$ is given by the $f=w=\max$ width generator function~\eqref{eq:vxik.prod},
which is increasing,
so if $\rho\in\mathcal{D}_{k\text{-prod}}$ for a given $k\in w(\pinv{P}_\text{I})$,
then we have the constraint $\max(\pinv{\xi})\leq k$ in~\eqref{eq:Fisherkf}.
Among such partitions $\pinv{\xi}$, by~\eqref{eq:s2useful}, we have to increase the parts $x$ as much as possible,
so the maximum of $s_2$ is taken for
$\pinv{\xi}=\mset{k,k,\dots,k,r}$, where we have $\lfloor n/k \rfloor$ parts of size $k$,
and one remainder part of size $r=n-\lfloor n/k \rfloor k$ (if not zero).
For this, the maximum value is $s_2(\mset{k,k,\dots,k,r})=\lfloor n/k \rfloor k^2 + (n-\lfloor n/k \rfloor k)^2$,
so we have the attainable bound~\eqref{eq:Fisherkf} on the quantum Fisher information
for $k$-producible states~\cite{Toth-2012,Hyllus-2012}
\begin{subequations}
\label{eq:FisherBoundprod}
\begin{equation}
\label{eq:FisherBoundprod.k}
\begin{split}
\max_{\rho\in\mathcal{D}_{k\text{-prod}}}
&F_\text{Q}(\rho,J^\text{z})
= b_\text{prod}(k) \\
&\quad= \Bigl\lfloor\frac{n}{k}\Bigr\rfloor k^2 + \Bigl(n-\Bigl\lfloor\frac{n}{k}\Bigr\rfloor k\Bigr)^2.
\end{split}
\end{equation}
This gives a necessary but not sufficient criterion of $k$-producible separability.
If $\rho$ is $k$-producibly separable then $F_\text{Q}(\rho,J^\text{z})\leq \lfloor n/k\rfloor k^2 + \bigl(n-\lfloor n/k\rfloor k\bigr)^2$, or, contrapositively,
if $F_\text{Q}(\rho,J^\text{z})>\lfloor n/k\rfloor k^2 + \bigl(n-\lfloor n/k\rfloor k\bigr)^2$ then $\rho$ is not $k$-producibly separable.
(For illustration, see Figure~\ref{fig:Bk.whr}.)
On the other hand, for all $\rho\in\mathcal{D}$,
we have the bound~\eqref{eq:FisherfDepth}
in terms of the entanglement depth $D\equiv D_\text{prod}$~\eqref{eq:Depth.prod} as
\begin{equation}
\label{eq:FisherBoundprod.D}
\begin{split}
&F_\text{Q}(\rho,J^\text{z})
\leq B_\text{prod}(\rho)\\
&\quad= \Bigl\lfloor\frac{n}{D(\rho)}\Bigr\rfloor D(\rho)^2 + 
\Bigl(n-\Bigl\lfloor\frac{n}{D(\rho)}\Bigr\rfloor D(\rho)\Bigr)^2.
\end{split}
\end{equation}
\end{subequations}
These recover the formerly known bounds 
for fully separable (that is, $1$-producible) states, $F_\text{Q}(\rho,J^\text{z})\leq n$,
and for all (that is, $n$-producible) states, $F_\text{Q}(\rho,J^\text{z})\leq n^2$~\cite{Giovannetti-2006,Pezze-2009}
as special cases.
 
On the other hand, since $\lfloor n/k \rfloor\leq n/k$, we have $s_2(\mset{k,k,\dots,k,r})\leq nk$,
and the resulting bound 
\begin{subequations}
\label{eq:FisherBoundprod2}
\begin{equation}
\label{eq:FisherBoundprod2.k}
\rho\in\mathcal{D}_{k\text{-prod}}:\quad
F_\text{Q}(\rho,J^\text{z})\leq nk
\end{equation}
is often used (see~\cite{Hyllus-2012,Apellaniz-2015}, also equation (8.2.12) in~\cite{Toth-2021} and as special case in~\cite{Ren-2021}),
being slightly weaker than~\eqref{eq:FisherBoundprod}, but easier to handle.
This gives a necessary but not sufficient criterion of $k$-producible separability.
If $\rho$ is $k$-producibly separable then $F_\text{Q}(\rho,J^\text{z})\leq nk$, or, contrapositively,
if $F_\text{Q}(\rho,J^\text{z})> nk$ then $\rho$ is not $k$-producibly separable,
however, in this case less states are identified,
since the bound is not attainable if $k$ does not divide $n$.
(For illustration, see Figure~\ref{fig:Bk.whr}.)
On the other hand, for all states $\rho\in\mathcal{D}$, we have the bound
in terms of its producibility entanglement depth~\eqref{eq:Depth.part} as
\begin{equation}
\label{eq:FisherBoundprod2.D}
F_\text{Q}(\rho,J^\text{z})\leq nD(\rho).
\end{equation}
\end{subequations}
(Indeed, noting that $\rho\in\mathcal{D}_{D(\rho)\text{-prod}}$ by~\eqref{eq:Depthf} or~\eqref{eq:DepthfDlevel},
this is just
$F_\text{Q}(\rho,J^\text{z})\leq\max_{\rho'\in\mathcal{D}_{D(\rho)\text{-prod}}} F_\text{Q}(\rho',J^\text{z})\leq nD(\rho)$,
by~\eqref{eq:FisherBoundprod2.k}. Note the slightly weaker bound here.)

For partitionability,
$\pinv{\vs{\xi}}_{k\text{-part}}$ is given by the $f=h=\abs{\cdot}$ height generator function~\eqref{eq:vxik.part},
which is decreasing,
so if $\rho\in\mathcal{D}_{k\text{-part}}$ for a given $k\in h(\pinv{P}_\text{I})$, 
then we have the constraint $\abs{\pinv{\xi}}\geq k$ in~\eqref{eq:Fisherkf}.
Among such partitions $\pinv{\xi}$, by~\eqref{eq:s2useful}, we have to increase the parts $x$ as much as possible,
while keeping at least $k$ parts,
so the maximum of $s_2$ is taken for
$\pinv{\xi}=\mset{n-k+1,1,1,\dots,1}$, where we have $k-1$ parts of size $1$,
and one part of size $n-(k-1)$.
For this, the maximum value is $s_2(\mset{n-k+1,1,1,\dots,1})=(n-k+1)^2+k-1=k^2-(2n+1)k+n^2+2n$,
so we have the attainable bound~\eqref{eq:Fisherkf} on the quantum Fisher information
for $k$-partitionable states
\begin{subequations}
\label{eq:FisherBoundpart}
\begin{equation}
\label{eq:FisherBoundpart.k}
\begin{split}
\max_{\rho\in\mathcal{D}_{k\text{-part}}} 
&F_\text{Q}(\rho,J^\text{z})
= b_\text{part}(k)\\
&\quad= k^2-(2n+1)k+n(n+2).
\end{split}
\end{equation}
This gives a necessary but not sufficient criterion of $k$-partitionable separability.
If $\rho$ is $k$-partitionably separable then $F_\text{Q}(\rho,J^\text{z})\leq k^2-(2n+1)k+n(n+2)$, or, contrapositively,
if $F_\text{Q}(\rho,J^\text{z})>k^2-(2n+1)k+n(n+2)$ then $\rho$ is not $k$-partitionably separable.
(For illustration, see Figure~\ref{fig:Bk.whr}.)
On the other hand, for all states $\rho\in\mathcal{D}$,
we have the bound~\eqref{eq:FisherfDepth}
in terms of the partitionability entanglement depth $D_\text{part}$~\eqref{eq:Depth.part} as
\begin{equation}
\label{eq:FisherBoundpart.D}
\begin{split}
&F_\text{Q}(\rho,J^\text{z})
\leq B_\text{part}(\rho)\\
&\quad= D_\text{part}(\rho)^2-(2n+1)D_\text{part}(\rho)+n(n+2).
\end{split}
\end{equation}
\end{subequations}
Note that this can also be seen from the combined producibility-partitionability (two-parameter) properties
also considered in the literature~\cite{Ren-2021}.

For stretchability,
$\pinv{\vs{\xi}}_{k\text{-str}}$ is given by the $f=r=\max - \abs{\cdot}$ rank generator function~\eqref{eq:vxik.str},
which is increasing,
so if $\rho\in\mathcal{D}_{k\text{-str}}$ for a given $k\in r(\pinv{P}_\text{I})$,
then we have the constraint $r(\pinv{\xi})\leq k$ in~\eqref{eq:Fisherkf}.
The maximization with respect to this in~\eqref{eq:Fisherkf} is rather involved, 
but still can be done explicitly~\cite{Ren-2021}, 
and we have the attainable bound~\eqref{eq:Fisherkf} on the quantum Fisher information
for $k$-stretchable states
\begin{equation}
\label{eq:FisherBoundstr.k}
\begin{split}
&\max_{\rho\in\mathcal{D}_{k\text{-str}}} F_\text{Q}(\rho,J^\text{z})=b_\text{str}(k)=\\
&\begin{cases}
n+24    &\begin{aligned}
&\text{if $n+k=10$}\\
&\quad\text{and $n\geq8$},
\end{aligned}\\
n+60    &\begin{aligned}
&\text{if $n+k=16$}\\
&\quad\text{and $n\geq12$},
\end{aligned}\\
\frac14(n+k)^2+\frac12(n-k)+2   &\text{if $n+k\in2\field{N}$},\\
\frac14(n+k+1)^2+\frac12(n-k-1) &\text{if $n+k\in2\field{N}+1$}.
\end{cases}
\end{split}
\end{equation}
This gives a necessary but not sufficient criterion of $k$-stretchable separability.
If $\rho$ is $k$-stretchably separable then $F_\text{Q}(\rho,J^\text{z})\leq b_\text{str}(k)$, or, contrapositively,
if $F_\text{Q}(\rho,J^\text{z})>b_\text{str}(k)$ then $\rho$ is not $k$-stretchably separable.
(For illustration, see Figure~\ref{fig:Bk.whr}.)
On the other hand,
the bound~\eqref{eq:FisherfDepth} for all states $\rho\in\mathcal{D}$
in terms of its stretchability entanglement depth $D_\text{str}$~\eqref{eq:Depth.str} can also be written similarly.

For toughness,
$\pinv{\vs{\xi}}_{k\text{-tgh}}:=\pinv{\vs{\xi}}_{k,t}$~\eqref{eq:vxir}
is given by the $f=t=\min$ toughness generator function~\eqref{eq:t},
which is increasing~\eqref{eq:mon.tm},
so if $\rho\in\mathcal{D}_{k\text{-tgh}}$ for a given $k\in t(\pinv{P}_\text{I})$,
then we have the constraint $\min(\pinv{\xi})\leq k$ in~\eqref{eq:Fisherkf}.
(Recall that the possible values are
$t(\pinv{\xi}) = 1,2,\dots,\lfloor n/2\rfloor$ for nontrivial partial separability,
and $t(\top)=n$ for the trivial partition.)
Among such partitions $\pinv{\xi}$, by~\eqref{eq:s2useful}, we have to increase the parts $x$ as much as possible,
while keeping at least one $x\leq k$,
so the maximum of $s_2$ is taken for
$\pinv{\xi}=\mset{n-1,1}$ if $k\neq n$.
For this, the maximum value is $s_2(\mset{n-1,1})=(n-1)^2+1=n^2-2n+2$,
so we have the attainable bound~\eqref{eq:Fisherkf} on the quantum Fisher information
for $k$-tough states for
$k=1,2,\dots,\lfloor n/2\rfloor$
\begin{equation}
\label{eq:FisherBoundtough.k}
\max_{\rho\in\mathcal{D}_{k\text{-tgh}}}
F_\text{Q}(\rho,J^\text{z}) = b_\text{tgh}(k)
= n^2-2n+2,
\end{equation}
which is $k$-independent, hence not useful.
Since $k\neq n$, this is a bound for biseparable states,
so genuine multipartite entanglement is the only class which could be detected by this bound.
However, for biseparability, which is $(n-1)$-producibility or $2$-partitionability,
the resulting bound~\eqref{eq:FisherBoundtough.k} agrees with~\eqref{eq:FisherBoundprod.k} and~\eqref{eq:FisherBoundpart.k},
so does not give anything stronger.
(For the notable state spaces and classes of toughness, see Section~\ref{sec:1param.maxmin}.)
Note that, although the bound~\eqref{eq:Fisherkf} is monotone in $k$~\eqref{eq:monB.gen} by construction,
this does not exclude $k$-independency.
This extreme case of $k$-independence expresses 
that the nested subsets~\eqref{eq:DsepIIpfmonk} of $k$-tough states $\mathcal{D}_{k\text{-tgh}}$
lie in such a `tilted' way in the whole state space $\mathcal{D}$
that $\mset{n-1,1}$-separable states, which are rather entangled, are $k$-tough for any $k$, 
that is, $\mathcal{D}_{\mset{n-1,1}\text{-sep}}\subseteq\mathcal{D}_{k\text{-tgh}}$.
We will elaborate on these kinds of issues later in Sections~\ref{sec:Metro.usefulness} and~\ref{sec:Metro.dommon},
but this can also be read off from Figure~\ref{fig:PpI23456PPS} in advance.

A kind of degenerate case is of squareability,
$\pinv{\vs{\xi}}_{k\text{-sq}}:=\pinv{\vs{\xi}}_{k,s_2}$~\eqref{eq:vxir} is given by the squareability $f=s_2$ generator function~\eqref{eq:sq2},
which is increasing~\eqref{eq:mon.sq},
so if $\rho\in\mathcal{D}_{k\text{-sq}}$ for a given $k\in s_2(\pinv{P}_\text{I})$, 
then we have the constraint $s_2(\pinv{\xi})\leq k$ in~\eqref{eq:Fisherkf}.
Then the maximization can be done without calculating the argmax, and
$\max_{\pinv{\xi}\in\pinv{P}_\text{I}:s_2(\pinv{\xi})\leq k} s_2(\pinv{\xi}) = k$,
so we have the attainable bound~\eqref{eq:Fisherkf} on the quantum Fisher information
for $k$-squareable states
\begin{subequations}
\label{eq:FisherBoundsq2}
\begin{equation}
\label{eq:FisherBoundsq2.k}
\max_{\rho\in\mathcal{D}_{k\text{-sq}}}
F_\text{Q}(\rho,J^\text{z}) = b_\text{sq}(k) = k.
\end{equation}
This gives a necessary but not sufficient criterion of $k$-squareable separability.
If $\rho$ is $k$-squareably separable then $F_\text{Q}(\rho,J^\text{z})\leq k$, or, contrapositively,
if $F_\text{Q}(\rho,J^\text{z})>k$ then $\rho$ is not $k$-squareably separable.
On the other hand, for all states $\rho\in\mathcal{D}$,
we have the bound~\eqref{eq:FisherfDepth}
in terms of the squareability entanglement depth $D_\text{sq}$~\eqref{eq:sq2ent.Depth} as
\begin{equation}
\label{eq:FisherBoundsq2.D}
F_\text{Q}(\rho,J^\text{z})\leq B_\text{sq}(\rho) = D_\text{sq}(\rho),
\end{equation}
since $b_\text{sq}\circ s_2 = s_2$.
\end{subequations}
These suggest that
squareability is the natural multipartite entanglement property from the point of view of quantum metrology;
it is what is bounded directly by the quantum Fisher information.
Note that, similarly, if $f=g\circ s_2$ is a monotone function of squareability,
then the bounds can simply be transformed by~\eqref{eq:gmeasuresfD},
so formulating bounds for, e.g., $2$-Tsallis or $2$-Rényi entanglement depths is straightforward
(see in Section~\ref{sec:Metro.remarks}).

Note that the bound by squareability depth~\eqref{eq:FisherBoundsq2.D}
is stronger than the bound by (producibility) depth~\eqref{eq:FisherBoundprod2.D}.
This is because
\begin{equation}
\label{eq:sq2w}
s_2(\pinv{\xi}) = \sum_{x\in\pinv{\xi}} x^2
\leq \sum_{x\in\pinv{\xi}} x \max(\pinv{\xi})
\equalsref{eq:hwr.w} n w(\pinv{\xi})
\end{equation}
holds for the generator functions $s_2$~\eqref{eq:sq2} and $w$~\eqref{eq:hwr.w},
so we have
\begin{equation}
\label{eq:Dsq2nD}
D_\text{sq}(\rho) \leq nD(\rho)
\end{equation}
for the corresponding depths by~\eqref{eq:Depthfmon},
since both $s_2$ and $w$ are increasing monotone~\eqref{eq:mon.wm},~\eqref{eq:mon.sq},
and by~\eqref{eq:gmeasuresfD}.
The bound~\eqref{eq:FisherBoundsq2.D} by the squareability depth
may be much stronger than the bound~\eqref{eq:FisherBoundprod2.D} by the (producibility) depth,
especially when the difference in~\eqref{eq:sq2w} is large,
that is, when many parts are much smaller than the maximal one.
For instance, suppose that we have $n=10$ particles, the system is described by a pure state $\pi=\proj{\psi}$,
and we measure $30\leq F_\text{Q}(\pi,J^\text{z})$ by the Cramér-Rao bound~\eqref{eq:QCramerRao}.
By the bound~\eqref{eq:FisherBoundprod2.D}
we have $3\leq F_\text{Q}(\pi,J^\text{z})/n\leq D(\pi)$, which allows 
also strictly $\mset{3,1,1,1,1,1,1,1}$-entangled states (or anything coarser).
On the other hand, by the bound~\eqref{eq:FisherBoundsq2.D}
we have $30\leq F_\text{Q}(\pi,J^\text{z})\leq D_\text{sq}(\pi)$, which excludes such disentangled subsystems,
and many others allowed by the weaker bound:
direct calculation shows that every pure state of entanglement depth $D(\pi)=3$ is excluded (so there has to be entangled subsystem of size at least $4$),
and also those which are $\pinv{\xi}$-entangled for $\pinv{\xi}\finereq\mset{4,3,1,1,1}$ or $\pinv{\xi}\finereq\mset{4,2,2,2}$,
since $s_2(\mset{4,3,1,1,1})=28<30$ and $s_2(\mset{4,2,2,2})=28<30$.

\subsection{Convex \texorpdfstring{$f$}{f}-entanglement criteria}
\label{sec:Metro.1paramoF}

The quantum Fisher information is the convex roof extension of the variance~\eqref{eq:FisherConvRoof},
which leads to \emph{convex} metrological bounds \emph{much} stronger than~\eqref{eq:FisherfDepth}.
For any $\rho\in\mathcal{D}$ we have the bound 
\begin{subequations}
\label{eq:superfrho}
\begin{equation}
\label{eq:superf}
F_\text{Q}(\rho,J^\text{z})\leq B^\text{oF}_f(\rho) \leq B_f(\rho),
\end{equation}
or, in terms of the depth and depth of formation of the induced $b_f\circ f$ property~\eqref{eq:BfDbff}
\begin{equation}
\label{eq:superfb}
F_\text{Q}(\rho,J^\text{z})\leq D^\text{oF}_{b_f\circ f}(\rho) \leq D_{b_f\circ f}(\rho).
\end{equation}
\end{subequations}
(Here $B^\text{oF}_f$ is the convex roof extension of $B_f$, which can be written, 
since $B_f=D_{b_f\circ f}$ is just a depth~\eqref{eq:BfDbff},
for which we have already seen that the convex roof extension can be written,
see~\eqref{eq:DepthOFf} and Appendix~\ref{app:Depthfs.DepthOFf}.)
Indeed, 
\begin{equation*}
\begin{split}
&F_\text{Q}(\rho,J^\text{z}) 
\equalsref{eq:FisherConvRoof}
\min_{\set{(p_j,\pi_j)}\decomp \rho}
 \sum_j p_j 4\Var(\pi_j,J^\text{z})\\ 
&\quad\lequals
\sum_j p'_j 4\Var(\pi'_j,J^\text{z})
\lequalsref{eq:FisherfDepth}
\sum_j p'_j B_f(\pi'_j)\\
&\quad\equalsref{eq:BfDbff}
\sum_j p'_j D_{b_f\circ f}(\pi'_j)
\equals
\min_{\set{(p_j,\pi_j)}\decomp \rho}
 \sum_j p_j D_{b_f\circ f}(\pi_j)\\
&\quad\equalsref{eq:DepthOFf}
 D^\text{oF}_{b_f\circ f}(\rho)
\lequalsref{eq:DepthfoFDepthf}
D_{b_f\circ f}(\rho),
\end{split}
\end{equation*}
where
the \emph{first equality} holds, because the Fisher information is the convex roof extension of (four times) the variance~\eqref{eq:FisherConvRoof};
the \emph{first inequality} holds for every particular decomposition $\set{(p'_j,\pi'_j)}\decomp \rho$;
the \emph{second inequality} is the inequality~\eqref{eq:FisherfDepth} for pure states $\pi'_j$ by~\eqref{eq:FisherPure};
the \emph{second equality} is the bound considered as depth of an induced property~\eqref{eq:BfDbff};
the \emph{third equality} holds, if we fix the decomposition $\set{(p'_j,\pi'_j)}\decomp \rho$ to be $D_{b_f\circ f}$-optimal,
that is, which minimizes $\sum_j p_j D_{b_f\circ f}(\pi_j)$;
the \emph{fourth equality} holds by the definition of the $f$-entanglement depth of formation~\eqref{eq:DepthOFf}
by nothing that $b_f$ is increasing/decreasing if and only if $f$ is increasing/decreasing~\eqref{eq:monB.gen}, so $b_f\circ f$ is always increasing;
the \emph{third inequality} is the inequality~\eqref{eq:DepthfoFDepthf}, since $b_f\circ f$ is increasing.

The convex bound~\eqref{eq:superfrho} gives a new type of multipartite entanglement criteria,
which are not about $(k,f)$-separability (the membership problem of $\mathcal{D}_{k,f}$),
but the ratios of pure states of different values of $f$-entanglement depth in the mixture.
This will be illustrated in Section~\ref{sec:Metro.meaning}.

\subsection{Examples of convex \texorpdfstring{$f$}{f}-entanglement criteria}
\label{sec:Metro.1paramoFxmpl}

Now let us see the convex bounds~\eqref{eq:superfrho}
resulting from the bounds~\eqref{eq:FisherfDepth}
for the case of some particular one-parameter partial entanglement properties.
Since the $b_f\circ f$ one-parameter properties,
the depths of which the bounds are given by,
are not particularly expressive in general,
we consider only the cases of squareability and a bound weaker than producibility.

The simplest case is that of squareability,
for which we have the bound $b_\text{sq}(k)=k$ in~\eqref{eq:FisherBoundsq2.k},
then~\eqref{eq:superfb} takes the form
\begin{equation}
\label{eq:supersq}
F_\text{Q}(\rho,J^\text{z})\leq D^\text{oF}_\text{sq}(\rho) \leq D_\text{sq}(\rho).
\end{equation}

The bound~\eqref{eq:superfb} for producibility
is not too expressive to use, it is 
$B^\text{oF}_\text{prod} = D^\text{oF}_{b_\text{prod}\circ w}
= D^\text{oF}_{\lfloor n/w\rfloor w^2 + (n-\lfloor n/w\rfloor w)^2}(\rho)$
by~\eqref{eq:FisherBoundprod.k}.
Instead of this, we make use of~\eqref{eq:Dsq2nD}
for the squareability entanglement depth and (producibility) entanglement depth,
and
\begin{equation}
\label{eq:Dsq2oFnDoF}
D^\text{oF}_\text{sq}(\rho) \leq nD^\text{oF}(\rho)
\end{equation}
for the corresponding depths of formations by~\eqref{eq:sq2w} and~\eqref{eq:DepthfoFmon}.
Using these, together with~\eqref{eq:DepthfoFDepthf} and~\eqref{eq:supersq},
we have the bound for all states $\rho\in\mathcal{D}$ 
in terms of its entanglement depth of formation~\eqref{eq:DepthOF.prod} as
\begin{equation}
\label{eq:superw}
F_\text{Q}(\rho,J^\text{z})\leq nD^\text{oF}(\rho) \leq nD(\rho).
\end{equation}
By~\eqref{eq:Dsq2oFnDoF}, this is weaker than~\eqref{eq:supersq},
but stronger than~\eqref{eq:FisherBoundprod2.D}.
Note that here we did not use the general result~\eqref{eq:superfb} directly, but did a detour to get a simpler formula.

\subsection{Meaning of the criteria}
\label{sec:Metro.meaning}

We have already seen that the  $s_q$ generator functions for the parameters $0\leq q$ have an entropic character, 
expressing the \emph{mixedness} of the size distribution,
but not a central moment character,
expressing the \emph{spread} of the size distribution
(see Sections~\ref{sec:1param.prob} and~\ref{sec:1param.remarks}).
Let us see now, what the metrological bounds tell us
about the characterization of the possible partial entanglement in terms of the size distribution,
by considering generator functions related to squareability.

Let us consider the first situation depicted in Section~\ref{sec:1param.prob} (random choice of elementary subsystems),
where the first moment~\eqref{eq:xipmom1.raw} of the subsystem size distribution $\pinv{\xi}/n$
is just 
\begin{equation}
\avg(\pinv{\xi}) :=\sum_{x\in\pinv{\xi}} \frac{x}{n} x = \frac{s_2(\pinv{\xi})}{n},
\end{equation}
being the average (center), and not the variance (spread) in this situation.
(Note that usually the variance is the quadratic expression,
however, in this situation the distribution and the values of the random variable are linked together, see Section~\ref{sec:1param.prob}.)
For this, we directly have
\begin{equation}
\label{eq:superavg}
\frac1n F_\text{Q}(\rho,J^\text{z}) \leq D^\text{oF}_\text{avg}(\rho)\leq D_{\avg}(\rho)
\end{equation}
with the depth of $\avg$-entanglement $D_\text{avg}=D_{sq}/n$
by~\eqref{eq:supersq},~\eqref{eq:gmeasuresfD} and~\eqref{eq:gmeasuresfDoF.affine}.
That is, $\frac1n F_\text{Q}(\rho,J^\text{z})$ puts a lower bound on the \emph{average size of entangled subsystems}
to which a (uniformly) randomly chosen elementary subsystem belongs.
Motivated by this, we call the depth of formation of $\avg$-entanglement 
\begin{equation}
D^\text{oF}_\text{avg}(\rho) = D^\text{oF}_\text{sq}(\rho)/n
\end{equation}
the \emph{average size of entangled subsystem} (ASES).
Note that we have two averages here,
first, with respect to the random choice of elementary subsystems,
second, with respect to the pure state in the decomposition.
To elaborate on this,
let us collect here the bounds~\eqref{eq:supersq} and~\eqref{eq:superw} with~\eqref{eq:Dsq2nD} and~\eqref{eq:Dsq2oFnDoF},
\begin{subequations}
\begin{equation}
\label{eq:hipersuper2}
\begin{array}{ccccc}
 & & D^\text{oF}(\rho) & \leq & D(\rho) \\[6pt]
 & & \rotatebox[origin=c]{90}{$\leq$} & & \rotatebox[origin=c]{90}{$\leq$} \\[6pt]
\frac1n F_\text{Q}(\rho,J^\text{z}) & \leq & D^\text{oF}_\text{avg}(\rho) & \leq & D_\text{avg}(\rho) 
\end{array}
\end{equation}
and also recall the measures involved.
The entanglement depth~\eqref{eq:Depth.prod},~\eqref{eq:Depthfdec}
\begin{equation}
\label{eq:hipersuper2.D}
D(\rho)
 = \min_{\set{(p_j,\pi_j)}\decomp \rho} \max_j D(\pi_j)
\end{equation}
is the \emph{minimum} (w.r.t.~decompositions) of the \emph{maximum} (in the decomposition) of the \emph{maximal} size of entangled subsystems.
The entanglement depth of formation~\eqref{eq:DepthOF.prod},~\eqref{eq:DepthOFf}
\begin{equation}
\label{eq:hipersuper2.DoF}
D^\text{oF}(\rho)
 = \min_{\set{(p_j,\pi_j)}\decomp \rho} \sum_j p_j D(\pi_j)
\end{equation}
is the \emph{minimum} (w.r.t.~decompositions) of the \emph{average} (in the decomposition) of the \emph{maximal} size of entangled subsystems.
The $\avg$-entanglement depth~\eqref{eq:Depthfdec} 
\begin{equation}
\label{eq:hipersuper2.Davg}
D_\text{avg}(\rho) = D_\text{sq}/n
 = \min_{\set{(p_j,\pi_j)}\decomp \rho} \max_j D_\text{avg}(\pi_j)
\end{equation}
is the \emph{minimum} (w.r.t.~decompositions) of the \emph{maximum} (in the decomposition) of the \emph{average} size (w.r.t.~picking elementary subsystems) of entangled subsystems.
The $\avg$-entanglement depth of formation, or ASES~\eqref{eq:DepthOFf}
\begin{equation}
\label{eq:hipersuper2.DavgoF}
D^\text{oF}_\text{avg}(\rho) = D^\text{oF}_\text{sq}/n
 = \min_{\set{(p_j,\pi_j)}\decomp \rho} \sum_j p_j D_\text{avg}(\pi_j)
\end{equation}
is the \emph{minimum} (w.r.t.~decompositions) of the \emph{average} (in the decomposition) of the \emph{average} size (w.r.t.~picking elementary subsystems) of entangled subsystems.
\end{subequations}
Then the vertical relations in~\eqref{eq:hipersuper2} mean
that $F_\text{Q}/n$ puts a lower bound on the average size of the entangled subsystems for random choice of elementary subsystems,
which is smaller than the the size of the largest entangled subsystem in any decomposition.

On the other hand,
the $2$-Rényi and $2$-Tsallis entropic generator functions, expressing the mixedness of the subsystem size distribution $\pinv{\xi}/n$,
related to squareability,
are also meaningful in this situation.
For the $2$-Rényi generator functions~\eqref{eq:Rq} we have
$R_2=g\circ s_2$ with the strictly monotone $g(u)=-\ln(u/n^2)$,
and applying this $g$ to the weaker bound in~\eqref{eq:supersq} leads to
\begin{subequations}
\begin{equation}
\label{eq:FisherUpperR2}
-\ln\Bigl(\frac1{n^2}F_\text{Q}(\rho,J^\text{z})\Bigr) \geq D_{R_2}(\rho)
\end{equation}
by~\eqref{eq:gmeasuresfD}.
So we have upper bound 
on the minimal mixedness (by $2$-Rényi entropy) of the subsystem size distribution
in every decomposition, see~\eqref{eq:Depthfdec}. 
Note that, since $g$ is not affine~\eqref{eq:gmeasuresfDoF.affine}, we cannot formulate bound for the depth of formation of $2$-Rényi property,
that is, the average $2$-Rényi property in the decompositions
($g$ is convex, so we have only~\eqref{eq:gmeasuresfDoF.conv}, and
$-\ln\bigl(\frac1{n^2}F_\text{Q}(\rho,J^\text{z})\bigr) \not\geq D_{R_2}^\text{oF}(\rho)$ in general).
For the $2$-Tsallis generator functions~\eqref{eq:Tq} we have
$T_2=g\circ s_2$ with the strictly monotone $g(u)=1-u/n^2$,
and applying this $g$ to the bounds in~\eqref{eq:supersq} leads to
\begin{equation}
\label{eq:FisherUpperT2}
1-\frac1{n^2}F_\text{Q}(\rho,J^\text{z}) \geq D^\text{oF}_{T_2}(\rho) \geq D_{T_2}(\rho)
\end{equation}
\end{subequations}
by~\eqref{eq:gmeasuresfD} and~\eqref{eq:gmeasuresfDoF.affine}, since $g$ is affine in this case.
So we have upper bound
on the the minimal \emph{and} also average mixedness (by $2$-Tsallis entropy) of the subsystem size distribution
in every decomposition, see~\eqref{eq:Depthfdec} and~\eqref{eq:DepthOFf}.
(It might be slightly contraintuitive to use decreasing generator functions,
which the $2$-Rényi and $2$-Tsallis entropies~\eqref{eq:mon.Tq}-\eqref{eq:mon.Rq} are.
They have \emph{upper bounds} on their depths,
but by a \emph{decreasing} function $g$ of the quantum Fisher information.
This is equivalent to have a \emph{lower bound} of a \emph{decreasing} function $g^{-1}$ 
of the depths by the quantum Fisher information.)

For the sake of completeness,
let us consider the second situation depicted in Section~\ref{sec:1param.prob} (random choice of composite subsystems) as well,
and let us denote the average and the variance of the subsystem sizes as
\begin{subequations}
\label{eq:xipmom}
\begin{align}
\label{eq:xipmom.avg}
\avg'(\pinv{\xi}) &:= \sum_{x\in\pinv{\xi}}\frac{1}{\abs{\pinv{\xi}}} x
 = \frac{n}{h(\pinv{\xi})},\\
\label{eq:xipmom.var}
\var'(\pinv{\xi})&:= \sum_{x\in\pinv{\xi}} \frac{1}{\abs{\pinv{\xi}}} \Bigl(x-\frac{n}{\abs{\pinv{\xi}}}\Bigr)^2
= \frac{1}{h(\pinv{\xi})} \Bigl( s_2(\pinv{\xi}) - \frac{n^2}{h(\pinv{\xi})}\Bigr),
\end{align}
\end{subequations}
being the first (raw) moment~\eqref{eq:xipmom2.raw} and the second central moment~\eqref{eq:xipmom2.central}
of the size of the randomly chosen subsystem.
Recall that $\var'(\pinv{\xi})$ is not a generator function,
so it does not define a one-parameter entanglement property, and its depth has no meaning.
However, we may write squareability in terms of these,
\begin{equation}
\label{eq:sqvarpavgp}
s_2(\pinv{\xi})/n=\var'(\pinv{\xi})/\avg'(\pinv{\xi}) + \avg'(\pinv{\xi}),
\end{equation}
and the quantum Fisher information gives a lower bound for that
in~\eqref{eq:supersq},
\begin{equation}
\label{eq:FisherUpperVarAvg}
\begin{split}
\frac1n F_\text{Q}(\rho,J^\text{z}) &\leq D^\text{oF}_{\var'/\avg'+\avg'}(\rho)\\
&\qquad\qquad\leq D_{\var'/\avg'+\avg'}(\rho),
\end{split}
\end{equation}
by~\eqref{eq:gmeasuresfD} and~\eqref{eq:gmeasuresfDoF.affine}.
That is, we do not have a lower bound on the variance itself, but on the combination of the variance and the average,
though it is not too expressive.
Note that the bounds~\eqref{eq:superavg} are formulated by $D_{\avg}$, not by $D_{\avg'}$, which can be much smaller.
This is because from \eqref{eq:sqvarpavgp} it directly follows that
\begin{equation}
\avg\geq\avg',
\end{equation}
expressing the difference between the two situations.
For example, for the partition $\pinv{\xi}=\mset{20,1,1,\dots,1}$, describing the state vector~\eqref{eq:state2},
we have $\avg(\pinv{\xi})=s_2(\pinv{\xi})/n=4.8$ and $\avg'(\pinv{\xi})=n/h(\pinv{\xi})=100/81\approx1.23$.

\subsection{Strength of convex \texorpdfstring{$f$}{f}-entanglement criteria}
\label{sec:Metro.1paramoFxmplStrength}

Now let us discuss the difference in the strength of the bounds in~\eqref{eq:hipersuper2}.

In the right column in~\eqref{eq:hipersuper2},
the bound by the avg-depth may be much stronger than the bound by the (producibility) depth,
especially when the difference in $\avg\leq w$ is large~\eqref{eq:sq2w},
that is, when the average size of the parts are much smaller than the maximal one.
This works in the same way as for the squareability (see at the end of Section~\ref{sec:Metro.1paramxmpl}),
but is more expressive.
For instance, suppose that we have $n=10$ particles, the system is described by a pure state $\pi=\proj{\psi}$,
and we measure $30\leq F_\text{Q}(\pi,J^\text{z})$ by the Cramér-Rao bound~\eqref{eq:QCramerRao}.
By bounding the (producibility) depth $D$,
we have $3\leq F_\text{Q}(\pi,J^\text{z})/n\leq D(\pi)$, which allows 
also strictly $\mset{3,1,1,1,1,1,1,1}$-entangled states (or anything coarser),
only the size of the largest entangled subsystem is bounded.
On the other hand, by bounding the avg-depth $D_\text{avg}$ we have bound on the average size of entangled subsystems 
in the pure state $3\leq F_\text{Q}(\pi,J^\text{z})/n\leq D_\text{avg}(\pi)$, which excludes such disentangled subsystems,
and many others allowed by the weaker bound:
direct calculation shows that every pure state of entanglement depth $D(\pi)=3$ is excluded (so there has to be entangled subsystem of size at least $4$),
and also those which are $\pinv{\xi}$-entangled for $\pinv{\xi}\finereq\mset{4,3,1,1,1}$ or $\pinv{\xi}\finereq\mset{4,2,2,2}$,
since the averages are $\avg(\mset{4,3,1,1,1})=2.8<3$ and $\avg(\mset{4,2,2,2})=2.8<3$.
(Note that the $3$-producible states are excluded also because
if the average size of the parts is at least $3$ then the size of the maximal part could be $3$ only if this would divide $n$,
which is not the case in this example.)

Now let us discuss in what sense
the first, convex bounds in the rows of~\eqref{eq:hipersuper2} are stronger than the second, original ones.
This is basically the same as the difference between
the~\eqref{eq:Depthfdec} characterization of the $f$-entanglement depth
and the~\eqref{eq:DepthOFf} characterization of the $f$-entanglement depth of formation.

Let us consider the upper row in~\eqref{eq:hipersuper2} first.
The weaker bound in the upper row in~\eqref{eq:hipersuper2} is
$F_\text{Q}(\rho,J^\text{z})/n\leq D(\rho)$.
This means that in all pure convex decompositions $\sum_jp_j\pi_j=\rho$,
there exists $\pi_j$ containing an entangled subsystem of size at least $F_\text{Q}(\rho,J^\text{z})/n$.
This result may be rather weak, since in principle it allows a mixture
of an infinitely small weight of 
strictly $k$-producible
pure state $\pi_k:=\proj{\psi_k}\in\mathcal{C}_{k\text{-prod}}\cap\mathcal{P}$
with fully separable state $\rho_1\in\mathcal{C}_{1\text{-prod}}$ orthogonal to it,
$\rho_\epsilon := (1-\epsilon)\rho_1 + \epsilon \pi_k$ for $\epsilon>0$.
Such state is $k$-producible and not 
$k'$-producible for $k'<k$ (by the same reasoning as at the beginning of Section~\ref{sec:Depthmeas.DepthOFfxmpl}),
so $D(\rho_\epsilon)=k$, but much less entangled as, e.g., $\pi_k$ itself, although $D(\pi_k)=k$.
A much lower quantum Fisher information $F_\text{Q}(\rho_\epsilon,J^\text{z})/n$ is expected from such a state $\rho_\epsilon$,
so the bound $F_\text{Q}(\rho,J^\text{z})/n\leq D(\rho)$ is rather weak.

The stronger bound in the upper row in~\eqref{eq:hipersuper2} is, however,
$F_\text{Q}(\rho,J^\text{z})/n\leq D^\text{oF}(\rho)$.
On the one hand, we have $D^\text{oF}(\rho_\epsilon)\leq (1-\epsilon)1 + \epsilon k$ for the state above by~\eqref{eq:hipersuper2.DoF},
so the bound is indeed much stricter.
On the other hand, which is even more important,
the meaning of this bound is that in all pure convex decompositions $\sum_jp_j\pi_j=\rho$,
the \emph{(decomposition-)average} size of the largest entangled subsystem is at least $F_\text{Q}(\rho,J^\text{z})/n$.
To elaborate on this, let us write
for all pure convex decompositions $\sum_jp_j\pi_j=\rho$
the $q_k:=\sum_{j:D(\pi_j)=k} p_j$ weights of the $k$-producible pure states in the decomposition,
by which
\begin{equation*}
\begin{split}
&F_\text{Q}(\rho,J^\text{z})/n\leq D^\text{oF}(\rho) \leq \sum_j p_j D(\pi_j) \\
&\quad= \sum_{k=1}^n q_k\sum_{j:D(\pi_j)=k} \frac{p_j}{q_k}D(\pi_j)=\sum_{k=1}^n q_k k.
\end{split}
\end{equation*}
For example, suppose that we have $n=10$ particles,
and we measure $30\leq F_\text{Q}(\rho,J^\text{z})$ by the Cramér-Rao bound~\eqref{eq:QCramerRao}.
Then $3\leq F_\text{Q}(\rho,J^\text{z})/n$, and we already know from the weaker bound in the upper row in~\eqref{eq:hipersuper2} by~\eqref{eq:Depthfdec} that 
in every decomposition there is a nonzero weight of $k$-producible pure states for at least one $k\geq3$;
however, now we also know from the stronger bound in the upper row in~\eqref{eq:hipersuper2} by~\eqref{eq:DepthOFf} that 
if in a decomposition there are $1$-producible (fully separable) states (of weight $q_1$),
then these have to be compensated by $k$-producible pure states for $k>3$.
For example, at least twice as much, $2q_1$ weight of $4$-producible states are needed for this
(indeed, $3\leq 1q_1+3q_3+4q_4=q_1+3(1-q_1-q_4)+4q_4$ leads to $2q_1\leq q_4$),
or at least the same $q_1$ weight of $5$-producible states
(indeed, $3\leq 1q_1+3q_3+5q_5=q_1+3(1-q_1-q_5)+5q_5$ leads to $q_1\leq q_5$).

Let us consider the lower row in~\eqref{eq:hipersuper2} similarly.
The weaker bound in the lower row in~\eqref{eq:hipersuper2} is 
$F_\text{Q}(\rho,J^\text{z})/n\leq D_\text{avg}(\rho)$.
This means that in all convex decompositions $\sum_jp_j\pi_j=\rho$,
there exists $\pi_j$ containing entangled subsystems,
the average of the sizes of which are at least $F_\text{Q}(\rho,J^\text{z})/n$.
This result may again be rather weak, since in principle it allows a mixture
of an infinitely small weight of
strictly $k$-avg
pure state $\pi_k:=\proj{\psi_k}\in\mathcal{C}_{k\text{-avg}}\cap\mathcal{P}$
with fully separable state $\rho_1\in\mathcal{C}_{1\text{-avg}}$ orthogonal to it,
$\rho_\epsilon := (1-\epsilon)\rho_1 + \epsilon \pi_k$ for $\epsilon>0$.
Such state is $k$-avg and not
$k'$-avg for $k'<k$ (by the same reasoning as at the beginning of Section~\ref{sec:Depthmeas.DepthOFfxmpl}),
so $D_\text{avg}(\rho_\epsilon)=k$, but much less entangled as, e.g., $\pi_k$ itself, although $D_\text{avg}(\pi_k)=k$.
A much lower quantum Fisher information $F_\text{Q}(\rho_\epsilon,J^\text{z})$ is expected from such a state $\rho_\epsilon$,
so the bound $F_\text{Q}(\rho,J^\text{z})/n\leq D_\text{avg}(\rho)$ is rather weak.

The stronger bound in the lower row in~\eqref{eq:hipersuper2} is, however,
$F_\text{Q}(\rho,J^\text{z})/n \leq D^\text{oF}_\text{avg}(\rho)$.
On the one hand, we have $D^\text{oF}_\text{avg}(\rho_\epsilon) \leq (1-\epsilon)1 + \epsilon k$ for the state above by~\eqref{eq:hipersuper2.DavgoF},
so the bound is indeed much stricter.
(Note that the possible avg values $k\in\avg(\pinv{P}_\text{I})$ are between $1$ and $n$,
but only some of those equal to some of the integers in that range.)
On the other hand, which is even more important,
the meaning of this bound is that in all pure convex decompositions $\sum_jp_j\pi_j=\rho$,
the \emph{(decomposition-)average} of the average ($\avg$) of the sizes of entangled subsystems is at least $F_\text{Q}(\rho,J^\text{z})/n$.
To elaborate on this, let us write
for all pure convex decompositions $\sum_jp_j\pi_j=\rho$
the $q_k:=\sum_{j:D_\text{avg}(\pi_j)=k} p_j$ weights of the $k$-avg pure states in the decomposition,
by which
\begin{equation*}
\begin{split}
&F_\text{Q}(\rho,J^\text{z})/n\leq D^\text{oF}_\text{avg}(\rho) \leq \sum_j p_j D_\text{avg}(\pi_j) \\
&\quad= \sum_{\substack{k\in\\\avg(\pinv{P}_\text{I})}} q_k\sum_{j:D_\text{avg}(\pi_j)=k} \frac{p_j}{q_k}D_\text{avg}(\pi_j)
= \sum_{\substack{k\in\\\avg(\pinv{P}_\text{I})}} q_k k.
\end{split}
\end{equation*}
For example, suppose that we have $n=10$ particles,
and we measure $30\leq F_\text{Q}(\rho,J^\text{z})$ by the Cramér-Rao bound~\eqref{eq:QCramerRao}.
Then $3\leq F_\text{Q}(\rho,J^\text{z})/n$, and we already know from the weaker bound in the lower row in~\eqref{eq:hipersuper2} by~\eqref{eq:Depthfdec} that
in every decomposition there is a nonzero weight of $k$-avg pure states for at least one $k\geq3$
(for example by $\mset{4,3,2,1}$-separable or $\mset{5,1,1,1,1,1}$-separable
or more entangled states, since $\avg(\mset{4,3,2,1})=\avg(\mset{5,1,1,1,1,1})=3$);
however, now we also know from the stronger bound in the lower row in~\eqref{eq:hipersuper2} by~\eqref{eq:DepthOFf} that
if in a decomposition there are $1$-avg (fully separable) states (of weight $q_1$),
then these have to be compensated by $k$-avg pure states for $k>3$.
For example 
(since the subsequent values of avg are $\avg(\mset{5,2,1,1,1})=3.2$ and $\avg(\mset{4,3,3})=\avg(\mset{4,4,1,1})=\avg(\mset{5,2,2,1})=3.4$),
at least ten times as much, $10q_1$ weight of $3.2$-avg states are needed for this
(indeed, $3\leq 1q_1+3q_3+3.2q_{3.2}=q_1+3(1-q_1-q_{3.2})+3.2q_{3.2}$ leads to $10q_1\leq q_{3.2}$),
or at least five times as much, $5q_1$ weight of $3.4$-squareable states 
(indeed, $3\leq 1q_1+3q_3+3.4q_{3.4}=q_1+3(1-q_1-q_{3.4})+3.4q_{3.4}$ leads to  $5q_1\leq q_{3.4}$).

\subsection{Metrological usefulness of \texorpdfstring{$f$}{f}-en\-tangle\-ment}
\label{sec:Metro.usefulness}

\begin{figure}\centering
\includegraphics{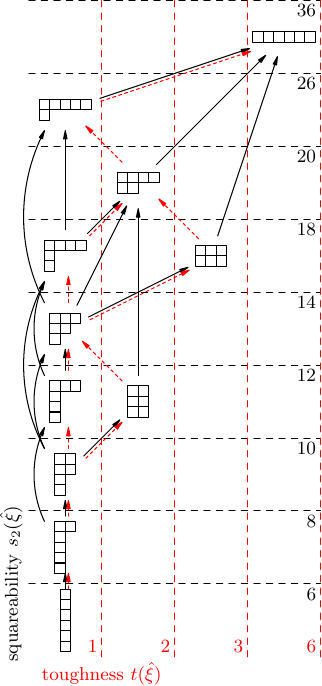}
\caption{The squareability~\eqref{eq:sq2} vs.~toughness~\eqref{eq:t} plot of the poset $\pinv{P}_\text{I}$ for $n=6$.
The inverse images~\eqref{eq:fset.level} of $t$ are in the columns,
and the sub-level sets~\eqref{eq:fset.sslevel} of $t$ are to the left of the vertical lines.
The refinement $\finereq$ and the dominance $\leq$ orders are denoted by \emph{consecutive} black and dashed red arrows, respectively.
The refinement implies dominance, which can be seen here as
two partitions connected by black arrows are also connected by red ones,
but the reverse does not hold.
The bound~\eqref{eq:FisherBoundtough.k} is already reached for $1$-tough states.}
\label{fig:PpI6SqTd}
\end{figure}

The case of entanglement toughness~\eqref{eq:FisherBoundtough.k} tells us that 
there are one-parameter partial entanglement properties which give not too many meaningful metrological bounds,
compared to the number of classes they lead to.
In other words,
not all the classes of some one-parameter properties are bounded by the quantum Fisher information.
To clarify this,
let us imagine how the bound $b_f(k)$ in~\eqref{eq:Fisherkf} can change
when larger and larger $k$ minimal values are allowed.
(See Figure~\ref{fig:PpI6SqTd} for illustration for the case of toughness.)
It is clear that the problem is that allowing partitions $\pinv{\xi}$ of larger and larger toughness
does not give larger squareability values, so the bound does not increase.
That is, the usefulness of a property depends on the \emph{strict} monotonicity of the bound $b_f$~\eqref{eq:Fisherkf}.

We have already seen 
that the bound $b_f(k)$ is monotone by construction~\eqref{eq:monB.gen}.
If, moreover,
the bound $b_f$ is \emph{strictly} monotone from the right/left in a given $k\in f(\pinv{P}_\text{I})$,
\begin{equation}
\label{eq:monB.str}
\forall k'\in f(\pinv{P}_\text{I}):\quad
k \lessgtr k' \dspthen b_f(k) < b_f(k'),
\end{equation}
then it can detect $(k,f)$-entanglement $\mathcal{D}_{k,f}$ for that $k$.
This is because~\eqref{eq:DsepIIpfmonk} implies also the strict inequalities
\begin{equation}
k\lessgtr k' \dspiff \mathcal{D}_{k,f} \subset \mathcal{D}_{k',f}
\end{equation}
for all $k,k'\in f(\pinv{P}_\text{I})$
(since $k=k'$ if and only if $k\leq k'$ and $k\geq k'$,
if and only if $\mathcal{D}_{k,f} \subseteq \mathcal{D}_{k',f}$ and $\mathcal{D}_{k,f} \supseteq \mathcal{D}_{k',f}$ by~\eqref{eq:DsepIIpfmonk},
if and only if $\mathcal{D}_{k,f}=\mathcal{D}_{k',f}$),
then the strict monotonicity~\eqref{eq:monB.str} results in
\begin{equation}
\mathcal{D}_{k,f} \subset \mathcal{D}_{k',f}
\dspthen
b_f(k) < b_f(k'),
\end{equation}
then it is possible to have $b_f(k) < F_\text{Q}(\rho,J^\text{z}) \leq b_f(k')$,
leading to $\rho\notin \mathcal{D}_{k,f}$.
If the strict monotonicity~\eqref{eq:monB.str} of the bound does not hold at $k$,
then the two bounds can only be violated simultaneously,
$b_f(k) = b_f(k')<F_\text{Q}(\rho,J^\text{z})$,
which gives only $\rho\notin \mathcal{D}_{k',f}$,
but does not exclude states in $\mathcal{D}_{k',f}\setminus\mathcal{D}_{k,f}$.

The overall metrological usefulness of a one-parameter property given by the generator function $f$
then can be characterized by the number of $k\in f(\pinv{P}_\text{I})$ values
for which the bound $b_f$ is strictly monotone~\eqref{eq:monB.str},
which is the $\abs{b_f(f(\pinv{P}_\text{I}))}-1$ number of steps
of the discrete function $b_f(k)$. 

A one-parameter property given by the generator function $f$
can be considered \emph{most suitable} for the purpose of formulating metrological bounds,
if it is as good as it can be,
that is, the bound $b_f(k)$ it induces 
is strictly monotone~\eqref{eq:monB.str} for all values $k\in f(\pinv{P}_\text{I})$.
This is necessary for the detection of $(k,f)$-entanglement $\mathcal{D}_{k,f}$ for all $k$.
Such properties are the producibility, partitionability, stretchability and squareability,
and also the $2$-Rényi, $2$-Tsallis etc.~properties, being related to squareability by strictly monotone functions.
(For squareability this is obvious, for the proof of the remaining ones, see Appendix~\ref{app:MetroUseful.hwr}.)

On the other hand, 
if $f$ is not constant (which would be meaningless even for the purpose of classification, see Section~\ref{sec:PSprops.remarks}),
then 
there is at least one step, that is, two values of the bounds $b_f(k)$,
leading to criteria for the detection of at least one $k$ value of $(k,f)$-entanglement $\mathcal{D}_{k,f}$.
(For the proof, see Appendix~\ref{app:MetroUseful.min}.)
This means that toughness is as bad as it can be,
it is among the one-parameter partial entanglement properties
\emph{most unsuitable} for the purpose of formulating metrological bounds.

All the one-parameter properties lie between these two extreme cases.
To check at which values $k\in f(\pinv{P}_\text{I})$ of a generator function $f$
is the bound $b_f(k)$ strictly monotone~\eqref{eq:monB.str},
and to find how many different values of the bound arise,
do not seem to be easy.
This could be done for $w$, $h$, $r$, $t$ and $s_2$ (then for $N_2$, $T_2$, $R_2$, $P_2$)
by the explicit formulas in Section~\ref{sec:Metro.PS},
however, explicit formulas could not be derived for the other generator functions.
Following these lines, in the subsequent subsection we identify
an important character of one-parameter properties,
playing role in the metrological usefulness.

\subsection{Dominance-monotone \texorpdfstring{$f$}{f}-entanglement}
\label{sec:Metro.dommon}

Motivated by the concrete method of finding the bounds in Section~\ref{sec:Metro.1paramxmpl},
we may consider generator functions $f$
which behave in accordance with $s_2$ in the sense of the~\eqref{eq:s2useful} property of $s_2$.
That is, if $f$ is an increasing/decreasing generator function,
then it also increases/decreases for moving a subsystem from a smaller part to a larger one,
that is, for $x_i\geq x_j$,
\begin{equation}
\label{eq:fuseful}
\begin{split}
&f(\set{\dots,x_i,\dots,x_j,\dots})\\
&\quad\lesseqgtr f(\set{\dots,x_i+1,\dots,x_j-1,\dots}).
\end{split}
\end{equation}
It is easy to check that such a move is described by the so called \emph{dominance relation}~\eqref{eq:poIpdom},
which we review in Appendix~\ref{app:1param.Gen} for the interested reader.
(The move in~\eqref{eq:fuseful} is along dominance, see~\eqref{eq:poIpdom.simpler},
and the covering relation of dominance~\eqref{eq:poIpdom.c} is such a move.)
Then~\eqref{eq:fuseful} means that $f$ is an increasing/decreasing monotone with respect to the dominance order~\eqref{eq:dommon}.
It can be proven that refinement implies dominance~\eqref{eq:poIs},
so refinement-monotonicity of a function is a weaker property than dominance-monotonicity,
that is, dominance-monotone functions are special generator functions.
(For more details on the partial orders appearing here, see Appendix~\ref{app:1param.Gen}.
The dominance-monotonicity of generator functions are given in Appendix~\ref{app:1param.dmon}.)

\begin{figure*}\centering
\includegraphics{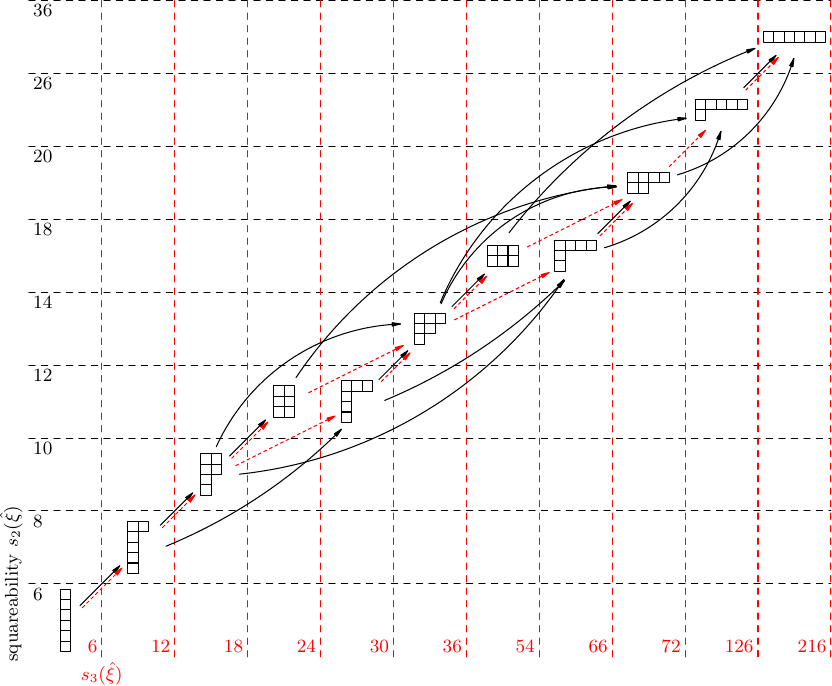}
\caption{The squareability~\eqref{eq:sq2} vs.~$s_3$~\eqref{eq:sq} plot of $\pinv{P}_\text{I}$ for $n=6$.
The inverse images~\eqref{eq:fset.level} of $s_3$ are in the columns,
and the sub-level sets~\eqref{eq:fset.sslevel} of $s_3$ are to the left of the vertical lines.
The refinement poset and the dominance lattice are denoted with black and dashed red arrows, respectively.}
\label{fig:PpI6SqScd}
\end{figure*}

First of all, toughness $t$~\eqref{eq:t} is not a dominance-monotone, neither $t_m$~\eqref{eq:extr.tm}.
(See Figure~\ref{fig:PpI6SqTd} for $n=6$ and $m=1$, which is the toughness~\eqref{eq:t};
although this can already be seen for $n=4$,
where the partitions 
$\mset{2,1,1}$, $\mset{2,2}$ and $\mset{3,1}$ are more and more dominant,
but $t(\mset{2,1,1})=1$, $t(\mset{2,2})=2$ and $t(\mset{3,1})=1$.)
So imposing dominance-monotonicity would exclude toughness, which is promising.
Comparing~\eqref{eq:mon} and~\eqref{eq:dmon} in Appendix~\ref{app:1param},
we have the generator functions (refinement-monotone) and those which are also dominance-monotone.
These are the $m$-widths $w_m$~\eqref{eq:extr.wm},
the height $h$~\eqref{eq:hwr.h},
the power sum $s_q$~\eqref{eq:sq} and related entropic quantities~\eqref{eq:qentr} for $q\geq0$,
the $q$-sum $N_q$~\eqref{eq:Nq} for $q>0$,
the $q$-mean $M_q$~\eqref{eq:Mq} for $q\geq1$,
and also
the entanglement dimensions $\Dim$ and $\Dim'$~\eqref{eq:Dim} and
the entanglement degree of freedoms $\DoF$ and $\DoF'$~\eqref{eq:DoF}.

The role of dominance-monotonicity~\eqref{eq:fuseful} of a generator function $f$
in the problem of the strict monotonicity of the metrological bound~\eqref{eq:monB.str} can be understood
by considering the dominance arrows in the $s_2$ vs.~$f$ plot.
For dominance-monotone $f$, these run always upwards and to the right/left if $f$ is increasing/decreasing
but never horizontally,
since $s_2$ is a strictly increasing dominance-monotone~\eqref{eq:s2useful}.
The refinement arrows do the same, since refinement implies dominance~\eqref{eq:poIs}
but dominance-monotonicity gives more constraints.

Now, the failure of strict monotonicity of the bound~\eqref{eq:monB.str} occurs when
there exist partitions $\pinv{\xi},\pinv{\xi}'\in\pinv{P}_\text{I}$
of $f(\pinv{\xi})=:k\lessgtr f(\pinv{\xi}')=:k'$, 
which maximize $s_2$ among those partitions
($s_2(\pinv{\xi})=\max_{\pinv{\upsilon}:f(\pinv{\upsilon})=k}s_2(\pinv{\upsilon})$,
$s_2(\pinv{\xi}')=\max_{\pinv{\upsilon}:f(\pinv{\upsilon}')=k'}s_2(\pinv{\upsilon}')$),
and
$b_f(k)=s_2(\pinv{\xi}) \geq s_2(\pinv{\xi}')$, and the bound does not increase, $b_f(k)=b_f(k')$ see~\eqref{eq:Fisherkf}.
This, however, is excluded if $f$ is dominance-monotone \emph{and}
$\pinv{\xi}$ and $\pinv{\xi}'$ are related by dominance,
since if $\pinv{\xi}$ is dominated by $\pinv{\xi}'$
then $s_2(\pinv{\xi})<s_2(\pinv{\xi}')$ and $f(\pinv{\xi})\lessgtr f(\pinv{\xi}')$ hold together,
and the bound does increase, $b_f(k)<b_f(k')$.
If, on the contrary, $\pinv{\xi}$ and $\pinv{\xi}'$ are related by dominance
\emph{but} $f$ is not dominance-monotone,
then it might happen that $\pinv{\xi}$ dominates $\pinv{\xi}'$ 
(while still $f(\pinv{\xi})=k\lessgtr f(\pinv{\xi}')=k'$ is the case)
then $b_f(k)=s_2(\pinv{\xi}) > s_2(\pinv{\xi}')$, and the bound does not increase, $b_f(k)=b_f(k')$,
which can be illustrated by case of toughness (see Figure~\ref{fig:PpI6SqTd}).
For $n=6$, we have $\pinv{\xi}:=\mset{5,1}$, dominating $\pinv{\xi}':=\mset{4,2}$,
so $s_2(\pinv{\xi})=26>s_2(\pinv{\xi}')=20$, while $t(\pinv{\xi})=1<t(\pinv{\xi}')=2$,
and these partitions are also of maximal squareability in the inverse images of $t$,
so $b_\text{tgh}(1)=b_\text{tgh}(2)=26$.
If $\pinv{\xi}$ and $\pinv{\xi}'$ are not related by dominance,
then imposing dominance-monotonicity of $f$ does not exclude the above possibility 
of the failure of the strict monotonicity~\eqref{eq:monB.str} of the bound.
That is, dominance-monotonicity is not sufficient for the bound to be strictly monotone~\eqref{eq:monB.str}
for all $k\in f(\pinv{P}_\text{I})$.
This can be illustrated by the case of $s_3$~\eqref{eq:sq} (see Figure~\ref{fig:PpI6SqScd}).
Although $s_3$ is dominance-monotone~\eqref{eq:dmon.sq},
but the bound $b_{s_3}(k)$ is not strictly monotone:
for $n=6$, we have $\pinv{\xi}:=\mset{2,2,2}$ and $\pinv{\xi}':= \mset{3,1,1,1}$,
not related by dominance,
and  $s_3(\pinv{\xi})=24<s_3(\pinv{\xi}')=30$,
but $s_2(\pinv{\xi})=s_2(\pinv{\xi}')=12$, and these partitions are also of maximal squareability in the inverse images of $s_3$,
so $b_{s_3}(24)=b_{s_3}(30)=12$.

One may object that dominance-monotonicity may not be necessary either for the bound to be strictly monotone~\eqref{eq:monB.str}
for all $k\in f(\pinv{P}_\text{I})$.
This is because the strict monotonicity of the bound~\eqref{eq:monB.str}
is determined by the maximal values of $s_2$ in the inverse images $f^{-1}(k)$,
while for other partitions $\pinv{\xi}\in f^{-1}(k)$ for which $s_2(\pinv{\xi})<b_f(k)$, 
the dominance-monotonicity could fail,
while not affecting the $b_f(k)$ values of the bounds.

Anyway, the main point is that 
the overall number of possibilities of the failure of the~\eqref{eq:monB.str} strict monotonicity of the bound
is greatly reduced by the imposing of dominance-monotonicity.
This is because the dominance lattice is significantly more `dense' than the refinement poset,
there are seemingly few pairs of partitions not related by dominance \emph{and}
which, in the same time, could break the~\eqref{eq:monB.str} strict monotonicity of the bound.
On the other hand,
the significance of dominance-monotonicity is that
it is the definitive property of generalized entropies,
expressing the mixedness of the subsystem size distribution
(see the discussions in Sections~\ref{sec:1param.entr} and~\ref{sec:1param.remarks}, Appendix~\ref{app:1param.Gen},
and further in Section~\ref{sec:Metro.remarks}).

\subsection{Remarks}
\label{sec:Metro.remarks}

Here we list some remarks 
on the connections of one-parameter partial entanglement properties and quantum metrology,
one paragraph each.

Note that
convex roof entanglement measures,
and, in fact, any proper entanglement measures for mixed states,
are infeasible to evaluate in general,
this is why bounds on those are so important.
In our case,
these bounds can not only be measured experimentally,
but also calculated by~\eqref{eq:Fisher}.
On the other hand, these are convex roof extensions of depths~\eqref{eq:superfb},
the meaning of which is easy to understand, especially in the particular cases of 
squareability~\eqref{eq:supersq} and producibility~\eqref{eq:superw},
see also the explanations in Section~\ref{sec:Metro.1paramoFxmpl}.

Note that
the partial separability is a dimension-independent classification,
while the metrological bounds are given only for $d=2$, qubits or spin-$1/2$ systems,
used directly in experiments.
The generalization for larger local Hilbert space dimension $d=\dim(\mathcal{H}_l)$ is, however, straigthforward.
Let us have the collective operator $A=\sum_{l=1}^n A_l\otimes \Id_{\cmpl{\set{l}}}$,
where the minimal and maximal eigenvalues of $A_l$ are $a_\text{min}$ and $a_\text{max}$ uniformly for all $l$.
The only proof affected is of~\eqref{eq:genFisherUpperpure},
where only~\eqref{eq:varJXboundqubit} has to be changed to~\eqref{eq:varAXbound}
(see Appendix~\ref{app:Metro}),
which leads to the
\begin{equation}
\max_{\pi\in\mathcal{P}_{\pinv{\xi}}} \frac{F_\text{Q}(\pi,A)}{(a_\text{max}-a_\text{min})^2}
= s_2(\pinv{\xi})
\end{equation}
generalization of~\eqref{eq:genFisherUpperpure}.
Then every bound in Section~\ref{sec:Metro} can be given for/by
the normalized quantum Fisher information $F_\text{Q}(\pi,A)/(a_\text{max}-a_\text{min})^2$
on the left-hand side.
Such bounds are independent of the dimension $d$ of the local Hilbert spaces,
and depend only on the uniform `spectral width' $a_\text{max}-a_\text{min}$ of the local operators.
The $J^\text{z}$ measurement of spin-$s$ systems can be of particular interest,
then $d=2s+1$, and $(a_\text{max}-a_\text{min})^2=4s^2$.

Note that entanglement can also be detected efficiently
by \emph{spin squeezing inequalities}
\cite{Sorensen-2001b,Toth-2007,Toth-2010b,Vitagliano-2011,Vitagliano-2024,Behbood-2014,Kong-2020}, and
even the entanglement depth can be detected
\cite{Sorensen-2001,Esteve-2008,Gross-2010,Leroux-2010,Hosten-2016,Duan-2011,Lucke-2014,Vitagliano-2017,Vitagliano-2018,Zou-2018,Xin-2023}.
A promising research direction is then to give spin squeezing bounds on other one-parameter multipartite entanglement properties.
For example, 
for macroscopic singlet states~\cite{Toth-2010b},
also demonstrated experimentally in both cold and hot atomic systems~\cite{Behbood-2014,Kong-2020},
we have the property $\pinv{\xi}=\mset{2,2,\dots,2}$;
and also the $s_{-\infty}$ number of spins unentangled with the rest~\eqref{eq:qlim.sninf}
can be bounded from above with the measurement results.

\section{Summary}
\label{sec:Summ}

We began this study with elaborating the general theory of
one-parameter families of partial entanglement properties (Section~\ref{sec:PSprops}),
and the resulting entanglement depth like quantities (Section~\ref{sec:Depthmeas}).
The main idea came by observing that the partial entanglement property
leading to entanglement depth is producibility,
which is given by sub-level sets of the width of the Young diagram of the partition of the system~\eqref{eq:vxik.prod}.
One-parameter partial entanglement properties, called \emph{$f$-entanglement},
could then be defined by sub- or super-level sets of \emph{generator functions} $f$~\eqref{eq:vxir},
for which the only requirement necessary for the construction was the \emph{monotonicity}~\eqref{eq:genf}.
The resulting one-parameter classification of quantum states is 
the chain hierarchy~\eqref{eq:DsepIIpfmonk} of \emph{nested state spaces}~\eqref{eq:DsepIIpf} of \emph{$(k,f)$-entangled states},
and the `layers' of \emph{disjoint classes}~\eqref{eq:CsepIIIpf} of \emph{strictly $(k,f)$-entangled states}.
We also had the
\emph{$f$-entanglement depth} and \emph{$f$-entanglement depth of formation},
being the generalization~\eqref{eq:Depthf} and~\eqref{eq:DepthOFf} 
of entanglement depth~\eqref{eq:Depth.prod} and entanglement depth of formation~\eqref{eq:DepthOF.prod}
with respect to these one-parameter partial entanglement properties,
for which we showed LOCC monotonicity.

We also provided an ample supply of one-parameter entanglement properties (Section~\ref{sec:1param})
by the corresponding generator functions,
beyond the \emph{entanglement partitionability}, \emph{producibility} and \emph{stretchability},
the most notable ones were 
the \emph{entanglement squareability},
defined by the sum of squares of the sizes of entangled subsystems~\eqref{eq:sq2},
the \emph{entanglement toughness},
defined by the size of the smallest entangled subsystem~\eqref{eq:t},
the \emph{entanglement degree of freedom},
defined by the effective number of elementary subsystems needed for the description of the state~\eqref{eq:DoF},
and the \emph{Rényi property of entanglement},
defined by the Rényi function of the subsystem size distribution~\eqref{eq:Rq},
nicely interpolating among producibility, partitionability and toughness.

The formulation of metrological multipartite entanglement criteria (Section~\ref{sec:Metro})
fits well into the framework of one-parameter entanglement properties.
We showed that for states of any given partial separability property,
the upper bound for the quantum Fisher information in collective spin-z measurement of qubits
is given by the squareability of that property~\eqref{eq:genFisherUpper},~\eqref{eq:Fisherkf}.
This could be evaluated for $f$-entanglement in some cases, 
and is given by a function of the $f$-entanglement depth,
which turns out to be the depth of an induced one-parameter property~\eqref{eq:FisherfDbff}.
This led to an identity if that property itself is the squareability~\eqref{eq:FisherBoundsq2},
suggesting that entanglement squareability is the natural multipartite entanglement property
from the point of view of quantum metrology.
Since the quantum Fisher information is the convex roof extension of the variance~\eqref{eq:FisherConvRoof},
thanks to the similar formulations of the $f$-entanglement depth~\eqref{eq:Depthfdec}
and the $f$-entanglement depth of formation~\eqref{eq:DepthOFf},
these bounds could be strengthened,
and the quantum Fisher information turned out to bound
the average of the depth (depth of formation~\eqref{eq:DepthOFf}) 
of the corresponding induced one-parameter property~\eqref{eq:superfb}
from below.
In particular, $F_\text{Q}/n$ puts a lower bound on the average size of the entangled subsystems for random choice of elementary subsystems,
which is much smaller than the size of the largest entangled subsystem in any decomposition.

While deriving the metrological bounds with respect to different one-parameter partial entanglement properties,
in the case of toughness~\eqref{eq:FisherBoundtough.k}
we were faced with the possibility that
a given one-parameter property may not lead to different bounds for different values of the parameter $k$.
This arises when the bound~\eqref{eq:Fisherkf} fails to be strictly monotone~\eqref{eq:monB.str},
in which case the one-parameter property cannot be detected for those particular values of $k$.
It turned out that the two extreme cases are those of producibility (or partitionability, etc.), where the bound is strictly monotone for all $k$,
and of toughness, where the bound is nowhere strictly monotone, apart from one $k$ value.
The possible failure of the strict monotonicity~\eqref{eq:monB.str} of the bound
comes from the interplay~\eqref{eq:Fisherkf} of the generator function of the given one-parameter property with squareability.
Imposing the dominance-monotonicity of the one-parameter property
significantly reduces the possibilities of the failure of strict monotonicity~\eqref{eq:monB.str} of the bound,
which singles out one-parameter properties given by entropies (dominance-monotone functions) of the subsystem size distributions.

\begin{acknowledgments}
Discussion with Manuel Gessner is gratefully acknowledged.
We acknowledge the support of the EU (QuantERA MENTA, QuantERA QuSiED, COST Action CA23115),
the Spanish MCIU (Grant No.~PCI2022-132947),
the Basque Government (Grant No.~IT1470-22),
the National Research, Development and Innovation Office of Hungary NKFIH (Grant Nos.~2019-2.1.7-ERA-NET-2021-00036 and K134983),
and the NKFIH within the Quantum Information National Laboratory of Hungary.
We acknowledge the support of the Grant No.~PID2021-126273NB-I00 funded by MCIN/AEI/10.13039/501100011033 and by `ERDF A way of making Europe'.
We thank the `Frontline' Research Excellence Programme of the NKFIH (Grant No. KKP133827).
We thank Project no. TKP2021-NVA-04, which has been implemented with the support provided by the Ministry of Innovation and Technology of Hungary from the National Research, Development and Innovation Fund, financed under the TKP2021-NVA funding scheme.
G.~T.~is thankful for a Bessel Research Award from the Humboldt Foundation.
Sz.~Sz.~happily acknowledges the support of the wonderful Bach performances of Marta Czech and Sir András Schiff.
\end{acknowledgments}

\appendix

\section{Usual measures of \texorpdfstring{$f$}{f}-entanglement}
\label{app:PSmeas}
In this section we recall the 
\emph{entanglement of formation} and \emph{relative entropy of entanglement}
of the permutation invariant partial entanglement properties~\cite{Szalay-2019},
and we also define those with respect to the one-parameter entanglement properties.

\subsection{Measures of partial entanglement properties}
\label{app:PSmeas.gen}

There are several possible ways for quantifying entanglement
(LOCC monotones, entanglement monotones~\cite{Bennett-1996a,Bennett-1996b,Vidal-2000,Horodecki-2001,Horodecki-2009}),
which, in some cases, can also be applied for the particular kinds of multipartite entanglement~\cite{Szalay-2015b,Szalay-2019}.
Here we recall the \emph{entanglement of formation}~\cite{Bennett-1996b,Uhlmann-2010} 
and the \emph{relative entropy of entanglement}~\cite{Vedral-1998,Chitambar-2019,Lami-2023},
which fit well to the hierarchy of properties~\eqref{eq:oisomDIIp},
in the sense that they
give measures for each permutation invariant multipartite entanglement property $\pinv{\vs{\xi}}\in\pinv{P}_\text{II}$,
and these measures are related in the same way as the properties (multipartite monotones~\cite{Szalay-2015b,Szalay-2019}).

\textit{On the first level,} for all integer partitions $\pinv{\xi}\in\pinv{P}_\text{I}$, we have the
\emph{$\pinv{\xi}$-entanglement of formation}, being the average $\pinv{\xi}$-entanglement of the optimal decomposition~\cite{Szalay-2019,Szalay-2015b},
and the
\emph{relative entropy of $\pinv{\xi}$-entanglement}, being the distinguishability from the $\pinv{\xi}$-separable states,
\begin{align}
\label{eq:EntOFI}
E^\text{oF}_{\pinv{\xi}}(\rho) 
&:= \min_{\set{(p_j,\pi_j)}\decomp \rho}
\sum_j p_j  \min_{\xi\in s^{-1}(\pinv{\xi})}  \sum_{X\in\xi} S_X(\pi_j) ,\\
\label{eq:REEntI}
E^\text{R}_{\pinv{\xi}}(\rho) 
&:= \min_{\sigma\in\mathcal{D}_{\pinv{\xi}}} S(\rho||\sigma),
\end{align}
where
$S_X(\rho) = S(\rho_X) = -\Tr(\rho_X\ln(\rho_X))$
is the \emph{von Neumann entropy}~\cite{Neumann-1927,Ohya-1993,Petz-2008,Wilde-2013}
of the state $\rho_X= \Tr_{\cmpl{X}}(\rho)$
of subsystem $X$ (where $\cmpl{X}=\set{1,2,\dots,n}\setminus X$),
and $S(\rho||\sigma)= \Tr(\rho\ln(\rho) - \rho\ln(\sigma))$
is the \emph{Umegaki relative entropy} or \emph{quantum Kullback-Leibler divergence}~\cite{Umegaki-1962,Ohya-1993,Petz-2008,Wilde-2013}.
(The first minimization in~\eqref{eq:EntOFI} is taken over all the $\rho=\sum_i p_i\pi_i$ pure convex decompositions of $\rho$,
for which we use the shorthand notation $\decomp$ above.
This is the so called \emph{convex roof extension}~\cite{Uhlmann-1998,Uhlmann-2010}
of the $\pinv{\xi}$-correlation~\cite{Szalay-2019,Szalay-2015b}.
The second minimization in~\eqref{eq:EntOFI} is taken over all the set partitions $\xi=\set{X_1,X_2,\dots,X_{\abs{\xi}}}$
which have parts of sizes $\pinv{\xi}=\set{x_1,x_2,\dots,x_{\abs{\pinv{\xi}}}}$.
For more details, see~\cite{Szalay-2019}.)
These are entanglement monotones~\cite{Szalay-2015b,Szalay-2019,Vedral-1998}
(convex and nonincreasing on average with respect to selective LOCC~\cite{Vidal-2000,Horodecki-2001,Horodecki-2009},
for the relative entropy of $\pinv{\xi}$-entanglement this follows from the general construction~\cite{Vedral-1998}),
faithful
($E^\text{oF/R}_{\pinv{\xi}}(\rho)=0$ if and only if  $\rho\in\mathcal{D}_{\pinv{\xi}}$),
and multipartite monotone~\cite{Szalay-2019,Szalay-2015b}, which is
\begin{equation}
\label{eq:EntOFREEntImmon}
\pinv{\upsilon}\finereq\pinv{\xi} \dspiff 
E^\text{oF/R}_{\pinv{\upsilon}}\geq E^\text{oF/R}_{\pinv{\xi}},
\end{equation}
expressing that any state is more entangled with respect to a finer partition.

\textit{On the second level,} for all down-sets of integer partitions $\pinv{\vs{\xi}}\in\pinv{P}_\text{II}$ we have the
\emph{$\vs{\pinv{\xi}}$-entanglement of formation}, being the average $\pinv{\xi}$-entanglement of the optimal decomposition~\cite{Szalay-2019,Szalay-2015b}, and the
\emph{relative entropy of $\vs{\pinv{\xi}}$-entanglement}, being the distinguishability from the $\pinv{\vs{\xi}}$-separable states,
\begin{align}
\label{eq:EntOFII}
E^\text{oF}_{\pinv{\vs{\xi}}}(\rho) 
&:= \min_{\set{(p_j,\pi_j)}\decomp \rho}
\sum_j p_j  \min_{\xi\in \vee s^{-1}(\pinv{\vs{\xi}})}  \sum_{X\in\xi} S_X(\pi_j) ,\\
\label{eq:REEntII}
E^\text{R}_{\pinv{\vs{\xi}}}(\rho)
&:= \min_{\sigma\in\mathcal{D}_{\pinv{\vs{\xi}}}} S(\rho||\sigma).
\end{align}
(The second minimization in~\eqref{eq:EntOFII} is taken over all the set partitions $\xi=\set{X_1,X_2,\dots,X_{\abs{\xi}}}$
which have parts of sizes $\pinv{\xi}\in \pinv{\vs{\xi}}$.
For more details, see~\cite{Szalay-2019}.)
These are entanglement monotones~\cite{Szalay-2015b,Szalay-2019,Vedral-1998},
faithful
($E^\text{oF/R}_{\pinv{\vs{\xi}}}(\rho)=0$ if and only if $\rho\in\mathcal{D}_{\pinv{\vs{\xi}}}$),
and multipartite monotone~\cite{Szalay-2019,Szalay-2015b}, which is
\begin{equation}
\label{eq:EntOFREEntIImmon}
\pinv{\vs{\upsilon}}\finereq\pinv{\vs{\xi}} \dspiff 
E^\text{oF/R}_{\pinv{\vs{\upsilon}}}\geq E^\text{oF/R}_{\pinv{\vs{\xi}}},
\end{equation}
expressing that any state is more entangled with respect to a finer multipartite entanglement property.

\subsection{Examples of \texorpdfstring{$f$}{f}-entanglement measures}
\label{app:PSmeas.1paramxmpl}

For the three notable one-parameter families of properties,
partitionability, producibility and stretchability~\eqref{eq:vxik},
we have by~\eqref{eq:EntOFII}
the \emph{$k$-partitionability entanglement of formation},
the \emph{$k$-producibility entanglement of formation},
and the \emph{$k$-stretchability entanglement of formation}~\cite{Szalay-2019,Szalay-2015b},
as well as by~\eqref{eq:REEntII}
the \emph{relative entropy of $k$-partitionability of entanglement},
the \emph{relative entropy of $k$-producibility of entanglement},
and the \emph{relative entropy of $k$-stretchability of entanglement},
for the respective ranges of $k$,
\begin{subequations}
\label{eq:EntOFREEntk}
\begin{align}
\label{eq:EntOFREEntk.part}
E^\text{oF/R}_{k\text{-part}}(\rho)  &:= E^\text{oF/R}_{\pinv{\vs{\xi}}_{k\text{-part}}}(\rho),\\
\label{eq:EntOFREEntk.prod}
E^\text{oF/R}_{k\text{-prod}}(\rho)  &:= E^\text{oF/R}_{\pinv{\vs{\xi}}_{k\text{-prod}}}(\rho),\\
\label{eq:EntOFREEntk.str}
E^\text{oF/R}_{k\text{-str}}(\rho)   &:= E^\text{oF/R}_{\pinv{\vs{\xi}}_{k\text{-str}}}(\rho),
\end{align}
\end{subequations}
with the monotonicity
\begin{subequations}
\label{eq:EntOFREEntmonk}
\begin{align}
\label{eq:EntOFREEntmonk.part}
k\geq k' &\dspiff E^\text{oF/R}_{k\text{-part}} \geq E^\text{oF/R}_{k'\text{-part}},\\
\label{eq:EntOFREEntmonk.prod}
k\leq k' &\dspiff E^\text{oF/R}_{k\text{-prod}} \geq E^\text{oF/R}_{k'\text{-prod}},\\
\label{eq:EntOFREEntmonk.str}
k\leq k' &\dspiff E^\text{oF/R}_{k\text{-str}}  \geq E^\text{oF/R}_{k'\text{-str}}.
\end{align}
\end{subequations}
by~\eqref{eq:vximonk} and~\eqref{eq:EntOFREEntIImmon}.

\subsection{\texorpdfstring{$f$}{f}-entanglement measures in general}
\label{app:PSmeas.1param}

\textit{In general,} for the one-parameter properties~\eqref{eq:vxir},
defined by the generator function $f$ over the permutation invariant properties~\eqref{eq:genf},
we have the \emph{$(k,f)$-entanglement of formation}
and the \emph{relative entropy of $(k,f)$-entanglement}
\begin{equation}
\label{eq:rfEoFR}
E^\text{oF/R}_{k,f}(\rho) := E^\text{oF/R}_{\pinv{\vs{\xi}}_{k,f}}(\rho)
\end{equation}
by~\eqref{eq:EntOFII} and~\eqref{eq:REEntII}.
These are entanglement monotones,
faithful ($E^\text{oF/R}_{k,f}(\rho)=0$ if and only if $\rho\in\mathcal{D}_{k,f}$),
by the corresponding properties of $E^\text{oF/R}_{\pinv{\vs{\xi}}}$,
and monotone in $k$
\begin{equation}
k\lesseqgtr k' \dspiff E^\text{oF/R}_{k,f} \geq E^\text{oF/R}_{k',f},
\end{equation}
by~\eqref{eq:vxirchain} and~\eqref{eq:EntOFREEntIImmon}.

Note that
if the generator functions $f_1$ and $f_2$ are both increasing,
then for all $k_2\in f_2(\pinv{P}_\text{I})$
\begin{equation}
f_1 \leq f_2 \dspthen E^\text{oF/R}_{k_2,f_2} \geq E^\text{oF/R}_{k_1(k_2),f_1}
\end{equation}
for $k_1(k_2):=\min\sset{k\in f_1(\pinv{P}_\text{I})}{k_2\leq k}$;
while if $f_1$ and $f_2$ are both decreasing, 
then for all $k_2\in f_2(\pinv{P}_\text{I})$
\begin{equation}
f_1 \leq f_2 \dspthen E^\text{oF/R}_{k_1,f_1} \geq E^\text{oF/R}_{k_2(k_1),f_2}
\end{equation}
for $k_2(k_1):=\max\sset{k\in f_2(\pinv{P}_\text{I})}{k_1\geq k}$.
(Indeed,
$\pinv{\vs{\xi}}_{k_2,f_2}\finereq \pinv{\vs{\xi}}_{k_1(k_2),f_1}$ and 
$\pinv{\vs{\xi}}_{k_1,f_1}\finereq \pinv{\vs{\xi}}_{k_2(k_1),f_2}$ are easy to check in the two cases
by definition~\eqref{eq:vxir},
then~\eqref{eq:EntOFREEntIImmon} with the definition~\eqref{eq:rfEoFR} leads to the claim.)

Note that, if a generator function $f:\pinv{P}_\text{I}\to\field{R}$ 
is composed with a \emph{monotone} function $g:\field{R}\to\field{R}$,
the measures~\eqref{eq:rfEoFR} may decrease with respect to this,
\begin{subequations}
\label{eq:gmeasuresfEoFR}
\begin{equation}
E^\text{oF/R}_{g(k),g\circ f} \leq E^\text{oF/R}_{k,f},
\end{equation}
while if $g$ is \emph{strictly} monotone,
then the measures~\eqref{eq:rfEoFR} are invariant with respect to this,
\begin{equation}
E^\text{oF/R}_{g(k),g\circ f} = E^\text{oF/R}_{k,f},
\end{equation}
\end{subequations}
by the result~\eqref{eq:gprops}, the monotonicity~\eqref{eq:EntOFREEntIImmon} and the definition~\eqref{eq:rfEoFR}.

For partitionability, producibility and stretchability,
we get back the measures~\eqref{eq:EntOFREEntk}-\eqref{eq:EntOFREEntmonk}
by the height, width and rank~\eqref{eq:hwr} as generator functions.

\section{\texorpdfstring{$f$}{f}-entanglement depth}
\label{app:Depthfs}

In this section we show the validity of the 
formulations~\eqref{eq:Depthf2},~\eqref{eq:Depthfdec} and~\eqref{eq:DepthOFf}
of $f$-entanglement depth and $f$-entanglement depth of formation
in Section~\ref{sec:Depthmeas}.

\subsection{\texorpdfstring{$f$}{f}-entanglement depth by membership}
\label{app:Depthfs.Depthf2}

Here we prove the formula~\eqref{eq:Depthf2}
for the case of increasing $f$,
\begin{equation*}
\min_{k\in f(\pinv{P}_\text{I})}\bigsset{k}{\rho\in \mathcal{D}_{k,f} } =
\min_{\pinv{\vs{\xi}}\in\pinv{P}_\text{II}}\bigsset{f(\pinv{\vs{\xi}})}{\rho\in \mathcal{D}_{\pinv{\vs{\xi}}} },
\end{equation*}
see~\eqref{eq:Depthf} and~\eqref{eq:Depthf2},
then similar reasoning works for the case of decreasing $f$.

\textit{First,}
if $\min\sset{k'}{\rho\in \mathcal{D}_{k',f} }=k$
then $\rho\in\mathcal{D}_{k,f}\equiv\mathcal{D}_{\pinv{\vs{\xi}}_{k,f}}$ by~\eqref{eq:DsepIIpf},
therefore 
\begin{equation*}
\min_{\pinv{\vs{\xi}}\in\pinv{P}_\text{II}}\bigsset{f(\pinv{\vs{\xi}})}{\rho\in \mathcal{D}_{\pinv{\vs{\xi}}} }\leq k \equiv
\min_{k'\in f(\pinv{P}_\text{I})}\bigsset{k'}{\rho\in \mathcal{D}_{k',f} }
\end{equation*}
by~\eqref{eq:fIIkf}.
\textit{Second,}
if $\min_{\pinv{\vs{\xi}}}\sset{f(\pinv{\vs{\xi}})}{\rho\in \mathcal{D}_{\pinv{\vs{\xi}}} }= k$
then there exists $\pinv{\vs{\xi}}\in\pinv{P}_\text{II}$ 
for which $f(\pinv{\vs{\xi}})\equiv \max_{\pinv{\xi}\in\pinv{\vs{\xi}}} f(\pinv{\xi})=k$ such that $\rho\in\mathcal{D}_{\pinv{\vs{\xi}}}$,
then $\pinv{\vs{\xi}}\finereq\pinv{\vs{\xi}}_{k,f}$ by~\eqref{eq:fII} and~\eqref{eq:vxir},
then $\rho\in \mathcal{D}_{\pinv{\vs{\xi}}_{k,f}}\equiv\mathcal{D}_{k,f}$ by~\eqref{eq:oisomDIIp} and~\eqref{eq:DsepIIpf},
therefore 
\begin{equation*}
\min_{k'\in f(\pinv{P}_\text{I})}\bigsset{k'}{\rho\in \mathcal{D}_{k',f} }\leq k \equiv
\min_{\pinv{\vs{\xi}}\in\pinv{P}_\text{II}}\bigsset{f(\pinv{\vs{\xi}})}{\rho\in \mathcal{D}_{\pinv{\vs{\xi}}} }
\end{equation*}
by~\eqref{eq:DsepIIpfmonk}.

\subsection{\texorpdfstring{$f$}{f}-entanglement depth by decomposition}
\label{app:Depthfs.Depthfdec}

Here we prove the formula~\eqref{eq:Depthfdec}
for the case of increasing $f$,
\begin{equation*}
\min_{k\in f(\pinv{P}_\text{I})}\bigsset{k}{\rho\in \mathcal{D}_{k,f} } =
\min_{\set{(p_i,\pi_i)}\decomp \rho}
 \max\limits_i\bigset{D_f(\pi_i)},
\end{equation*}
see~\eqref{eq:Depthf} and~\eqref{eq:Depthfdec},
then similar reasoning works for the case of decreasing $f$.
(Note that the minimum with respect to the decompositions is attained
for a given finite length of decompositions,
since $D_f$ takes finite number of discrete values $f(\pinv{P}_\text{I})$,
and so is $\max_i D_f(\pi_i)$.)

\textit{First,}
if $\min\sset{k'}{\rho\in \mathcal{D}_{k',f} }=k$
then $\rho\in\mathcal{D}_{k,f}=\Conv(\mathcal{P}_{k,f})$ by~\eqref{eq:DsepIIpf},
that is, there exists a pure decomposition $\set{(p_i,\pi_i)}$ of $\rho$,
such that $\pi_i\in\mathcal{P}_{k,f}$ for all $i$,
that is, $D_f(\pi_i)\leq k$ for all $i$ by~\eqref{eq:PsepIIpf} and~\eqref{eq:Depthf},
then $\max_i D_f(\pi_i)\leq k$ for that decomposition,
therefore 
\begin{equation*}
\min_{\set{(p_i,\pi_i)}\decomp \rho}
 \max_i D_f(\pi_i)\leq k \equiv
\min_{k'\in f(\pinv{P}_\text{I})}\bigsset{k'}{\rho\in \mathcal{D}_{k',f} }.
\end{equation*}
\textit{Second,}
if $\min_{\set{(p_i,\pi_i)}\decomp \rho}
 \max_i D_f(\pi_i) = k$,
then there exists a pure decomposition $\set{(p_i,\pi_i)}$ of $\rho$
for which $D_f(\pi_i)\leq\max_i D_f(\pi_i)= k$ for all $i$,
then $\pi_i\in\mathcal{P}_{D_f(\pi_i),f}\subseteq\mathcal{P}_{k,f}$ for all $i$ by~\eqref{eq:DepthfD} and~\eqref{eq:PsepIIpfmonk},
then $\rho = \sum_i p_i\pi_i\in\Conv(\mathcal{P}_{k,f})\equiv\mathcal{D}_{k,f}$ by~\eqref{eq:DsepIIpf},
therefore 
\begin{equation*}
\min\bigsset{k'}{\rho\in \mathcal{D}_{k',f} }\leq k \equiv
\min_{\set{(p_i,\pi_i)}\decomp \rho}
 \max_i D_f(\pi_i)
\end{equation*}
by~\eqref{eq:DsepIIpfmonk}.

\subsection{\texorpdfstring{$f$}{f}-entanglement depth of formation}
\label{app:Depthfs.DepthOFf}

Here 
first we recall and slightly extend some results on convex roof construction~\cite{Uhlmann-1998,Uhlmann-2010},
then prove the validity of the formula~\eqref{eq:DepthOFf}.

In general, for a function $g:\mathcal{P}\to\field{R}$,
the functions $g^{\cup/\cap}:\mathcal{D}\to\field{R}$ given as
\begin{subequations}
\begin{align}
g^\cup(\rho) := \inf_{\set{(p_i,\pi_i)}\decomp \rho}
 \sum_i p_i g(\pi_i),\\
g^\cap(\rho) := \sup_{\set{(p_i,\pi_i)}\decomp \rho}
 \sum_i p_i g(\pi_i),
\end{align}
\end{subequations}
are the (unique) \emph{largest convex} and \emph{smallest concave extensions} of $g$ to $\mathcal{D}$~\cite{Uhlmann-1998,Uhlmann-2010}.
These are clearly extensions, $g^\cup(\pi)=g^\cap(\pi)=g(\pi)$ for all pure states $\pi\in\mathcal{P}$.
To see that $g^{\cup/\cap}$ are indeed the largest convex/smallest concave extensions,
we have that for all convex/concave extensions $G:\mathcal{D}\to\field{R}$
for all pure convex decompositions $\rho=\sum_ip_i\pi_i$ of all $\rho\in\mathcal{D}$,
\begin{equation*}
G(\rho)\lesseqgtr\sum_ip_iG(\pi_i)=\sum_ip_ig(\pi_i),
\end{equation*}
then $G(\rho)\lesseqgtr g^{\cup/\cap}(\rho)$.
To see that $g^{\cup/\cap}$ are indeed convex/concave,
we have that 
for all pure convex decompositions 
$\rho_j=\sum_i p_{j,i}\pi_{j,i}$
of all $\rho_j\in\mathcal{D}$,
such that 
for all $\epsilon>0$,
$0\leq\sum_i p_{j,i}g(\pi_{j,i})-g^\cup(\rho_j)<\epsilon$,
or
$0\leq g^\cap(\rho_j)-\sum_ip_{j,i}g(\pi_{j,i})<\epsilon$,
in the two cases, respectively,
and
for weights $0\leq w_j$, $\sum_j w_j=1$,
\begin{equation*}
\begin{split}
&g^{\cup/\cap}\Bigl(\sum_j w_j\rho_j\Bigr)
\lesseqgtr \sum_j \sum_i w_j p_{j,i}g^{\cup/\cap}(\pi_{j,i}) \\
&\quad= \sum_j w_j \sum_i p_{j,i}g(\pi_{j,i})\\
&\quad\lesseqgtr \sum_j w_j g^{\cup/\cap}(\rho_j) + \sum_j w_j \epsilon,
\end{split}
\end{equation*}
which holds for arbitrarily small $\epsilon$.

If $g:\mathcal{P}\to\field{R}$ is lower/upper semicontinuous,
then the infimum in the definition of $g^\cup$, respectively the supremum in the definition of $g^\cap$ is attained.
(Then these are convex/concave \emph{roof} extensions.
The proof of the continuous case~\cite{Uhlmann-1998,Uhlmann-2010} is generalized here.)
To see this,
let us consider the graph of the function $g$
in $\Lin_\text{SA}(\mathcal{H})\oplus\field{R}$,
\begin{equation*}
\mathcal{G}(g):=\bigsset{(\pi,g(\pi))}{\pi\in\mathcal{P}},
\end{equation*}
and for $y\in\field{R}$ its restriction (sub/super-level graph)
\begin{equation*}
\mathcal{G}_y(g):=\bigsset{(\pi,g(\pi))}{\pi\in\mathcal{P}, g(\pi)\lesseqgtr y}\subset\mathcal{G}(g).
\end{equation*}
It is known that a function $g:\mathcal{P}\to\field{R}$ is lower/upper semicontinuous
if and only if its sub/super-level sets 
\begin{equation*}
\mathcal{P}_y(g):=\bigsset{\pi\in\mathcal{P}}{g(\pi)\lesseqgtr y}
\end{equation*}
are closed for all $y\in\field{R}$.
Then $\mathcal{G}_y(g)$ is closed, and also compact, since $\mathcal{P}$ is compact.
Then the convex hull 
$\Conv\bigl(\mathcal{G}_y\bigr)\subset\Lin_\text{SA}(\mathcal{H})\oplus\field{R}$
is also compact.
On the other hand, its extremal points are $\Extr\bigl(\Conv(\mathcal{G}_y)\bigr)=\mathcal{G}_y$.
(Indeed, 
the points $(\pi,g(\pi))\in\mathcal{G}_y$ are extremal, since 
the elements of $\Conv(\mathcal{G}_y)$ are of the form 
$\sum_ip_i(\pi_i,g(\pi_i))=\bigl(\sum_ip_i\pi_i,\sum_ip_ig(\pi_i)\bigr)$ with $\pi_i\in\mathcal{P}_y(g)\subseteq\mathcal{P}$,
which equals to a $(\pi,g(\pi))$ with $\pi\in\mathcal{P}_y(g)\subseteq\mathcal{P}$ if and only if the mixture is trivial.
So we have $\Extr\bigl(\Conv(\mathcal{G}_y)\bigr) \supseteq \mathcal{G}_y$,
and the other inclusion $\Extr\bigl(\Conv(\mathcal{G}_y)\bigr) \subseteq \mathcal{G}_y$ is obvious by definition.)
For a $\rho\in\Conv\bigl(\mathcal{P}_y(g)\bigr)$,
we form the intersection
$\sset{(\rho,y')}{y'\in\field{R}} \cap \Conv\bigl(\mathcal{G}_y(g)\bigr)$,
which is compact (since $\Conv\bigl(\mathcal{G}_y(g)\bigr)$ is compact), so it is a line segment $\sset{(\rho,y')}{y'\in[y_\text{min},y_\text{max}]}$.
Since this segment is in $\Conv\bigl(\mathcal{G}_y(g)\bigr)$,
all of its points can be written as $(\rho,y') = \sum_ip_i(\pi_i,g(\pi_i)) = \bigl(\rho,\sum_ip_ig(\pi_i)\bigr)$
for $\pi_i\in\mathcal{P}_y(g)$,
so  $g^\cup(\rho)=y_\text{min}=\sum_ip_ig(\pi_i)$ in the lower semicontinuous case
and $g^\cap(\rho)=y_\text{max}=\sum_ip_ig(\pi_i)$ in the upper semicontinuous case.
Note that $\dim\bigl(\Conv\bigl(\mathcal{G}_y(g)\bigr)\bigr)=\dim(\mathcal{D})+1=(d^{2n}-1)+1=d^{2n}$,
and the points $(\rho,y_\text{min})$ or $(\rho,y_\text{max})$ belong to its boundary, being a face of dimension at most $d^{2n}-1$,
so, due to Carathéodory's theorem, the length of the decomposition above is at most $(d^{2n}-1)+1=d^{2n}$.

Now, 
to see the validity of the formula~\eqref{eq:DepthOFf},
we need that 
$D_f|_\mathcal{P}:\mathcal{P}\to\field{R}$ are
lower/upper semicontinuous for increasing/decreasing generator functions $f$.
A function is lower/upper semicontinuous
if and only if its sub/super-level sets are closed.
In our case,
let us have for $y\in\field{R}$
\begin{equation*}
k(y)=\begin{cases}
\max \sset{k'\in f(\pinv{P}_\text{I})}{k'\leq y},\\
\min \sset{k'\in f(\pinv{P}_\text{I})}{k'\geq y},
\end{cases}
\end{equation*}
for increasing and decreasing $f$, respectively,
then the sub/super-level sets of $D_f|_\mathcal{P}$ for increasing/decreasing $f$ are just
$\mathcal{P}_y(D_f|_\mathcal{P})=\bigsset{\pi\in\mathcal{P}}{D_f(\pi)\lesseqgtr y} = \mathcal{D}_{k(y),f}\cap\mathcal{P}=\mathcal{P}_{k(y),f}$
see~\eqref{eq:DepthfDlevel},
which are closed~\eqref{eq:PsepIIpf}.

\section{Coarsenings and symmetries}
\label{app:trafg}

In this section we consider the natural transformations of
the one-parameter entanglement properties, given in Section~\ref{sec:PSprops},
and of the resulting depths, given in Section~\ref{sec:Depthmeas}.

\subsection{Coarsenings and symmetries of \texorpdfstring{$f$}{f}-en\-tan\-gle\-ment}
\label{app:trafg.prop}

If a generator function $f:\pinv{P}_\text{I}\to\field{R}$
is composed with a \emph{monotone} function $g:\field{R}\to\field{R}$,
then the composition $g\circ f$ is also a generator function,
and it leads to the (possibly coarser) properties.
\begin{subequations}
\label{eq:gprops}
\begin{equation}
\label{eq:gprops.gen}
\pinv{\vs{\xi}}_{g(k),g\circ f} \coarsereq \pinv{\vs{\xi}}_{k,f},
\end{equation}
while if $g$ is \emph{strictly} monotone, then
the resulting generator function $g\circ f$,
although takes different values,
leads to the same properties (same sub- or super-level sets~\eqref{eq:vxir}),
\begin{equation}
\label{eq:gprops.strict}
\pinv{\vs{\xi}}_{g(k),g\circ f} = \pinv{\vs{\xi}}_{k,f},
\end{equation}
\end{subequations}
as the original generator function $f$.

To see~\eqref{eq:gprops.gen},
for all $k'\in g(f(\pinv{P}_\text{I}))$,
for all $k\in f(\pinv{P}_\text{I})$ such that $g(k)=k'$,
we have
\begin{subequations}
\label{eq:gpropsp}
\begin{equation}
\label{eq:gpropsp.gen}
\pinv{\vs{\xi}}_{k',g\circ f} \coarsereq \pinv{\vs{\xi}}_{k,f}.
\end{equation}
Indeed, by the definition~\eqref{eq:vxir},
for all $\pinv{\xi}\in\pinv{\vs{\xi}}_{k,f}$
we have $f(\pinv{\xi})\lesseqgtr k$,
then $g(f(\pinv{\xi}))\lesseqgtr g(k)$ for increasing $g$
or $g(f(\pinv{\xi}))\gtreqless g(k)$ for decreasing $g$,
both lead to $\pinv{\xi}\in\pinv{\vs{\xi}}_{g(k),g\circ f}=\pinv{\vs{\xi}}_{k',g\circ f}$~\eqref{eq:vxir}.

To see~\eqref{eq:gprops.strict},
we have the inverse function $g^{-1}$
in the case of strictly monotone $g$, 
so using~\eqref{eq:gprops.gen} with $g$ and also with $g^{-1}$
we have 
$\pinv{\vs{\xi}}_{k,f} = \pinv{\vs{\xi}}_{g^{-1}(g(k)),g^{-1}\circ g \circ f} \coarsereq  \pinv{\vs{\xi}}_{g(k),g\circ f} \coarsereq \pinv{\vs{\xi}}_{k,f}$,
then the antisymmetry of the partial order $\finereq$ leads to the claim.

Note that the main point here is that the effect of the transformation $g$
is just a possible coarsening of the classification.
For all $k'\in g(f(\pinv{P}_\text{I}))$,
we have
\begin{equation}
\label{eq:gpropsp.eq}
\pinv{\vs{\xi}}_{k',g\circ f} = \pinv{\vs{\xi}}_{k,f} \dspiff 
k=\begin{cases}
\max\bigl( g^{-1}(\set{k'}) \bigr),\\
\min\bigl( g^{-1}(\set{k'}) \bigr),
\end{cases}
\end{equation}
\end{subequations}
for increasing or decreasing $f$, respectively.
(Here $g^{-1}(\set{k'})=\sset{k\in f(\pinv{P}_\text{I})}{g(k)=k'}$ is the inverse image of $g$,
since $g$ is not necessarily strictly monotone, that is, not invertible in general.)
Indeed, 
to see the \textit{`if' direction}, 
\begin{equation*}
\pinv{\vs{\xi}}_{k',g\circ f} = \pinv{\vs{\xi}}_{k,f} \dspif 
k=\begin{cases}
\max\bigl( g^{-1}(\set{k'}) \bigr),\\
\min\bigl( g^{-1}(\set{k'}) \bigr),
\end{cases}
\end{equation*}
we have the $\coarsereq$ relation in~\eqref{eq:gpropsp.gen},
and we need to see the $\finereq$ relation,
that is,
for all $k'\in g(f(\pinv{P}_\text{I}))$, for all $\pinv{\xi}\in\pinv{P}_\text{I}$,
assuming increasing $g$,
\begin{equation*}
g(f(\pinv{\xi}))\lesseqgtr k' \dspthen f(\pinv{\xi})\lesseqgtr
\begin{cases}
\max\bigl( g^{-1}(\set{k'}) \bigr),\\
\min\bigl( g^{-1}(\set{k'}) \bigr),
\end{cases}
\end{equation*}
by~\eqref{eq:vxir},
which is equivalent by contraposition to that
\begin{equation*}
g(f(\pinv{\xi}))\gtrless k' \dspif f(\pinv{\xi})\gtrless
\begin{cases}
\max\bigl( g^{-1}(\set{k'}) \bigr),\\
\min\bigl( g^{-1}(\set{k'}) \bigr),
\end{cases}
\end{equation*}
which holds, because from the right-hand side
$g(f(\pinv{\xi}))\gtreqless k'$ follows by the increasing monotonicity of $g$,
and also $f(\pinv{\xi})\notin g^{-1}(\set{k'})$ which is equivalent to $g(f(\pinv{\xi}))\neq k'$
(this is where we used that not any element of the inverse image $g^{-1}(\set{k'})$ is used, but only the maximal/minimal one).
The proof of this point goes analogously for decreasing $g$.
To see the \textit{`only if' direction},
we use the contraposition
\begin{equation*}
\pinv{\vs{\xi}}_{k',g\circ f} \neq \pinv{\vs{\xi}}_{k'',f} \dspif 
k''\neq\begin{cases}
\max\bigl( g^{-1}(\set{k'}) \bigr),\\
\min\bigl( g^{-1}(\set{k'}) \bigr),
\end{cases}
\end{equation*}
which can be seen
by already having that $\pinv{\vs{\xi}}_{k',g\circ f}=\pinv{\vs{\xi}}_{k,f}$ 
for $k=\max/\min(g^{-1}(\set{k'}))$,
so it cannot be equal to $\pinv{\vs{\xi}}_{k'',f}$ for another $k''\neq\max/\min(g^{-1}(\set{k'}))$,
since
for all $k,k''\in f(\pinv{P}_\text{I})$ we have
$\pinv{\vs{\xi}}_{k,f}\neq\pinv{\vs{\xi}}_{k'',f}$ if and only if $k\neq k''$
by the order isomorphism~\eqref{eq:vxirchain} and standard order theory.

Having the two chains~\eqref{eq:vxirchain} of one-parameter entanglement properties,
the meaning of~\eqref{eq:gpropsp} is that $(k',g\circ f)$-entanglement properties are 
$(k,f)$-entanglement properties for specific $k$ values (which are maximal or minimal among those which are mapped to $k'$ by $g$),
the $(k,f)$-entanglement properties for the other $k$ values do not appear among the $(k',g\circ f)$-entanglement properties.

\subsection{Coarsenings and symmetries of \texorpdfstring{$f$}{f}-en\-tan\-gle\-ment state spaces and classes}
\label{app:trafg.DC}

If a generator function $f:\pinv{P}_\text{I}\to\field{R}$ 
is composed with a \emph{monotone} function $g:\field{R}\to\field{R}$,
then the composition $g\circ f$ 
leads to the (possibly larger) state spaces~\eqref{eq:DPsepIIpf} and (possibly coarser) classes~\eqref{eq:CsepIIIpf}
\begin{subequations}
\label{eq:gstates}
\begin{align}
\label{eq:gstates.genD}
\mathcal{D}_{g(k),g\circ f} &\supseteq \mathcal{D}_{k,f},\\
\label{eq:gstates.genC}
\mathcal{C}_{k',g\circ f} &= \bigcup_{k\in g^{-1}(\set{k'})} \mathcal{C}_{k,f}
\end{align}
for all $k'\in g(f(\pinv{P}_\text{I}))$,
while if $g$ is \emph{strictly} monotone, then
the resulting generator function $g\circ f$,
although taking different values,
leads to the same state spaces~\eqref{eq:DPsepIIpf} and classes~\eqref{eq:CsepIIIpf},
\begin{align}
\label{eq:gstates.strictD}
\mathcal{D}_{g(k),g\circ f} &= \mathcal{D}_{k,f},\\
\label{eq:gstates.strictC}
\mathcal{C}_{g(k),g\circ f} &= \mathcal{C}_{k,f},
\end{align}
\end{subequations}
as the original generator function $f$.

We have~\eqref{eq:gstates.genD} by
\eqref{eq:gprops.gen},~\eqref{eq:oisomDIIp} and the definition~\eqref{eq:DsepIIpf}.

To see~\eqref{eq:gstates.genC}, 
let us denote 
$g^{-1}(\set{k'})=\set{k_1,k_2,\dots k_m}$, where $k_i \gtrless k_j$ if $i<j$,
$k_1=\max/\min\bigl(g^{-1}(\set{k'})\bigr)$
and also the next value $k_{m+1}=\max/\min\bigl(g^{-1}(\set{k'_\mp})\bigr)$,
then we have 
\begin{equation*}
\mathcal{C}_{k',g\circ f} = \mathcal{D}_{k',g\circ f}\setminus \mathcal{D}_{k'_\mp,g\circ f}
= \mathcal{D}_{k_1,f}\setminus \mathcal{D}_{k_{m+1},f}
\end{equation*}
by definition~\eqref{eq:CsepIIIpf} and the result~\eqref{eq:gpropsp.eq},
then we have $\mathcal{D}_{k_i,f}\supseteq \mathcal{D}_{k_j,f}$ if $i<j$ by~\eqref{eq:DsepIIpfmonk},
so the above is
\begin{equation*}
\begin{split}
= 
(\mathcal{D}_{k_1,f}\setminus \mathcal{D}_{k_2,f}) &\cup
(\mathcal{D}_{k_2,f}\setminus \mathcal{D}_{k_3,f}) \cup \dots\\
&\qquad\qquad\dots\cup
(\mathcal{D}_{k_{m-1},f}\setminus \mathcal{D}_{k_m,f}),
\end{split}
\end{equation*}
which can be seen by the consecutive application of the set identity
$(A_1 \setminus A_2)\cup(A_2 \setminus A_3) = A_1 \setminus A_3$
for the nested sets $A_1 \supseteq A_2 \supseteq A_3$.

We have~\eqref{eq:gstates.strictD} by
\eqref{eq:gprops.strict},~\eqref{eq:oisomDIIp} and the definition~\eqref{eq:DsepIIpf}.

We have~\eqref{eq:gstates.strictC} by
\eqref{eq:gstates.strictD} and the definition~\eqref{eq:CsepIIIpf}.

\subsection{Coarsenings and symmetries of \texorpdfstring{$f$}{f}-en\-tan\-gle\-ment depth}
\label{app:trafg.fDepth}

Here we prove the formula in~\eqref{eq:gmeasuresfD}.
For all $k'\in g(f(\pinv{P}_\text{I}))$
we have $D_{g\circ f}(\rho)=k'$ if and only if $\rho\in\mathcal{C}_{k',g\circ f}$ by~\eqref{eq:DepthfC},
then $\rho\in \mathcal{C}_{k,f}$ for a $k\in f(\pinv{P}_\text{I})$ for which $g(k)=k'$ by~\eqref{eq:gstates.genC},
which holds if and only if $D_f(\rho)=k$ by~\eqref{eq:DepthfC} again,
then $g(D_f(\rho))=g(k)=k'=D_{g\circ f}(\rho)$.

\subsection{Coarsenings and symmetries of \texorpdfstring{$f$}{f}-en\-tan\-gle\-ment depth of formation}
\label{app:trafg.fDepthoF}

Here we prove the formulas in~\eqref{eq:gmeasuresfDoF}.

To see the case~\eqref{eq:gmeasuresfDoF.conv} of convex $g$,
for increasing generator function $f:\pinv{P}_\text{I}\to\field{R}$~\eqref{eq:genf}
and increasing convex function $g:\field{R}\to\field{R}$,
we have
\begin{equation*}
\begin{split}
&D^\text{oF}_{g\circ f}
\equalsref{eq:DepthOFf} \min\limits_{\set{(p_j,\pi_j)}\decomp \rho} \sum_j p_j D_{g\circ f}(\pi_j) \\
&\quad\equalsref{eq:gmeasuresfD} \min\limits_{\set{(p_j,\pi_j)}\decomp \rho} \sum_j p_j g\bigl( D_f(\pi_j) \bigr) \\
&\quad\gequals \min\limits_{\set{(p_j,\pi_j)}\decomp \rho} g\Bigl(\sum_j p_j D_f(\pi_j) \Bigr) \\
&\quad\equals g\Bigl(\min\limits_{\set{(p_j,\pi_j)}\decomp \rho} \sum_j p_j D_f(\pi_j) \Bigr)
\equalsref{eq:DepthOFf} g\bigl( D^\text{oF}_f\bigr),
\end{split}
\end{equation*}
where
the \textit{first equality} holds by definition~\eqref{eq:DepthOFf}, because $g\circ f$ is increasing if both $f$ and $g$ are increasing;
the \textit{second equality} holds because~\eqref{eq:gmeasuresfD} holds for all monotone $g$;
the \textit{inequality} holds by the convexity of $g$, and because a pointwise smaller function has smaller minimum;
the \textit{third equality} holds because $g$ is increasing;
and
the \textit{last equality} holds by definition~\eqref{eq:DepthOFf}, because $f$ is increasing.
For decreasing $f$ and $g$, we have increasing $g\circ f$,
and the derivation above holds with slight changes:
after the \textit{third equality} $\min$ is changed to $\max$, because $g$ is decreasing;
and
but the \textit{last equality} still holds by definition~\eqref{eq:DepthOFf}, because $f$ is now decreasing.
For decreasing $f$ and increasing $g$, we have decreasing $g\circ f$,
and the derivation above holds with slight changes:
after the \textit{first equality} $\min$ is changed to $\max$, because $g\circ f$ is decreasing;
and
the \textit{last equality} still holds by definition~\eqref{eq:DepthOFf}, because $f$ is now decreasing.
For increasing $f$ and decreasing $g$, we have decreasing $g\circ f$,
and the derivation above holds with slight changes:
after the \textit{first equality} $\min$ is changed to $\max$, because $g\circ f$ is decreasing;
but after the \textit{third equality} $\max$ is changed back to $\min$, because $g$ is decreasing;
and
the \textit{last equality} holds again by definition~\eqref{eq:DepthOFf}, because $f$ is increasing.

The case~\eqref{eq:gmeasuresfDoF.conc} for concave $g$ can be seen similarly,
only the inequality is flipped in the derivation above.

The case~\eqref{eq:gmeasuresfDoF.affine} for affine $g$ follows from~\eqref{eq:gmeasuresfDoF.conv} and~\eqref{eq:gmeasuresfDoF.conc},
by noting that an affine function is both convex \emph{and} concave.

\section{General properties of power-based generator functions}
\label{app:powerf}

In this section we show some properties
of power-based functions in Section~\ref{sec:1param}
together with proof for the convenience of the reader.

\subsection{Some basic tools for power functions}
\label{app:powerf.misc}

Here we recall some basic facts about power functions,
with emphasis on the precise ranges.

For $0< x\in\field{R}$, $q,q'\in \field{R}$, we have
\begin{equation}
\label{eq:xqmonq}
1 \lesseqgtr x: \quad q\leq q' \dspthen x^q \lesseqgtr x^{q'}.
\end{equation}
(This can easily be seen by taking the logarithm, or by the derivative $\partial x^q/\partial q = x^q\ln(x)$.)
For $0< x,x'\in\field{R}$, $q\in \field{R}$, we have
\begin{equation}
\label{eq:xqmonx}
0 \lesseqgtr q: \quad x\leq x' \dspthen x^q \lesseqgtr (x')^q.
\end{equation}
(This can easily be seen by taking the logarithm, or by the derivative $\partial x^q/\partial x = qx^{q-1}$.)
For $0< x_i\in\field{R}$, $q\in \field{R}$, we also have for the finite sums
\begin{equation}
\label{eq:sumq}
1 \lesseqgtr q: \quad \sum_{i=1}^m x_i^q \lesseqgtr \Bigl(\sum_{i=1}^m x_i\Bigr)^q.
\end{equation}
(This is simply because
$\frac{\sum_i x_i^q}{\bigl(\sum_j x_j\bigr)^q} 
= \sum_i\bigl(\frac{x_i}{\sum_j x_j}\bigr)^q
\lesseqgtr \sum_i\bigl(\frac{x_i}{\sum_j x_j}\bigr)^1 = 1$,
where $0 < \frac{x_i}{\sum_j x_j}\leq1$, so the inequalities in the two cases $1 \lesseqgtr q$
arise by applying the lower relation in~\eqref{eq:xqmonq}.)

\subsection{\texorpdfstring{$q$}{q}-limits of power based generator functions}
\label{app:powerf.Genfqlim}

Here we list the limits of the generator functions 
studied in Section~\ref{sec:1param}
with respect to the parameter $q$.
Although many of these are more or less well known,
we collect these together with the proofs for the convenience of the reader.
\begin{subequations}
\label{eq:qlim}
\begin{align}
\label{eq:qlim.s0}
s_0(\pinv{\xi}) &= h(\pinv{\xi}) \equiv \abs{\pinv{\xi}},\\
\label{eq:qlim.s1}
s_1(\pinv{\xi}) &= n,\\
\label{eq:qlim.spinf}
\lim_{q\to+\infty} s_q(\pinv{\xi}) 
&= \begin{cases}
n & \text{if $\pinv{\xi}=\bot$},\\
\infty  & \text{else},
\end{cases}\\
\label{eq:qlim.sninf}
s_{-\infty}(\pinv{\xi}) &:= \lim_{q\to-\infty} s_q(\pinv{\xi}) 
= \bigabs{\smset{x\in\pinv{\xi}}{x=1}}\\
\label{eq:qlim.N0p}
\lim_{q\to 0^-}N_q(\pinv{\xi}) &= \begin{cases}
n& \text{if $\pinv{\xi}=\top$},\\
0& \text{else},
\end{cases}\\ 
\label{eq:qlim.N0m}
\lim_{q\to 0^+}N_q(\pinv{\xi}) &= \begin{cases}
n & \text{if $\pinv{\xi}=\top$},\\
\infty & \text{else},
\end{cases}\\ 
\label{eq:qlim.N1}
N_1(\pinv{\xi}) &= n,\\ 
\label{eq:qlim.Npinf}
N_{+\infty}(\pinv{\xi}) &:= \lim_{q\to+\infty} N_q(\pinv{\xi}) = \max(\pinv{\xi}),\\
\label{eq:qlim.Nninf}
N_{-\infty}(\pinv{\xi}) &:= \lim_{q\to-\infty} N_q(\pinv{\xi}) = \min(\pinv{\xi}),\\
\label{eq:qlim.M0}
M_0(\pinv{\xi}) &:= \lim_{q\to 0}M_q(\pinv{\xi}) = \Bigl(\prod\pinv{\xi}\Bigr)^{1/\abs{\pinv{\xi}}},\\
\label{eq:qlim.M1}
M_1(\pinv{\xi}) &:= n/\abs{\pinv{\xi}},\\
\label{eq:qlim.Mpinf}
M_{+\infty}(\pinv{\xi}) &:= \lim_{q\to+\infty} M_q(\pinv{\xi}) = \max(\pinv{\xi}),\\
\label{eq:qlim.Mninf}
M_{-\infty}(\pinv{\xi}) &:= \lim_{q\to-\infty} M_q(\pinv{\xi}) = \min(\pinv{\xi}),\\
\label{eq:qlim.T0}
T_0(\pinv{\xi}) &= \abs{\pinv{\xi}}-1,\\
\label{eq:qlim.T1}
T_1(\pinv{\xi}) &:= \lim_{q\to 1} T_q(\pinv{\xi}) = S(\pinv{\xi}),\\
\label{eq:qlim.Tpinf}
T_{+\infty}(\pinv{\xi}) &:= \lim_{q\to+\infty} T_q(\pinv{\xi}) = 0,\\
\label{eq:qlim.Tninf}
\lim_{q\to-\infty} T_q(\pinv{\xi})
&= \begin{cases}
0 & \text{if $\pinv{\xi}=\top$},\\
\infty  & \text{else},
\end{cases}\\
\label{eq:qlim.R0}
R_0(\pinv{\xi}) &= \ln(\abs{\pinv{\xi}}),\\
\label{eq:qlim.R1}
R_1(\pinv{\xi}) &:= \lim_{q\to 1} R_q(\pinv{\xi}) = S(\pinv{\xi}),\\
\label{eq:qlim.Rpinf}
R_{+\infty}(\pinv{\xi}) &:= \lim_{q\to+\infty} R_q(\pinv{\xi}) = \ln(n)-\ln(\max(\pinv{\xi})),\\
\label{eq:qlim.Rninf}
R_{-\infty}(\pinv{\xi}) &:= \lim_{q\to-\infty} R_q(\pinv{\xi}) = \ln(n)-\ln(\min(\pinv{\xi})),\\
\label{eq:qlim.P0}
P_0(\pinv{\xi}) &= \abs{\pinv{\xi}},\\
\label{eq:qlim.P1}
P_1(\pinv{\xi}) &:= \lim_{q\to 1} P_q(\pinv{\xi}) = \ee^{S(\pinv{\xi})},\\
\label{eq:qlim.Ppinf}
P_{+\infty}(\pinv{\xi}) &:= \lim_{q\to+\infty} P_q(\pinv{\xi}) = n/\max(\pinv{\xi}),\\
\label{eq:qlim.Pninf}
P_{-\infty}(\pinv{\xi}) &:= \lim_{q\to-\infty} P_q(\pinv{\xi}) = n/\min(\pinv{\xi}).
\end{align}
\end{subequations}

We have~\eqref{eq:qlim.s0} and~\eqref{eq:qlim.s1} by definition~\eqref{eq:sq}.

To see~\eqref{eq:qlim.spinf} for~\eqref{eq:sq},
we have that 
$\lim_{q\to\infty}x^q=1$, or $\infty$ if $x=1$ or $1<x$, respectively,
and the subsystem sizes $x\in\pinv{\xi}$ are $1\leq x$,
and all of those are $1$ if and only if $\pinv{\xi}=\bot$.

To see~\eqref{eq:qlim.sninf} for~\eqref{eq:sq},
we have that 
$\lim_{q\to-\infty}x^q=1$, or $0$ if $x=1$ or $1<x$, respectively.

To see~\eqref{eq:qlim.N0p} and~\eqref{eq:qlim.N0m} for~\eqref{eq:Nq},
if $\pinv{\xi}=\top$ then we have $N_q(\top)=n$ without respect to $q$.
If $\pinv{\xi}\neq\top$ then we have 
$\lim_qN_q(\pinv{\xi})
=\lim_q\ee^{\ln(N_q(\pinv{\xi}))}
=\ee^{\lim_q\ln(N_q(\pinv{\xi}))}$,
for which
$\lim_{q\to0^\pm}\ln(N_q(\pinv{\xi}))
=\lim_{q\to0^\pm}\frac1q\ln\bigl(\sum_{x\in\pinv{\xi}}x^q\bigr)
=\bigl(\lim_{q\to0^\pm}\frac1q\bigr)\bigl(\lim_{q\to0^\pm}\ln(\sum_{x\in\pinv{\xi}}x^q)\bigr)
=\bigl(\lim_{q\to0^\pm}\frac1q\bigr)\ln(\abs{\pinv{\xi}})
=\pm\infty$
where $\ln(\abs{\pinv{\xi}})>0$ since $\abs{\pinv{\xi}}>1$ since $\pinv{\xi}\neq\top$,
leading to the claim. 

We have~\eqref{eq:qlim.N1} by definition~\eqref{eq:Nq}.

To see~\eqref{eq:qlim.Npinf} for~\eqref{eq:Nq},
we have $x\leq x_\text{max}:=\max(\pinv{\xi})$,
so, for $q\geq0$,
we have $x^q\leq x_\text{max}^q$ by~\eqref{eq:xqmonx},
which leads to the second inequality in
\begin{subequations}
\label{eq:cucc}
\begin{equation}
\label{eq:cucc.max}
0\leq q: \quad 
x_\text{max}^q\leq\sum_{x\in\pinv{\xi}}x^q\leq \abs{\pinv{\xi}} x_\text{max}^q.
\end{equation}
Applying the $q$-th root for $q>0$ to this
gives
$x_\text{max}\leq N_q(\pinv{\xi})
\leq \abs{\pinv{\xi}}^{1/q}x_\text{max}$ by~\eqref{eq:xqmonx}.
Taking the limit $q\to+\infty$ leads to the claim.

To see~\eqref{eq:qlim.Nninf} for~\eqref{eq:Nq},
we have $x\geq x_\text{min}:=\min(\pinv{\xi})$,
so, for $q\leq0$,
we have $x^q\leq x_\text{min}^q$ by~\eqref{eq:xqmonx},
which leads to the second inequality in
\begin{equation}
\label{eq:cucc.min}
0\geq q: \quad 
x_\text{min}^q\leq\sum_{x\in\pinv{\xi}}x^q\leq \abs{\pinv{\xi}} x_\text{min}^q.
\end{equation}
\end{subequations}
Applying the $q$-th root for $q<0$ to this gives
$x_\text{min}\geq N_q(\pinv{\xi})
\geq \abs{\pinv{\xi}}^{1/q}x_\text{min}$ by~\eqref{eq:xqmonx}.
Taking the limit $q\to-\infty$ leads to the claim.

To see~\eqref{eq:qlim.M0} for~\eqref{eq:Mq},
we have
$\lim_qM_q(\pinv{\xi})
=\lim_q\ee^{\ln M_q(\pinv{\xi})}
=\ee^{\lim_q\ln M_q(\pinv{\xi})}$,
for which we have
$\lim_{q\to0}\ln(M_q(\pinv{\xi}))
=\lim_{q\to0}\frac1q\ln\bigl(\frac1{\abs{\pinv{\xi}}}\sum_{x\in\pinv{\xi}}x^q\bigr)$,
which, using L'H\^{o}pital rule for the $0/0$-type limit, is
$\lim_{q\to0}\frac{\sum_{x\in\pinv{\xi}}x^q\ln(x)}{\sum_{x\in\pinv{\xi}}x^q}
=\frac1{\abs{\pinv{\xi}}}\sum_{x\in\pinv{\xi}}\ln(x)
=\ln((\prod_{x\in\pinv{\xi}} x)^{1/\abs{\pinv{\xi}}})$,
leading to the claim.

We have~\eqref{eq:qlim.M1} by definition~\eqref{eq:Mq}.

To see~\eqref{eq:qlim.Mpinf} for~\eqref{eq:Mq},
we recall~\eqref{eq:cucc.max}.
Multiplying this with $\abs{\pinv{\xi}}^{-1}$ and 
applying the $q$-th root for $q>0$ to this
gives
$\abs{\pinv{\xi}}^{-1/q}x_\text{max}\leq M_q(\pinv{\xi})
\leq x_\text{max}$ by~\eqref{eq:xqmonx}.
Taking the limit $q\to+\infty$ leads to the claim.

To see~\eqref{eq:qlim.Mninf} for~\eqref{eq:Mq},
we recall~\eqref{eq:cucc.min}.
Multiplying this with $\abs{\pinv{\xi}}^{-1}$ and 
applying the $q$-th root for $q<0$ to this
gives
$\abs{\pinv{\xi}}^{-1/q}x_\text{min}\geq M_q(\pinv{\xi})
\geq x_\text{min}$ by~\eqref{eq:xqmonx}.
Taking the limit $q\to-\infty$ leads to the claim.

We have~\eqref{eq:qlim.T0} by definition~\eqref{eq:Tq}.

To see~\eqref{eq:qlim.T1} for~\eqref{eq:Tq},
we use L'H\^{o}pital rule for the $0/0$-type limit,
$\lim_{q\to1}\frac1{1-q}(\sum_{x\in\pinv{\xi}}(x/n)^q-1)
=\lim_{q\to1}-\sum_{x\in\pinv{\xi}}(x/n)^q\ln(x/n)
= -\sum_{x\in\pinv{\xi}}(x/n)\ln(x/n)$,
which is~\eqref{eq:S}.

To see~\eqref{eq:qlim.Tpinf} for~\eqref{eq:Tq},
we have that $x/n\leq1$,
so the nominator is bounded, while the denominator grows to $-\infty$,
leading to the claim.

To see~\eqref{eq:qlim.Tninf} for~\eqref{eq:Tq},
we have that if $\pinv{\xi}\neq\top$ then $x/n<1$,
so the nominator grows exponentially, while the denominator grows only linearly,
leading to the claim.
If $\pinv{\xi}=\top=\set{n}$ then there is only one summand,
$x/n=1$, leading to the claim.

We have~\eqref{eq:qlim.R0} by definition~\eqref{eq:Rq}.

To see~\eqref{eq:qlim.R1} for~\eqref{eq:Rq},
we use L'H\^{o}pital rule for the $0/0$-type limit,
$\lim_{q\to1}\frac1{1-q}\ln(\sum_{x\in\pinv{\xi}}(x/n)^q)
=\lim_{q\to1}\frac{-\sum_{x\in\pinv{\xi}}(x/n)^q\ln(x/n)}{\sum_{x\in\pinv{\xi}}(x/n)^q}
= -\sum_{x\in\pinv{\xi}}(x/n)\ln(x/n)$,
which is~\eqref{eq:S}.

To see~\eqref{eq:qlim.Rpinf} for~\eqref{eq:Rq},
we recall~\eqref{eq:cucc.max}.
Multiplying this with $1/n^q$, applying $\ln()$ and multiplying with $1/(1-q)$
for $q>1$ gives
$\frac{q}{1-q}\ln(x_\text{max}/n)\geq R_q(\pinv{\xi})
\geq \frac{\ln(\abs{\pinv{\xi}})}{1-q} + \frac{q}{1-q}\ln(x_\text{max}/n)$.
Taking the limit $q\to+\infty$ leads to the claim.

To see~\eqref{eq:qlim.Rninf} for~\eqref{eq:Rq},
we recall~\eqref{eq:cucc.min}.
Multiplying this with $1/n^q$, applying $\ln()$ and multiplying with $1/(1-q)$
for $q<0$ gives
$\frac{q}{1-q}\ln(x_\text{min}/n)\leq R_q(\pinv{\xi})
\leq \frac{\ln(\abs{\pinv{\xi}})}{1-q} + \frac{q}{1-q}\ln(x_\text{min}/n)$.
Taking the limit $q\to-\infty$ leads to the claim.

To see~\eqref{eq:qlim.P0},~\eqref{eq:qlim.P1},~\eqref{eq:qlim.Ppinf} and~\eqref{eq:qlim.Pninf} for~\eqref{eq:Rq},
note that $P_q=\ee^{R_q}$ is a monotone function of $R_q$, for which we have
\eqref{eq:qlim.R0},~\eqref{eq:qlim.R1},~\eqref{eq:qlim.Rpinf} and~\eqref{eq:qlim.Rninf}.

\subsection{\texorpdfstring{$q$}{q}-monotonicity of power based generator functions}
\label{app:powerf.Genfqmon}

Here we list the monotonicity of the power based generator functions 
studied in Section~\ref{sec:1param}
with respect to the parameter $q$.
Although many of these are more or less well known,
we collect these together with the proofs for the convenience of the reader.
\begin{subequations}
\label{eq:qmon}
\begin{align}
\label{eq:qmon.sq}
q < q'     &\dspthen s_q(\pinv{\xi}) \leq s_{q'}(\pinv{\xi}),\\
\label{eq:qmon.Nq1}
0 < q < q' &\dspthen N_q(\pinv{\xi}) \geq N_{q'}(\pinv{\xi}),\\
\label{eq:qmon.Nq2}
q < 0 < q' &\dspthen N_q(\pinv{\xi}) \leq N_{q'}(\pinv{\xi}),\\
\label{eq:qmon.Nq3}
q < q'< 0  &\dspthen N_q(\pinv{\xi}) \geq N_{q'}(\pinv{\xi}),\\
\label{eq:qmon.Mq}
q < q'     &\dspthen M_q(\pinv{\xi}) \leq M_{q'}(\pinv{\xi}),\\
\label{eq:qmon.Tq}
q < q'     &\dspthen T_q(\pinv{\xi}) \geq T_{q'}(\pinv{\xi}),\\
\label{eq:qmon.Rq}
q < q'     &\dspthen R_q(\pinv{\xi}) \geq R_{q'}(\pinv{\xi}),\\
\label{eq:qmon.Pq}
q < q'     &\dspthen P_q(\pinv{\xi}) \geq P_{q'}(\pinv{\xi}).
\end{align}
\end{subequations}

To see~\eqref{eq:qmon.sq} for~\eqref{eq:sq},
we have~\eqref{eq:xqmonq} for each summand, $1\leq x$.

To see~\eqref{eq:qmon.Nq1}-\eqref{eq:qmon.Nq3} for~\eqref{eq:Nq},
we note that $N_q(\pinv{\xi})$ is not continuous in $q=0$, see~\eqref{eq:qlim.N0p}-\eqref{eq:qlim.N0m},
so first we consider the cases $0<q\leq q'$ and $q\leq q'<0$.
For these, for $0 \lessgtr q$
we have $x\lesseqgtr N_q(\pinv{\xi})$ for all $x\in\pinv{\xi}$
(by~\eqref{eq:xqmonx}, since $x^q\leq\sum_{x\in\pinv{\xi}}x^q$),
so $x/N_q(\pinv{\xi})\lesseqgtr 1$.
Then $\bigl(x/N_q(\pinv{\xi})\bigr)^q\gtreqless\bigl(x/N_q(\pinv{\xi})\bigr)^{q'}$
by~\eqref{eq:xqmonq}.
Summing up,
$1=\sum_{x\in\pinv{\xi}}\bigl(x/N_q(\pinv{\xi})\bigr)^q\gtreqless
\sum_{x\in\pinv{\xi}}\bigl(x/N_q(\pinv{\xi})\bigr)^{q'}$,
that is,
$N_q^{q'}(\pinv{\xi})\gtreqless\sum_{x\in\pinv{\xi}}x^{q'}$,
which is
$N_q(\pinv{\xi})\geq N_{q'}(\pinv{\xi})$
in the two cases $0 \lessgtr q$
by~\eqref{eq:xqmonx}.
Now for the case $q<0<q'$ we have the limits~\eqref{eq:qlim.Npinf}-\eqref{eq:qlim.Nninf},
by which $ N_q \leq N_{-\infty}=\min \leq \max=N_{+\infty} \leq N_{q'}$.

To see~\eqref{eq:qmon.Mq} for~\eqref{eq:Mq} (power mean inequality, with equal weights),
first we consider the cases $0<q\leq q'$ and $q\leq q'<0$.
For these, for $0 \lessgtr q$,
we have
$0<q/q' \lesseqgtr 1$, so the function
$x\mapsto x^{q/q'}$ is concave/convex in the two cases,
so
$\sum_{x\in\pinv{\xi}}\frac1{\abs{\pinv{\xi}}}x^q = 
\sum_{x\in\pinv{\xi}}\frac1{\abs{\pinv{\xi}}}(x^{q'})^{q/q'} \lesseqgtr
\bigl(\sum_{x\in\pinv{\xi}}\frac1{\abs{\pinv{\xi}}} x^{q'} \bigr)^{q/q'}$,
which, taking the $q$-th root, is
$M_q(\pinv{\xi})\leq M_{q'}(\pinv{\xi})$
by~\eqref{eq:xqmonx}.
Now for the case $q<0<q'$ we have the limit~\eqref{eq:qlim.M0},
by which
$\ln(M_0(\pinv{\xi}))=\ln\bigl(\bigl(\prod_{x\in\pinv{\xi}} x\bigr)^{1/\abs{\pinv{\xi}}}\bigr)
=\sum_{x\in\pinv{\xi}}\frac1{\abs{\pinv{\xi}}} \ln(x)
=\frac1q\sum_{x\in\pinv{\xi}}\frac1{\abs{\pinv{\xi}}} \ln(x^q)
\lesseqgtr 
 \frac1q\ln\bigl(\sum_{x\in\pinv{\xi}}\frac1{\abs{\pinv{\xi}}}x^q \bigr)
=\ln\bigl(\bigl(\sum_{x\in\pinv{\xi}}\frac1{\abs{\pinv{\xi}}}x^q \bigr)^{1/q}\bigr)
=\ln(M_q(\pinv{\xi}))$
for the cases $0 \lesseqgtr q$,
where the concavity of the logarithm was used.

To see~\eqref{eq:qmon.Tq} for~\eqref{eq:Tq},
we differentiate the $q$-logarithm $\ln_{(q)}(u)=\frac{u^{1-q}-1}{1-q}$ for $u>0$ as
$\frac{\partial}{\partial q} \ln_{(q)}(u)=\frac1{(1-q)^2} \bigl(u^{1-q}-1 -(1-q)u^{1-q}\ln(u)\bigr)
=\frac1{(1-q)^2} \bigl(u^{1-q}-1 -u^{1-q}\ln(u^{1-q})\bigr)
=\frac1{(1-q)^2} \bigl(v-1-v\ln(v)\bigr)\leq0$
using the new variable $v=u^{1-q}$.
This holds because $-v\ln(v)\leq1-v$, which follows from that
the two sides, as well as their derivatives are equal in $v=1$,
and the derivative of the left-hand side ($-1-\ln(v)$) is larger/smaller than that of the right-hand side ($-1$)
for $v\leq1$ and $v\geq1$, respectively.
So $\ln_{(q)}(u)$ is decreasing in $q$,
then $T_q(\pinv{\xi})=\frac1{1-q}\Bigl(\sum_{x\in\pinv{\xi}}\bigl(\frac{x}{n}\bigr)^q -1 \Bigr)
=\sum_{x\in\pinv{\xi}} \frac{x}{n}\ln_{(q)}\bigl(\frac{n}{x}\bigr)$
is also decreasing in $q$.

To see~\eqref{eq:qmon.Rq} for~\eqref{eq:Rq},
we differentiate $R_q$ as
\begin{equation*}
\begin{split}
&(1-q)^2\frac{\partial}{\partial q} R_q(\pinv{\xi})
=(1-q)^2\frac{\partial}{\partial q} \frac1{1-q} \ln\Bigl(\sum_{x\in \pinv{\xi}} \frac{x^q}{n^q} \Bigr)\\
&\quad= \ln\Bigl(\sum_{x} \frac{x^q}{n^q} \Bigr) 
+ \frac{(1-q)}{\sum_{x'}\frac{{x'}^q}{n^q}} \sum_{x} \frac{x^q}{n^q}\ln\bigl(\frac{x}{n}\bigr) \\
&\quad= \ln\Bigl(\sum_{x''} \frac{{x''}^q}{n^q} \Bigr)
+ \sum_x \frac{\frac{x^q}{n^q}}{\sum_{x'}\frac{{x'}^q}{n^q}} \Bigl(\ln\bigl(\frac{x}{n}\bigr)-\ln\bigl(\frac{x^q}{n^q}\bigr)\Bigr)\\
&\quad= \sum_x \frac{\frac{x^q}{n^q}}{\sum_{x'}\frac{{x'}^q}{n^q}}
\Bigl(
\ln\Bigl(\sum_{x''} \frac{{x''}^q}{n^q} \Bigr)
+ \ln\bigl(\frac{x}{n}\bigr)-\ln\bigl(\frac{x^q}{n^q}\bigr) \Bigr)\\
&\quad=- \sum_x \frac{\frac{x^q}{n^q}}{\sum_{x'}\frac{{x'}^q}{n^q}}
\Bigl(
\ln\Bigl( \frac{\frac{x^q}{n^q}}{\sum_{x''}\frac{{x''}^q}{n^q}}\Bigr) - \ln\bigl(\frac{x}{n}\bigr) 
\Bigr),
\end{split}
\end{equation*}
so $-(1-q)^2\frac{\partial}{\partial q} R_q(\pinv{\xi}$
is the Kullback-Leibler divergence $D(p,r)=\sum_i p_i(\ln(p_i)-\ln(q_i))$, or relative entropy~\cite{Wilde-2013} of the distributions
$\bigsset{ \frac{(x_i/n)^q}{s_q(\pinv{\xi}/n)}}{i=1,2,\dots,\abs{\pinv{\xi}}}$
and 
$\pinv{\xi}/n=\sset{x_i/n}{i=1,2,\dots,\abs{\pinv{\xi}}}$
(for any indexing of the elements $\pinv{\xi}$).
This is nonnegative,
which can be seen by the convexity of the logarithm: $-D(p,r)= \sum_i p_i\ln(\frac{q_i}{p_i})\leq \ln(\sum_ip_i\frac{q_i}{p_i}=\ln(\sum_i q_i)=0$.

To see~\eqref{eq:qmon.Pq} for~\eqref{eq:Pq},
note that $P_q=\ee^{R_q}$ is a strictly increasing function of $R_q$,
for which we have~\eqref{eq:qmon.Rq}.

\section{Construction of \texorpdfstring{$f$}{f}-entanglement}
\label{app:1param}

In this section we provide some tools for the refinement and dominance orders used in the main text,
and show the monotonicity of the generator functions~\eqref{eq:genf} in Section~\ref{sec:1param}
with respect to these.

\subsection{Orders}
\label{app:1param.Gen}

Here we recall the partial orders 
applied in the construction in the main text~\cite{Szalay-2015b,Szalay-2019},
and consider their relationship.

An \emph{extremal stochastic $m'\times m$ matrix} is given 
as a matrix of elements $a_{ij}\in\set{0,1}$,
such that $\sum_{i=1}^{m'} a_{ij}=1$ for all $j=1,2,\dots,m$.

A \emph{partition} of the set $\set{1,2,\dots,n}$ is its disjoint co\-ve\-ring
$\xi=\set{X_1,X_2,\dots,X_{\abs{\xi}}}$,
and the elements $X\in\xi$ are called \emph{parts}.
The set of partitions of $\set{1,2,\dots,n}$ is denoted with ${P}_\text{I}$.
The partial order \emph{refinement}~\cite{Davey-2002,Stanley-2012} is given for the partitions $\upsilon$ and $\xi$ as
\begin{subequations}
\label{eq:poI}
\begin{equation}
\label{eq:poI.p}
\upsilon\finereq\xi \dspdef \forall Y\in\upsilon, \exists X\in\xi \dispt{s.t.} Y\subseteq X.
\end{equation}
It is easy to check that for the partitions
$\upsilon=\set{Y_1,Y_2,\dots,Y_{\abs{\upsilon}}}$ and $\xi=\set{X_1,X_2,\dots,X_{\abs{\xi}}}$
we have
\begin{equation}
\label{eq:poI.x}
\begin{split}
\upsilon\finereq\xi \dspiff  
\text{$\exists [a_{ij}]$ extremal stochastic}\\
\text{$\abs{\xi}\times\abs{\upsilon}$ matrix, s.t. } X_i = \bigcup_{j=1}^{\abs{\upsilon}} {a_{ij}} Y_j,
\end{split}
\end{equation}
(where we use the notation $0Y=\emptyset$, $1Y=Y$),
which informally means that
$\xi\majors\upsilon$ can be obtained from $\upsilon$
by forming partial unions.
Note that the set partitions $P_\text{I}$ with respect to the refinement order forms a lattice~\cite{Roman-2008,Stanley-2012},
that is, there exist unique least upper and greatest lower bounds for all pairs of set partitions.
The covering relation (arrow) of $\finereq$ is
\begin{equation}
\label{eq:poI.c}
\upsilon\finerc\xi \dspiff
\upsilon\finer\xi\;\text{and}\;\abs{\upsilon}=\abs{\xi}-1,
\end{equation}
which informally means that
only two parts from $\upsilon$ are joined to obtain $\pinv{\xi}$.
\end{subequations}

The \emph{type} $\pinv{\xi}$ of a partition $\xi$ is given as the \emph{multiset}
$\pinv{\xi}=\smset{\abs{X}}{X\in\xi}$, which is an integer partition of $n$~\cite{Davey-2002,Stanley-2012,Szalay-2019}.
Let $s$ denote the elementwise action of $\abs{\;}$,
so $\pinv{\xi}=s(\xi)$, and 
let $s^{-1}$ denote its inverse image,
$s^{-1}(\pinv{\xi})=\sset{\xi\in P_\text{I}}{s(\xi)=\pinv{\xi}}$.
The set of integer partitions of $n$ is denoted with $\pinv{P}_\text{I}$.
The partial order \emph{refinement}~\cite{Szalay-2019} is given for the integer partitions $\pinv{\upsilon}$ and $\pinv{\xi}$ as
\begin{subequations}
\label{eq:poIp}
\begin{equation}
\label{eq:poIp.p}
\pinv{\upsilon}\finereq\pinv{\xi} \dspdef \exists \upsilon\in s^{-1}(\pinv{\upsilon}),\xi\in s^{-1}(\pinv{\xi}) \dispt{s.t.} \upsilon\finereq\xi.
\end{equation}
It is easy to check that for the integer partitions
$\pinv{\upsilon}=\set{y_1,y_2,\dots,y_{\abs{\pinv{\upsilon}}}}$ and $\pinv{\xi}=\set{x_1,x_2,\dots,x_{\abs{\pinv{\xi}}}}$
we have
\begin{equation}
\label{eq:poIp.x}
\begin{split}
\pinv{\upsilon}\finereq\pinv{\xi} \dspiff 
\text{$\exists [a_{ij}]$ extremal stochastic}\\
\text{$\abs{\pinv{\xi}}\times\abs{\pinv{\upsilon}}$ matrix, s.t. } x_i = \sum_{j=1}^{\abs{\pinv{\upsilon}}} a_{ij} y_j,
\end{split}
\end{equation}
which informally means that
$\pinv{\xi}\majors\pinv{\upsilon}$ can be obtained from $\pinv{\upsilon}$
by partial summation.
Note that the integer partitions $\pinv{P}_\text{I}$ with respect to the refinement order does not form a lattice,
that is, there are pairs of integer partitions for which there exists no unique least upper or greatest lower bound
(for $n\geq5$, see Figure~\ref{fig:PpI23456PPS}).
The covering relation (arrow, see Figure~\ref{fig:PpI23456PPS}) of $\finereq$ is
\begin{equation}
\label{eq:poIp.c}
\pinv{\upsilon}\finerc\pinv{\xi} \dspiff
\pinv{\upsilon}\finer\pinv{\xi}\;\text{and}\;\abs{\pinv{\upsilon}}=\abs{\pinv{\xi}}-1,
\end{equation}
which informally means that
only two parts from $\pinv{\upsilon}$ are added to obtain $\pinv{\xi}$.
\end{subequations}

Note that, although the definition of extremal stochastic matrices allows $0$ rows,
such matrices do not play role in the characterizations~\eqref{eq:poI.x} and~\eqref{eq:poIp.x}.

\begin{figure}\centering
\includegraphics{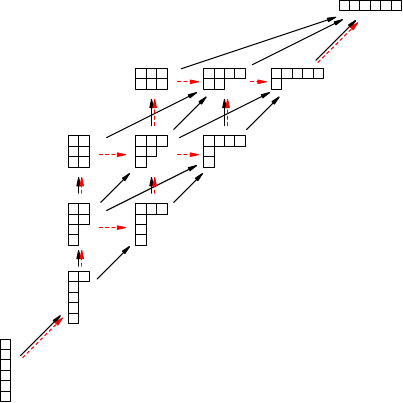}
\caption{The covering relations of the refinement poset (black arrows,~\eqref{eq:poI.c})
and the dominance lattice (dashed red arrows,~\eqref{eq:poIpdom.c}) on $\pinv{P}_\text{I}$
illustrated for $n=6$ in the height~\eqref{eq:hwr.h} vs.~width~\eqref{eq:hwr.w} plot.}
\label{fig:PpI6d}
\end{figure}

The \emph{dominance order} (also called \emph{majorization}~\cite{Bengtsson-2006,Marshall-2010,Sagawa-2022} for the continuous variable case),
is a partial order given for the integer partitions~\cite{Brylawski-1973,Stanley-2012}
$\pinv{\upsilon}=\set{y_1,y_2,\dots,y_{\abs{\pinv{\upsilon}}}}$ and $\pinv{\xi}=\set{x_1,x_2,\dots,x_{\abs{\pinv{\xi}}}}$,
indexed in weakly decreasing order ($x_i\geq x_j$ and $y_i\geq y_j$, $i<j$), as
\begin{subequations}
\label{eq:poIpdom}
\begin{equation}
\label{eq:poIpdom.p}
\begin{split}
\pinv{\upsilon}\majordeq\pinv{\xi} \dspdef 
&\sum_{i=1}^m y_i \leq \sum_{i=1}^m x_i \\
&\quad\text{for all $m\in\set{1,2,\dots,n}$},
\end{split}
\end{equation}
where $y_i=0$ or $x_i=0$ is set if $i>\abs{\pinv{\upsilon}}$ or $i>\abs{\pinv{\xi}}$, respectively.
Note that the integer partitions $\pinv{P}_\text{I}$ with respect to the dominance order forms a lattice~\cite{Brylawski-1973},
that is, there exist unique least upper and greatest lower bounds for all pairs of integer partitions.
The covering relation (arrow) of $\majordeq$ is also known~\cite{Brylawski-1973},
\begin{equation}
\label{eq:poIpdom.c}
\pinv{\upsilon}\majordc\pinv{\xi} \dspiff
\begin{cases}
x_i=y_i+1,\\
x_j=y_j-1,\\
x_k=y_k, \forall k\neq i,j,\\
\text{when $j=i+1$ or $y_i=y_j$}
\end{cases}
\end{equation}
which informally means that
the Young diagram of $\pinv{\xi}\majorsc\pinv{\upsilon}$
can be obtained from the Young diagram of $\pinv{\upsilon}$
by removing the last box of row $j$ and appending it 
either to the end of the immediately preceding row $j-1$,
or to the end of row $i<j$ if the rows $i$ through $j$ of the Young diagram of $\pinv{\upsilon}$ all have the same length,
$y_{i-1}<y_i=y_{i+1}=\dots=y_{j-1}=y_j<y_{j+1}$.
(Note that we drop the possible $x_j=y_j-1=0$ values.)
(For illustration, see Figure~\ref{fig:PpI6d}.)
We will also use the following condition
\begin{equation}
\label{eq:poIpdom.simpler}
\pinv{\upsilon}\majordeq\pinv{\xi} \dspif
\exists i<j \;\text{s.t.}\;
\begin{cases}
x_i=y_i+1,\\
x_j=y_j-1,\\
x_k=y_k, \forall k\neq i,j,
\end{cases}
\end{equation}
\end{subequations}
which is easy to prove by the definition~\eqref{eq:poIpdom.p}.

Now we show that
refinement $\finereq$ implies dominance $\majordeq$,
that is, for the integer partitions $\pinv{\upsilon},\pinv{\xi}\in \pinv{P}_\text{I}$,
\begin{equation}
\label{eq:poIs}
\pinv{\upsilon}\finereq\pinv{\xi} \dspthen \pinv{\upsilon}\leq\pinv{\xi}.
\end{equation}
To see this, we assume~\eqref{eq:poIp.x}
with weakly decreasingly ordered elements $y_j$ and $x_i$.
Before we show the inequality in~\eqref{eq:poIpdom.p} for general $m$,
first let us check the $m=1$ and $2$ cases as illustration.
For the case $m=1$,
let $i\in\set{1,2,\dots,\abs{\pinv{\xi}}}$ be the index for which $a_{i1}=1$
(this is \emph{unique}, because $a_{ij}$ are elements of an extremal stochastic matrix),
then
\begin{equation*}
y_1 \leq \sum_{j=1}^{\abs{\pinv{\upsilon}}} a_{ij} y_j = x_i \leq x_1,
\end{equation*}
where the assumption~\eqref{eq:poIp.x} and the decreasing ordering of the elements $x_i$ are also used.
For the case $m=2$,
let $i_1,i_2\in\set{1,2,\dots,\abs{\pinv{\xi}}}$ be the \emph{unique} indices for which $a_{i_11}=1$ and $a_{i_22}=1$,
then
\begin{equation*}
y_1 + y_2 \leq \sum_{j=1}^{\abs{\pinv{\upsilon}}} a_{i_1j} y_j + \sum_{j=1}^{\abs{\pinv{\upsilon}}} a_{i_2j} y_j = x_{i_1} + x_{i_2} \leq x_1 + x_2,
\end{equation*}
where the assumption~\eqref{eq:poIp.x} and the decreasing ordering of the elements $x_i$ are also used:
the sum of any two $x$s are less than or equal to the sum of the two largest ones.
In general, for all $l\in\set{1,2,\dots,m}$,
let $i_l\in\set{1,2,\dots,\abs{\pinv{\xi}}}$ be the \emph{unique} indices for which $a_{i_ll}=1$,
then
\begin{equation*}
\sum_{l=1}^m y_l \leq \sum_{l=1}^m\sum_{j=1}^{\abs{\pinv{\upsilon}}} a_{i_lj} y_j = \sum_{l=1}^m x_{i_l} \leq \sum_{l=1}^m x_l,
\end{equation*}
where the assumption~\eqref{eq:poIp.x} and the decreasing ordering of the elements $x_i$ are also used:
the sum of any $l$ $x$-s are less than or equal to the sum of the $l$ largest ones.
Then~\eqref{eq:poIpdom.p} leads to the claim.
Note that the reverse implication does not hold in~\eqref{eq:poIs},
as can be shown by the simple counterexample
$\pinv{\upsilon}=\mset{2,2}$ and $\pinv{\xi}=\mset{3,1}$,
for which $\pinv{\upsilon}\majordeq\pinv{\xi}$
but $\pinv{\upsilon}\not\finereq\pinv{\xi}$ and $\pinv{\upsilon}\not\coarsereq\pinv{\xi}$.

There are functions $f:\pinv{P}_\text{I}\to\field{R}$
increasing/decreasing monotone with respect to dominance~\eqref{eq:poIpdom.p}
\begin{equation}
\label{eq:dommon}
\pinv{\upsilon}\majordeq\pinv{\xi} \dspthen f(\pinv{\upsilon})\lesseqgtr f(\pinv{\xi}).
\end{equation}
(These are called \emph{Schur-convex/concave} for the case of probability distributions or continuous variables in general.)
Then, putting together~\eqref{eq:genf.mon} and~\eqref{eq:dommon} with~\eqref{eq:poIs},
we have that any $\majordeq$-monotone (dominance-monotone) function $\pinv{P}_\text{I}\to\field{R}$
is $\finereq$-monotone (refinement-monotone),
hence proper generator function~\eqref{eq:genf}.
In the continuous variable case, decreasing dominance-monotonicity,
or Schur-concavity is the defining property of \emph{(generalized) entropies}~\cite{Bengtsson-2006},
so any entropy of the normalized integer partitions $\pinv{\xi}/n\equiv\sset{x/n}{x\in\pinv{\xi}}$
is a proper generator function. 
Note, however, that for integer partitions, as is being considered here,
we have a slightly different domain of the functions.
Usually integer partitions of $n$ are represented as $n$-tuples of nonnegative integers $\set{0,1,2,\dots,n}\subset \field{N}_0$,
by adding zeroes
$\pinv{\xi}=\mset{x_1,x_2,\dots,x_{\abs{\pinv{\xi}}}}\mapsto (x_1,x_2,\dots,x_{\abs{\pinv{\xi}}},0,\dots,0)$
and then the dominance-monotonicity or Schur-convexity/concavity of $n$-variable functions over $\set{0,1,2,\dots,n}^n$ or $[0,n]_\field{R}^n$ can be defined.
Here we represent partitions of $n$ as multisets of strictly positive integers $\set{1,2,\dots,n}\subset \field{N}$,
so zero is not allowed, which is the natural way in our construction.
Then the generator functions acting on partitions
can be valid even if the analogously defined $n$-variable functions would not
(see for example the power based generator functions $s_q$, $N_q$, $M_q$, $T_q$, $R_q$, $P_q$ for $q<0$ in Appendix~\ref{app:1param.Mon}).
This also prevents the use of the Schur-Ostrowski criterion~\cite{Marshall-2010} for $n$-variable functions
to check Schur-convexity/concavity, 
so we will give direct proofs (see in Appendix~\ref{app:1param.dmon}) based on the covering relation~\eqref{eq:poIpdom.c}.

\subsection{Monotonicity of generator functions}
\label{app:1param.Mon}

Here we show the monotonicity properties of the generator functions
studied in Section~\ref{sec:1param}.
\begin{subequations}
\label{eq:mon}
\begin{align}
\label{eq:mon.wm}
m\in\field{N}: \quad
\pinv{\upsilon}\finereq\pinv{\xi} &\dspthen  w_m(\pinv{\upsilon}) \leq w_m(\pinv{\xi}),\\
\label{eq:mon.tm}
m\in\field{N}: \quad
\pinv{\upsilon}\finereq\pinv{\xi} &\dspthen  t_m(\pinv{\upsilon}) \leq t_m(\pinv{\xi}),\\
\label{eq:mon.sq}
1 \lesseqgtr q: \quad 
\pinv{\upsilon}\finereq\pinv{\xi} &\dspthen  s_q(\pinv{\upsilon}) \lesseqgtr s_q(\pinv{\xi}),\\
\label{eq:mon.Nq}
\left.\begin{aligned}
q < 0, 1\leq q\\
0 < q\leq 1
\end{aligned}\right\} \quad 
\pinv{\upsilon}\finereq\pinv{\xi} &\dspthen  N_q(\pinv{\upsilon}) \lesseqgtr N_q(\pinv{\xi}),\\
\label{eq:mon.Mq}
1 \leq q: \quad 
\pinv{\upsilon}\finereq\pinv{\xi} &\dspthen  M_q(\pinv{\upsilon}) \leq M_q(\pinv{\xi}),\\
\label{eq:mon.Tq}
q\in\field{R}: \quad 
\pinv{\upsilon}\finereq\pinv{\xi} &\dspthen  T_q(\pinv{\upsilon}) \geq T_q(\pinv{\xi}),\\
\label{eq:mon.Rq}
q\in\field{R},\pm\infty: \quad 
\pinv{\upsilon}\finereq\pinv{\xi} &\dspthen  R_q(\pinv{\upsilon}) \geq R_q(\pinv{\xi}),\\
\label{eq:mon.S}
\pinv{\upsilon}\finereq\pinv{\xi} &\dspthen  S(\pinv{\upsilon}) \geq S(\pinv{\xi}),\\
\label{eq:mon.Pq}
q\in\field{R},\pm\infty: \quad 
\pinv{\upsilon}\finereq\pinv{\xi} &\dspthen  P_q(\pinv{\upsilon}) \geq P_q(\pinv{\xi}),\\
\label{eq:mon.Dimd}
\left.\begin{aligned}
2 \leq b\\
0<b\leq1
\end{aligned}\right\} \quad 
\pinv{\upsilon}\finereq\pinv{\xi} &\dspthen  \Dim_b(\pinv{\upsilon}) \lesseqgtr \Dim_b(\pinv{\xi}),\\
\label{eq:mon.Dimpd}
0<b: \quad
\pinv{\upsilon}\finereq\pinv{\xi} &\dspthen  \Dim'_b(\pinv{\upsilon}) \leq \Dim'_b(\pinv{\xi}),\\
\label{eq:mon.DoFd}
\left.\begin{aligned}
0<b<1\\
2 \leq b
\end{aligned}\right\} \quad 
\pinv{\upsilon}\finereq\pinv{\xi} &\dspthen  \DoF_b(\pinv{\upsilon}) \leq \DoF_b(\pinv{\xi}),\\
\label{eq:mon.DoFpd}
1< b: \quad
\pinv{\upsilon}\finereq\pinv{\xi} &\dspthen  \DoF'_b(\pinv{\upsilon}) \leq \DoF'_b(\pinv{\xi}).
\end{align}
\end{subequations}
Note that for the entanglement dimension~\eqref{eq:Dimd}-\eqref{eq:Dimpd} 
and the entanglement degree of freedom~\eqref{eq:DoFd}-\eqref{eq:DoFpd}
we considered more general parameter ranges $b\in\field{R}$ for the sake of mathematical completeness.
These functions bear physical motivation for $b=d\in\set{2,3,4,\dots}$, being Hilbert space dimensions.

To see~\eqref{eq:mon.wm} for~\eqref{eq:extr.wm},
note that $w_m$ is the sum of the largest $m$ parts,
and $w_m(\pinv{\upsilon}) \leq w_m(\pinv{\xi})$ for all $m$ is just the condition~\eqref{eq:poIpdom.p} for dominance,
which has already been shown to follow from the refinement~\eqref{eq:poIs}.

To see~\eqref{eq:mon.tm} for~\eqref{eq:extr.tm},
let us have $\pinv{\xi}=\mset{x_1,x_2,\dots,x_{\abs{\pinv{\xi}}}}$
and $\pinv{\upsilon}=\mset{y_1,y_2,\dots,y_{\abs{\pinv{\upsilon}}}}$
with increasingly ordered indexing, $x_i\leq x_j$ and $y_i\leq y_j$ for $i<j$.
Then $t_m$ is the sum of the $m$ smallest parts,
which for $\pinv{\upsilon}\finereq\pinv{\xi}$ is
\begin{equation*}
t_m(\pinv{\xi})
= \sum_{i=1}^m x_i
\equalsref{eq:poIp.x}
\sum_{i=1}^m \sum_{j=1}^{\abs{\pinv{\upsilon}}} a_{ij} y_j
\geq
\sum_{j=1}^m y_j
= t_m(\pinv{\upsilon}),
\end{equation*}
where the inequality follows from that on the left-hand side the number of summands is at least $m$,
while on the right-hand side there is the sum of the $m$ smallest parts.
(Here $x_i$ and $y_j$ are set to zero if $i>\abs{\pinv{\xi}}$ and $j>\abs{\pinv{\upsilon}}$, respectively.)
Note that, contrary to the proof of~\eqref{eq:mon.wm},
here we could not use the majorization,
since zeroes are not allowed in this formalism.

To see~\eqref{eq:mon.sq} for~\eqref{eq:sq},
\begin{equation*}
\begin{split}
&s_q(\pinv{\xi}) 
\equalsref{eq:sq}
  \sum_{i=1}^{\abs{\pinv{\xi}}} x_i^q 
 \equalsref{eq:poIp.x}
  \sum_{i=1}^{\abs{\pinv{\xi}}} \Bigl( \sum_{j=1}^{\abs{\pinv{\upsilon}}} a_{ij} y_j \Bigr)^q \\
&\quad\equals
  \sum_{i=1}^{\abs{\pinv{\xi}}} \Bigl( \sum_{\substack{j=1,\\a_{ij}\neq0}}^{\abs{\pinv{\upsilon}}}y_j \Bigr)^q
 \gtreqlesssref{eq:sumq} 
  \sum_{i=1}^{\abs{\pinv{\xi}}} \sum_{\substack{j=1,\\a_{ij}\neq0}}^{\abs{\pinv{\upsilon}}} y_j^q \\
&\quad\equals
  \sum_{i=1}^{\abs{\pinv{\xi}}} \sum_{j=1}^{\abs{\pinv{\upsilon}}} a_{ij} y_j^q
 \equals
  \sum_{j=1}^{\abs{\pinv{\upsilon}}} \Bigl( \sum_{i=1}^{\abs{\pinv{\xi}}} a_{ij} \Bigr) y_j^q \\
&\quad\equals
  \sum_{j=1}^{\abs{\pinv{\upsilon}}} y_j^q
 \equalsref{eq:sq}
  s_q(\pinv{\upsilon}),
\end{split}
\end{equation*}
where at the unlabeled equalities we used the properties of the extremal stochastic matrix,
arising from~\eqref{eq:poIp.x} by the assumption.

To see~\eqref{eq:mon.Nq} for~\eqref{eq:Nq}, note that 
the~\eqref{eq:mon.sq} monotonicity of $s_q$ is flipped 
by the monotonicity of $u\mapsto u^{1/q}$ if $q<0$, see~\eqref{eq:xqmonx}.

To see~\eqref{eq:mon.Mq} for~\eqref{eq:Mq}, note that
$\abs{\pinv{\xi}}=s_0(\pinv{\xi})$ is decreasing~\eqref{eq:mon.sq},
so $1/s_0$ is increasing,
$s_q$ is increasing for $1\leq q$~\eqref{eq:mon.sq},
so $s_q/s_0$ is increasing,
so $M_q$ is increasing for $1\leq q$~\eqref{eq:Mq}
by the monotonicity of $u\mapsto u^{1/q}$, see~\eqref{eq:xqmonx}.
We note that 
counterexamples show that $M_q$ is not monotone for $q<1$. 

To see~\eqref{eq:mon.Rq} for~\eqref{eq:Rq}, note that
$u\mapsto\ln(u)$ is increasing monotone,
so $\pinv{\xi}\mapsto\ln(s_q(\pinv{\xi}))$ is increasing/decreasing for $1 \lesseqgtr q$ by~\eqref{eq:mon.sq};
on the other hand, 
$0\gtrless\frac{1}{1-q}$ for $1 \lessgtr q$,
so the multiplication with that flips the monotonicity for $1<q$,
leading to the claim for $q\neq1$,
which also holds for $q=1$ by the continuity of $q\mapsto R_q$,~\eqref{eq:qlim.R1}.

We have~\eqref{eq:mon.Tq} for~\eqref{eq:Tq}
by similar reasoning as for~\eqref{eq:mon.Rq}.

We have~\eqref{eq:mon.S} for~\eqref{eq:S},
since this is just the $q=1$ case 
of~\eqref{eq:mon.Tq} by~\eqref{eq:qlim.T1} or 
of~\eqref{eq:mon.Rq} by~\eqref{eq:qlim.R1}.

To see~\eqref{eq:mon.Pq} for~\eqref{eq:Pq}, note that
the~\eqref{eq:mon.Rq} monotonicity of $R_q$ is not changed
by the monotonicity of $u\mapsto\ee^u$.

To see~\eqref{eq:mon.Dimd} for~\eqref{eq:Dimd},
\begin{equation*}
\begin{split}
&\Dim_b(\pinv{\xi}) 
\equalsref{eq:Dimd}
  \sum_{i=1}^{\abs{\pinv{\xi}}} b^{x_i}
 \equalsref{eq:poIp.x}
  \sum_{i=1}^{\abs{\pinv{\xi}}} b^{\sum_{j=1}^{\abs{\pinv{\upsilon}}} a_{ij} y_j}\\
&\quad\gtreqlesssref{eq:ssadd}
  \sum_{i=1}^{\abs{\pinv{\xi}}}\sum_{j=1}^{\abs{\pinv{\upsilon}}} a_{ij} b^{y_j}
 \equals
  \sum_{j=1}^{\abs{\pinv{\upsilon}}} \Bigl( \sum_{i=1}^{\abs{\pinv{\xi}}} a_{ij} \Bigr)b^{y_j}\\
&\quad\equals
  \sum_{j=1}^{\abs{\pinv{\upsilon}}} b^{y_j}
\equalsref{eq:Dimd}
  \Dim_b(\pinv{\upsilon}),
\end{split}
\end{equation*}
where at the unlabeled equalities we used the properties of the extremal stochastic matrix,
arising from~\eqref{eq:poIp.x} by the assumption.
The inequality is from the super/subadditivity of the exponential function for this case
\begin{equation}
\label{eq:ssadd}
b^{\sum_{j=1}^{\abs{\pinv{\upsilon}}} a_{ij} y_j}
\gtreqlesss
\sum_{j=1}^{\abs{\pinv{\upsilon}}} a_{ij} b^{y_j},
\end{equation}
which holds if and only if
$b^{z_1+z_2} \gtreqlesss b^{z_1} + b^{z_2}$
for any $1\leq z_1,z_2\in\field{N}$,
which is equivalent to
$1 \gtreqlesss 1/b^{z_1}+1/b^{z_2}$.
If $2\leq b$, we have $1\geq 1/b^{z_1}+1/b^{z_2}$,
since it holds for $b=2$, $z_1=z_2=1$, and the right-hand side
is decreasing with $z_1,z_2$, see~\eqref{eq:xqmonq},
and decreasing with $b$, see~\eqref{eq:xqmonx};
if $0<b\leq1$, we have $1 \leq 1/b^{z_1}+1/b^{z_2}$,
since it holds for $b=1$, $z_1=z_2=1$, and the right-hand side
is increasing with $z_1,z_2$, see~\eqref{eq:xqmonq},
and increasing with the decreasing of $b$, see~\eqref{eq:xqmonx};
for the intermediate case $1<b<2$, both inequalities can be violated for suitable values of $z_1,z_2$.

To see~\eqref{eq:mon.Dimpd} for~\eqref{eq:Dimpd},
\begin{align*}
&\Dim'_b(\pinv{\xi}) 
\equalsref{eq:Dimpd}
  \sum_{i=1}^{\abs{\pinv{\xi}}} (b^{x_i}-1) +1 \\
&\quad\equalsref{eq:poIp.x}
  \sum_{i=1}^{\abs{\pinv{\xi}}} (b^{\sum_{j=1}^{\abs{\pinv{\upsilon}}} a_{ij} y_j}-1)+1\\
&\quad\gequalsref{eq:ssaddp}
  \sum_{i=1}^{\abs{\pinv{\xi}}}\sum_{j=1}^{\abs{\pinv{\upsilon}}} a_{ij} (b^{y_j}-1)+1\\
&\quad\equals
  \sum_{j=1}^{\abs{\pinv{\upsilon}}} \Bigl( \sum_{i=1}^{\abs{\pinv{\xi}}} a_{ij} \Bigr)(b^{y_j}-1)+1\\
&\quad\equals
  \sum_{j=1}^{\abs{\pinv{\upsilon}}} (b^{y_j}-1)+1
\equalsref{eq:Dimpd}
  \Dim'_b(\pinv{\upsilon}),
\end{align*}
where at the unlabeled equalities we used the properties of the extremal stochastic matrix,
arising from~\eqref{eq:poIp.x} by the assumption.
The inequality is from the superadditivity of the modified exponential function
\begin{equation}
\label{eq:ssaddp}
b^{\sum_{j=1}^{\abs{\pinv{\upsilon}}} a_{ij} y_j}-1
\geq
\sum_{j=1}^{\abs{\pinv{\upsilon}}} a_{ij} (b^{y_j}-1),
\end{equation}
which holds if and only if
$(b^{z_1+z_2}-1) \geq (b^{z_1}-1) + (b^{z_2}-1)$
for any $1\leq z_1,z_2\in\field{N}$,
which is equivalent to
$0\leq(b^{z_1}-1)(b^{z_2}-1)$,
which holds in both the $0<b\leq1$ and $1\leq b$ cases.
(Note that the monotonicity for $2\leq b$ would simply follow
from that of $\Dim_b$ and $h$ by definition~\eqref{eq:Dimpd},
however, the direct proof here works for the larger range $0<b$.)

To see~\eqref{eq:mon.DoFd} for~\eqref{eq:DoFd},
note that $u\mapsto\log_b(u)$ is increasing/decreasing monotone for $1\lessgtr b$,
which flips the monotonicity of $\Dim_b$ for which we have~\eqref{eq:mon.Dimd}.

To see~\eqref{eq:mon.DoFpd} for~\eqref{eq:DoFpd},
note that $u\mapsto\log_b(u)$ is increasing for $1< b$,
and for $\Dim'_b$ we have~\eqref{eq:mon.Dimpd}.
(Note that the parameter range is constrained not to substitute negative number into the logarithm.
Since $\Dim'_b$ is increasing monotone~\eqref{eq:mon.Dimpd},
we need that $0\leq\Dim'_b(\bot)= n(b-1)+1$.
This holds for $b\geq1$,
but it is violated for $0<b<1$ for $n\geq 1/(1-b)$.)

\subsection{Dominance-monotonicity of generator functions}
\label{app:1param.dmon}

Here we show the monotonicity~\eqref{eq:dommon} of the generator functions, studied in Section~\ref{sec:1param},
with respect to the dominance order.
\begin{subequations}
\label{eq:dmon}
\begin{align}
\label{eq:dmon.wm}
m\in\field{N}: \quad
\pinv{\upsilon}\majordeq\pinv{\xi} &\dspthen  w_m(\pinv{\upsilon}) \leq w_m(\pinv{\xi}),\\
\label{eq:dmon.h}
\pinv{\upsilon}\majordeq\pinv{\xi} &\dspthen  h(\pinv{\upsilon}) \geq h(\pinv{\xi}),\\
\label{eq:dmon.r}
\pinv{\upsilon}\majordeq\pinv{\xi} &\dspthen  r(\pinv{\upsilon}) \leq r(\pinv{\xi}),\\
\label{eq:dmon.sq}
\left.\begin{aligned}
1\leq q \\
0\leq q \leq 1
\end{aligned}\right\} \quad 
\pinv{\upsilon}\majordeq\pinv{\xi} &\dspthen  s_q(\pinv{\upsilon}) \lesseqgtr s_q(\pinv{\xi}),\\
\label{eq:dmon.Nq}
\left.\begin{aligned}
1\leq q\\
0 < q\leq 1
\end{aligned}\right\} \quad 
\pinv{\upsilon}\majordeq\pinv{\xi} &\dspthen  N_q(\pinv{\upsilon}) \lesseqgtr N_q(\pinv{\xi}),\\
\label{eq:dmon.Mq}
1 \leq q: \quad 
\pinv{\upsilon}\majordeq\pinv{\xi} &\dspthen  M_q(\pinv{\upsilon}) \leq M_q(\pinv{\xi}),\\
\label{eq:dmon.Tq}
0\leq q: \quad 
\pinv{\upsilon}\majordeq\pinv{\xi} &\dspthen  T_q(\pinv{\upsilon}) \geq T_q(\pinv{\xi}),\\
\label{eq:dmon.Rq}
0\leq q, q=\infty: \quad 
\pinv{\upsilon}\majordeq\pinv{\xi} &\dspthen  R_q(\pinv{\upsilon}) \geq R_q(\pinv{\xi}),\\
\label{eq:dmon.S}
\pinv{\upsilon}\majordeq\pinv{\xi} &\dspthen  S(\pinv{\upsilon}) \geq S(\pinv{\xi}),\\
\label{eq:dmon.Pq}
0\leq q, q=\infty: \quad 
\pinv{\upsilon}\majordeq\pinv{\xi} &\dspthen  P_q(\pinv{\upsilon}) \geq P_q(\pinv{\xi}),\\
\label{eq:dmon.Dimd}
2 \leq b:  \quad 
\pinv{\upsilon}\majordeq\pinv{\xi} &\dspthen  \Dim_b(\pinv{\upsilon}) \leq \Dim_b(\pinv{\xi}),\\
\label{eq:dmon.Dimpd}
0 < b: \quad
\pinv{\upsilon}\majordeq\pinv{\xi} &\dspthen  \Dim'_b(\pinv{\upsilon}) \leq \Dim'_b(\pinv{\xi}),\\
\label{eq:dmon.DoFd}
2 \leq b:  \quad
\pinv{\upsilon}\majordeq\pinv{\xi} &\dspthen  \DoF_b(\pinv{\upsilon}) \leq \DoF_b(\pinv{\xi}),\\
\label{eq:dmon.DoFpd}
1 < b: \quad
\pinv{\upsilon}\majordeq\pinv{\xi} &\dspthen  \DoF'_b(\pinv{\upsilon}) \leq \DoF'_b(\pinv{\xi}).
\end{align}
\end{subequations}
Note that the dominance-monotonicity
is enough to be proven for arrows~\eqref{eq:poIpdom.c},
however, sometimes it is easier to use more general steps, 
for which the right-hand side of~\eqref{eq:poIpdom.simpler} holds,
since these include the arrows. 
Note that throughout this section
the integer partitions $\pinv{\upsilon}=\set{y_1,y_2,\dots,y_{\abs{\pinv{\upsilon}}}}$ and $\pinv{\xi}=\set{x_1,x_2,\dots,x_{\abs{\pinv{\xi}}}}$
are indexed in weakly decreasing order ($x_i\geq x_j$ and $y_i\geq y_j$, $i<j$).

To see~\eqref{eq:dmon.wm} for~\eqref{eq:extr.wm},
note that $w_m$ is the sum of the largest $m$ parts,
and $w_m(\pinv{\upsilon}) \leq w_m(\pinv{\xi})$ for all $m$
is just the condition~\eqref{eq:poIpdom.p} for the dominance order.

To see~\eqref{eq:dmon.h} for~\eqref{eq:hwr.h},
we need to check the monotonicity of $s_q$
with respect to the steps~\eqref{eq:poIpdom.simpler} of the dominance order.
For the case $2\leq y_j$, we have $\abs{\pinv{\upsilon}}=\abs{\pinv{\xi}}$,
and for the case $y_j=1$, we have $\abs{\pinv{\upsilon}}=\abs{\pinv{\xi}}+1$. 

We have~\eqref{eq:dmon.r} for~\eqref{eq:hwr.r},
since rank is the difference of the width and the height,
being dominance-increasing~\eqref{eq:dmon.wm} and dominance-decreasing~\eqref{eq:dmon.h}, respectively.

To see~\eqref{eq:dmon.sq} for~\eqref{eq:sq},
we need to check the monotonicity of $s_q$
with respect to the steps~\eqref{eq:poIpdom.simpler} of the dominance order.
Writing $y:=y_j$ and $\delta:=y_i-y_j\in\set{0,1,2,\dots}$, this is 
\begin{subequations}
\label{eq:dmon.sq.arrow}
\begin{align}
\label{eq:dmon.sq.arrow.2}
(y+\delta)^q + y^q &\lesseqgtr (y+\delta+1)^q + (y-1)^q,\quad\text{for $2\leq y$},\\
\label{eq:dmon.sq.arrow.1}
(1+\delta)^q + 1^q &\lesseqgtr (2+\delta)^q,
\end{align}
\end{subequations}
where we had to treat the $y=1$ case separately,
since such a step results in $x_j=y_j-1=y-1=0$ row~\eqref{eq:poIpdom.simpler},
for which we have different form of the generator function.
\textit{First}, consider~\eqref{eq:dmon.sq.arrow.2},
which is equivalent to $y^q-(y-1)^q\lesseqgtr (y+(\delta+1))^q-(y+(\delta+1)-1)^q=(y+\Delta)^q-(y+\Delta-1)^q$,
which follows from the monotonicity of $\tau_q(y):=y^q-(y-1)^q$ in $y$.
This is by $0\lesseqgtr(\partial\tau_q/\partial y)(y)=q(y^{q-1}-(y-1)^{q-1})$,
which is nonnegative for $q\leq 0$ and $1\leq q$,
and nonpositive for $0\leq q\leq 1$ by~\eqref{eq:xqmonx}.
\textit{Second}, note that we do not need to check~\eqref{eq:dmon.sq.arrow.1} directly,
since the $y\equiv y_j=1$ case means a step 
where the only one box of the $j$-th row of $\pinv{\upsilon}$ is moved to the end of the $i$-th~\eqref{eq:poIpdom.simpler},
so  $x_j=y_j-1=0$
and $x_i=y_i+1=y_i+y_j$,
that is, $\pinv{\xi}$ is obtained by partial summation of $\pinv{\upsilon}$,
which is an arrow of the refinement~\eqref{eq:poIp.c}, $\pinv{\upsilon}\finerc\pinv{\xi}$,
for which we have the monotonicity~\eqref{eq:mon.sq}.
\textit{Finally}, the two cases agree for $0\leq q$, where
$s_q$ is increasing/decreasing with respect to dominance for $1\leq q$ or $0\leq q\leq 1$, respectively,
and we do not have dominance-monotonicity for $q<0$.

To see~\eqref{eq:dmon.Nq} for~\eqref{eq:Nq}, note that
the~\eqref{eq:dmon.sq} dominance-monotonicity of $s_q$ is not changed
by the monotonicity of $u\mapsto u^{1/q}$ if $0<q$, see~\eqref{eq:xqmonx}.

To see~\eqref{eq:dmon.Mq} for~\eqref{eq:Mq}, note that
$\abs{\pinv{\xi}}=s_0(\pinv{\xi})$ is dominance-decreasing~\eqref{eq:dmon.sq},
so $1/s_0$ is dominance-increasing,
$s_q$ is dominance-increasing for $1\leq q$~\eqref{eq:dmon.sq},
so $s_q/s_0$ is dominance-increasing,
so $M_q$ is dominance-increasing for $1\leq q$~\eqref{eq:Mq}
by the monotonicity of $u\mapsto u^{1/q}$, see~\eqref{eq:xqmonx}.
We have already seen that $M_q$ is not even a generator function for $q<1$,~\eqref{eq:mon.Mq},
so cannot be dominance-monotone~\eqref{eq:poIs} either.

To see~\eqref{eq:dmon.Rq} for~\eqref{eq:Rq}, note that
$u\mapsto\ln(u)$ is increasing monotone,
so in the range $0\leq q$,  $\pinv{\xi}\mapsto\ln(s_q(\pinv{\xi}))$ is dominance-increasing/decreasing for $1 \lesseqgtr q$ by~\eqref{eq:dmon.sq};
on the other hand,
$0\gtrless\frac{1}{1-q}$ for $1 \lessgtr q$,
so the multiplication with that flips the dominance-monotonicity for $1<q$,
leading to the claim for $q\neq1$,
which also holds for $q=1$ by the continuity of $q\mapsto R_q$,~\eqref{eq:qlim.R1}.

We have~\eqref{eq:dmon.Tq} for~\eqref{eq:Tq}
by similar reasoning as for~\eqref{eq:dmon.Rq}.

We have~\eqref{eq:dmon.S} for~\eqref{eq:S},
since this is just the $q=1$ case
of~\eqref{eq:dmon.Tq} by~\eqref{eq:qlim.T1} or
of~\eqref{eq:dmon.Rq} by~\eqref{eq:qlim.R1}.

To see~\eqref{eq:dmon.Pq} for~\eqref{eq:Pq}, note that
the~\eqref{eq:dmon.Rq} dominance-monotonicity of $R_q$ is not changed
by the monotonicity of $u\mapsto\ee^u$.

To see~\eqref{eq:dmon.Dimd} for~\eqref{eq:Dimd},
we need to check the monotonicity of $\Dim_b$
with respect to the steps~\eqref{eq:poIpdom.simpler} of the dominance order.
Writing $y:=y_j$ and $\delta:=y_i-y_j\in\set{0,1,2,\dots}$, this is 
\begin{subequations}
\label{eq:dmon.Dimd.arrow}
\begin{align}
\label{eq:dmon.Dimd.arrow.2}
b^{y+\delta} + b^y &\lesseqgtr b^{y+\delta+1} + b^{y-1},\quad\text{for $2\leq y$},\\
\label{eq:dmon.Dimd.arrow.1}
b^{1+\delta} + b &\lesseqgtr b^{2+\delta},
\end{align}
\end{subequations}
where we had to treat the $y=1$ case separately,
since such a step results in $x_j=y_j-1=y-1=0$ row~\eqref{eq:poIpdom.simpler},
for which we have different form of the generator function. 
\textit{First,} consider~\eqref{eq:dmon.Dimd.arrow.2},
which is equivalent to $0\lesseqgtr (b-1)(b^{y+\delta}-b^{y-1}) = b^{y-1}(b-1)(b^{\delta+1}-1)$,
which is nonnegative for $0<b$, since the two parentheses are positive or negative simultaneously.
\textit{Second,} note that we do not need to check~\eqref{eq:dmon.Dimd.arrow.1} directly,
since the $y\equiv y_j=1$ case means a step
where the only one box of the $j$-th row of $\pinv{\upsilon}$ is moved to the end of the $i$-th~\eqref{eq:poIpdom.simpler},
so  $x_j=y_j-1=0$
and $x_i=y_i+1=y_i+y_j$,
that is, $\pinv{\xi}$ is obtained by partial summation of $\pinv{\upsilon}$,
which is an arrow of the refinement~\eqref{eq:poIp.c}, $\pinv{\upsilon}\finerc\pinv{\xi}$,
for which we have the monotonicity~\eqref{eq:mon.Dimd}.
\textit{Finally}, the two cases agree for  $2\leq b$, where
$\Dim_q$ is increasing with respect to dominance,
and we do not have dominance-monotonicity for $0<b<2$.

To see~\eqref{eq:dmon.Dimpd} for~\eqref{eq:Dimpd},
we need to check the monotonicity of $\Dim'_b$
with respect to the steps~\eqref{eq:poIpdom.simpler} of the dominance order.
We have to treat the $y_j=1$ case separately,
since such a step results in $x_j=y_j-1=0$ row~\eqref{eq:poIpdom.simpler},
for which we have different form of the generator function.
\textit{First,} for the case $2\leq y_j$, we have $\abs{\pinv{\upsilon}}=\abs{\pinv{\xi}}$,
so, by~\eqref{eq:Dimpd}, we get the same equation~\eqref{eq:dmon.Dimd.arrow.2} as for $\Dim_b$,
and have the upper relation sign for $0<b$.
\textit{Second,} for the case $y\equiv y_j=1$,
we again have that the step~\eqref{eq:poIpdom.simpler} of the dominance order 
is an arrow of the refinement~\eqref{eq:poIp.c}, $\pinv{\upsilon}\finerc\pinv{\xi}$,
for which we have the monotonicity~\eqref{eq:mon.Dimpd}.
\textit{Finally}, the two cases agree for $0<b$, where
$\Dim'_q$ is increasing with respect to dominance.

To see~\eqref{eq:dmon.DoFd} for~\eqref{eq:DoFd}, note that 
the~\eqref{eq:dmon.Dimd} dominance-monotonicity of $\Dim_b$ is not changed
by the monotonicity of $u\mapsto\log_b(u)$ if $2\leq b$.

To see~\eqref{eq:dmon.DoFpd} for~\eqref{eq:DoFpd},
note that $u\mapsto\log_b(u)$ is increasing for $1< b$,
and for $\Dim'_b$ we have~\eqref{eq:dmon.Dimpd}.
(Note that the parameter range is constrained not to substitute negative number into the logarithm,
see at the proof of~\eqref{eq:mon.DoFpd}.)

\section{Bounds on the variance}
\label{app:Metro}

In this section we write out some more or less well-known calculations
related to the variance in multipartite systems
for the convenience of the reader.

\subsection{Upper bound for the variance}
\label{app:MetroVar.Abound}

For any finite dimensional Hilbert space $\mathcal{H}$,
for the self-adjoint operator $A\in\Lin(\mathcal{H})$ given by its spectral decomposition as
$A=\sum_{i=1}^{\dim(\mathcal{H})} a_i\proj{\alpha_i}$,
with $a_\text{min}:=\min_i a_i$ and $a_\text{max}:=\max_i a_i$,
we have
\begin{equation}
\label{eq:varAbound}
\Var(\rho,A) \leq \frac14(a_\text{max}-a_\text{min})^2.
\end{equation}

Indeed, consider the operators
$(a_\text{max}\Id-A)$ and $(A-a_\text{min}\Id)$,
which are positive semidefinite and commuting,
so $(a_\text{max}\Id-A)(A-a_\text{min}\Id)$ is also positive semidefinite,
so 
\begin{equation*}
\begin{split}
0&\leq\Tr\bigl(\rho(a_\text{max}\Id-A)(A-a_\text{min}\Id)\bigr)\\
&= -\Tr(\rho A^2) + (a_\text{max}+a_\text{min})\Tr(\rho A) - a_\text{max}a_\text{min}.
\end{split}
\end{equation*}
Then we have
\begin{equation*}
\begin{split}
&\Var(\rho,A)   
= \Tr(\rho A^2)-\Tr(\rho A)^2\\ 
&\quad\leq (a_\text{max}+a_\text{min})\Tr(\rho A) - a_\text{max}a_\text{min} - \Tr(\rho A)^2 \\
&\quad= \bigl(a_\text{max}-\Tr(\rho A)\bigr)\bigl(\Tr(\rho A)-a_\text{min}\bigr),
\end{split}
\end{equation*}
(this inequality between the first and last term is called Bhatia–Davis inequality~\cite{Bhatia-2000})
then the inequality between the arithmetic and geometric mean
($q=0$ and $q'=1$ in~\eqref{eq:qmon.Mq})
leads to~\eqref{eq:varAbound}.
 
The bound~\eqref{eq:varAbound} can be attained by the pure state $\rho=\proj{\psi}$
with $\cket{\psi}=\frac1{\sqrt{2}}\bigl(\cket{\alpha_\text{max}}+\ee^{i\vartheta}\cket{\alpha_\text{min}}\bigr)$ for any phase $\vartheta$,
or by the mixed state $\rho=\frac12\bigl(\proj{\alpha_\text{max}}+\proj{\alpha_\text{min}}\bigr)$,
or by convex combinations of these,
leading to $\frac12\bigl( \proj{\alpha_\text{max}} + \proj{\alpha_\text{min}}
 + c      \cket{\alpha_\text{max}}\bra{\alpha_\text{min}}
 + \bar{c}\cket{\alpha_\text{min}}\bra{\alpha_\text{max}}\bigr)$ for any $c\in\field{C}$, $\abs{c}\leq1$.

The classical version of the bound~\eqref{eq:varAbound}
is a standard result in classical probability theory, called Popoviciu's inequality~\cite{Popoviciu-1935}.
Here we have seen the widely used straightforward generalization of that
for the quantum case~\cite{Bhatia-2000,Giovannetti-2006}.

\subsection{Upper bound for the variance of collective operators}
\label{app:MetroVar.JzXbound}

Let us have the collective operator
$A_X=\sum_{l\in X} A_l\otimes \Id_{\cmpl{\set{l}}}$
of subsystem $X$,
where
$A_l=\sum_{i=1}^{\dim(\mathcal{H}_l)} a_i\proj{\alpha_{l,i}}$
have the same spectrum
for all elementary subsystems $l\in X$.
If $a_\text{min}:=\min_i\set{a_i}$ and $a_\text{max}:=\max_i\set{a_i}$,
then the minimal and maximal eigenvalues of $A_X$ are
$a_{X,\text{min}}=\abs{X}a_\text{min}$ and
$a_{X,\text{max}}=\abs{X}a_\text{max}$,
and the bound~\eqref{eq:varAbound} takes the form
\begin{subequations}
\begin{equation}
\label{eq:varAXbound}
\Var(\rho_X,A_X) \leq \frac{\abs{X}^2}{4}(a_\text{max}-a_\text{min})^2.
\end{equation}
This bound can be attained by the pure state $\rho_X=\proj{\psi_X}$
with $\cket{\psi_X}=\frac1{\sqrt{2}}
\bigl(\bigotimes_{l\in X}\cket{\alpha_{l,\text{max}}}+\ee^{i\vartheta}\bigotimes_{l\in X}\cket{\alpha_{l,\text{min}}}\bigr)$,
or by the mixed state $\rho_X=\frac12\bigl(\bigotimes_{l\in X}\proj{\alpha_{l,\text{max}}}+\bigotimes_{l\in X}\proj{\alpha_{l,\text{min}}}\bigr)$,
or by convex combinations of these,
leading to $\frac12\bigl( \bigotimes_{l\in X}\proj{\alpha_{l,\text{max}}} + \bigotimes_{l\in X}\proj{\alpha_{l,\text{min}}}
 + c      \bigotimes_{l\in X}\cket{\alpha_{l,\text{max}}}\bra{\alpha_{l,\text{min}}}
 + \bar{c}\bigotimes_{l\in X}\cket{\alpha_{l,\text{min}}}\bra{\alpha_{l,\text{max}}}\bigr)$ for any $c\in\field{C}$, $\abs{c}\leq1$.

In particular,
if the elementary subsystems are of spin $s\in\frac12\field{N}_0$,
the collective spin operator $A_X:=J_X$ arises
with $A_l:= J_l = v^\text{x}J^\text{x}_l + v^\text{y}J^\text{y}_l + v^\text{z}J^\text{z}_l$
for the direction $v=(v^\text{x},v^\text{y},v^\text{z})\in\field{R}^3$, $\norm{v}=1$.
For this we have $a_\text{min}=-s$ and $a_\text{max}=s$,
and the bound~\eqref{eq:varAXbound} takes the form
\begin{equation}
\label{eq:varJXbound}
\Var(\rho_X,J_X) \leq \abs{X}^2 s^2,
\end{equation}
which is 
\begin{equation}
\label{eq:varJXboundqubit}
\Var(\rho_X,J_X) \leq \abs{X}^2 /4
\end{equation}
\end{subequations}
for $s=1/2$ (qubits).

\subsection{Variance of local operators in uncorrelated states}
\label{app:MetroVar.sum}

For $\xi$-uncorrelated state $\rho=\bigotimes_{X\in\xi}\rho_X$
and $\xi$-local operator $A=\sum_{X\in\xi} A_X\otimes \Id_{\cmpl{X}}$
(here $\cmpl{X} = \set{1,2,\dots,n}\setminus X$),
we have
\begin{equation}
\label{eq:sumvar}
\Var(\rho,A)=\sum_{X\in\xi}\Var(\rho_X,A_X).
\end{equation}
(This, together with the concavity of the variance
has also been used for entanglement detection~\cite{Hofmann-2003,Guhne-2004,Toth-2004}.)

Indeed,
\begin{equation*}
\begin{split}
&\Var(\rho,A)
= \Tr(\rho A^2) - \Tr(\rho A)^2\\
&\quad=
  \Tr\Bigl(\rho \Bigl(\sum_{X\in\xi} A_X\otimes \Id_{\cmpl{X}}\Bigr)^2\Bigr)
- \Tr\Bigl(\rho \sum_{X\in\xi} A_X\otimes \Id_{\cmpl{X}}\Bigr)^2\\
&\quad=
  \Tr\Bigl(\rho \sum_{\substack{X,X'\in\xi\\X\neq X'}} A_X\otimes A_{X'}\otimes \Id_{\cmpl{X\cup X'}} \Bigr)\\
&\quad + \Tr\Bigl(\rho\sum_{X\in\xi} A_X^2\otimes \Id_{\cmpl{X}} \Bigr)
  - \Tr\Bigl(\rho\sum_{X\in\xi} A_X\otimes \Id_{\cmpl{X}}\Bigr)^2\\
&\quad=
  \sum_{\substack{X,X'\in\xi\\X\neq X'}} \Tr\bigl((\rho_X\otimes\rho_{X'})(A_X\otimes A_{X'})\bigr) \\
&\quad   + \sum_{X\in\xi}\Tr(\rho_X A_X^2)
  - \Bigl( \sum_{X\in\xi} \Tr(\rho_X A_X)\Bigr)^2\\
&\quad=
  \sum_{\substack{X,X'\in\xi\\X\neq X'}} \Tr(\rho_X A_X)\Tr(\rho_{X'}A_{X'}) 
 + \sum_{X\in\xi}\Tr(\rho_X A_X^2)\\
&\quad  - \sum_{X\in\xi}\Tr(\rho_X A_X)^2 
   - \sum_{\substack{X,X'\in\xi\\X\neq X'}} \Tr(\rho_X A_X)\Tr(\rho_{X'}A_{X'})\\
&\quad=
  \sum_{X\in\xi} \bigl(\Tr(\rho_X A_X^2) - \Tr(\rho_X A_X)^2\bigr)\\
&\quad=\sum_{X\in\xi}\Var(\rho_X,A_X),
\end{split}
\end{equation*}
where we make use of that the state is $\xi$-uncorrelated at the \textit{fourth equality}.

\section{Metrological usefulness}
\label{app:MetroUseful}
In this section we
consider extremal cases of metrological usefulness,
that is, the strictness of the monotonicity~\eqref{eq:monB.str} of the bounds $b_f(k)$.

\subsection{Metrological usefulness of some one-parameter properties}
\label{app:MetroUseful.hwr}

Here we show that the $b_f(k)$
bounds are \emph{strictly} monotone~\eqref{eq:monB.str} for all $k\in f(\pinv{P}_\text{I})$
in the case of producibility, partitionability and stretchability.

For producibility, given by the $w$ width generator function~\eqref{eq:hwr.w},
the bound $b_\text{prod}(k)$ is given in~\eqref{eq:FisherBoundprod.k},
as $b_\text{prod}(k)=\lfloor\frac{n}{k}\rfloor k^2 + \bigl(n-\lfloor\frac{n}{k}\rfloor k\bigr)^2$.
Extending this from $[1,n]_\field{N}=w(\pinv{P}_\text{I})$
to the reals $[1,n]_\field{R}$,
we claim that this function is continuous, and its derivative is strictly positive, although not continuous.
To see these, let us consider the intervals $(\frac{n}{m+1},\frac{n}{m}]$ for $m\in\field{N}$,
since
if $\frac{n}{m+1}<k\leq\frac{n}{m}$
then $m+1>\frac{n}{k}\geq m$, so
$\lfloor\frac{n}{k}\rfloor=m$.
\textit{First}, the continuity is obvious inside the intervals, it has to be checked at the boundaries of each two neighbouring intervals.
For $\frac{n}{m+1}<k\leq\frac{n}{m}$ we have $\lfloor\frac{n}{k}\rfloor=m$,
and $b_\text{prod}(k)=mk^2+(n-mk)^2$;
for $\frac{n}{m}<k\leq\frac{n}{m-1}$ we have $\lfloor\frac{n}{k}\rfloor=m-1$
and $b_\text{prod}(k)=(m-1)k^2+(n-(m-1)k)^2$;
these agree for the value $k=\frac{n}{m}$. 
(The continuity can be seen intuitively as well,
imagining the sums of squares of the rows of 
the `Young diagrams' of not rows of discrete boxes,
but of rows of continuously variable stripes.)
\textit{Second}, the monotonicity has to be checked for every interval.
If $k\in(\frac{n}{m+1},\frac{n}{m}]$ then $\lfloor\frac{n}{k}\rfloor=m$,
and $\frac{\partial b_\text{prod}(k)}{\partial k} = \frac{\partial}{\partial k} (mk^2 + (n-mk)^2)=2m(mk+k-n)>0$
which is $k>\frac{n}{m+1}$, which holds in the given interval.

For partitionability, given by the $h$ height generator function~\eqref{eq:hwr.h},
the bound $b_\text{part}(k)$ is given in~\eqref{eq:FisherBoundpart.k},
as $b_\text{part}(k)=k^2-(2n+1)k+n(n+2)$.
Extending this from $[1,n]_\field{N}=h(\pinv{P}_\text{I})$
to the reals $[1,n]_\field{R}$, and differentiating by $k$ we have strictly negative derivative for all $k<n+1/2$.

For stretchability, given by the $r$ rank generator function~\eqref{eq:hwr.r},
the bound $b_\text{str}(k)$ is given in~\eqref{eq:FisherBoundstr.k}.
By changing to the new variable $l:=n+k\in [1,2n-1]_\field{N}\setminus\set{2,2n-2}$
the bound~\eqref{eq:FisherBoundstr.k} takes the form
\begin{equation}
\begin{cases}
n+24    &\text{if $l=10$ ($n\geq8$)},\\
n+60    &\text{if $l=16$ ($n\geq12$)},\\
(\frac{l}{2})^2 -\frac{l}{2} +n+2   &\text{if $l\in2\field{N}$},\\
(\frac{l+2}{2})^2 -\frac{l+1}{2}+n &\text{if $l\in2\field{N}+1$},
\end{cases}
\end{equation}
the strict monotonicity of which is straightforward to check.

\subsection{Minimal metrological usefulness}
\label{app:MetroUseful.min}

Here we show that
for a (nonconstant) generator function $f$
there exist at least two different values of the bound $b_f(k)$.

Indeed, if $f$ is not constant then we have
$f(\bot)\neq f(\top)$ by~\eqref{eq:frange},
and $f^\lesseqgtr\bigl(f(\top)\bigr)=\pinv{P}_\text{I}$
while $f^\lesseqgtr\bigl(f(\bot)\bigr)\not\ni\top$ by definition~\eqref{eq:fset.sslevel}.
Then in the two cases
\begin{subequations}
\begin{equation}
b_f\bigl(f(\top)\bigr)=\max_{\pinv{\xi}\in f^\lesseqgtr(f(\top))} s_2(\pinv{\xi})
= \max_{\pinv{\xi}\in\pinv{P}_\text{I}}s_2(\pinv{\xi})=s_2(\top),
\end{equation}
and
\begin{equation}
b_f\bigl(f(\bot)\bigr)=\max_{\pinv{\xi}\in f^\lesseqgtr(f(\bot))} s_2(\pinv{\xi})
<s_2(\top),
\end{equation}
\end{subequations}
where we used that
the maximum of $s_2$ is taken on the top uniquely, 
$s_2(\pinv{\xi})=n^2$ if and only if $\pinv{\xi}=\top$.

\bibliographystyle{quantum}
\bibliography{depths}{}

\end{document}